\documentclass[a4paper,11pt]{article}
\pdfoutput=1 

\usepackage{jheppub} 

\usepackage[T1]{fontenc} 
\usepackage{lineno}
\usepackage{overpic}
\usepackage{physics}
\usepackage{subfigure}
\usepackage{siunitx}
\usepackage{xspace}
\usepackage{relsize}
\usepackage{orcidlink}
\input{belle2-symbols}

\subheader{}

\def\NB{\ensuremath{\mathcal{C}}\xspace}

\title{Measurement of branching-fraction ratios and \CP asymmetries
  in $\Bpm \to \PD_{\CP\pm}\Kpm$ decays at Belle and Belle~II}

\collaboration{The Belle and Belle II Collaboration}
  \author{I.~Adachi\,\orcidlink{0000-0003-2287-0173},} 
  \author{L.~Aggarwal\,\orcidlink{0000-0002-0909-7537},} 
  \author{H.~Aihara\,\orcidlink{0000-0002-1907-5964},} 
  \author{N.~Akopov\,\orcidlink{0000-0002-4425-2096},} 
  \author{A.~Aloisio\,\orcidlink{0000-0002-3883-6693},} 
  \author{N.~Anh~Ky\,\orcidlink{0000-0003-0471-197X},} 
  \author{D.~M.~Asner\,\orcidlink{0000-0002-1586-5790},} 
  \author{H.~Atmacan\,\orcidlink{0000-0003-2435-501X},} 
  \author{T.~Aushev\,\orcidlink{0000-0002-6347-7055},} 
  \author{V.~Aushev\,\orcidlink{0000-0002-8588-5308},} 
% ...

\emailAdd{coll-publications@belle2.org}

\abstract{

  We report results from a study of $B^\pm \rightarrow DK^\pm$ decays followed by $D$ decaying to \CP~eigenstates, where $D$ indicates a $D^0$ or \Dzb\ meson.  
  These decays are sensitive to the Cabibbo-Kobayashi-Maskawa unitarity-triangle angle $\phi_{3}$.
  The results are based on a combined analysis of the final data set of $772 \times 10^6~B\bar{B}$ pairs collected by the Belle experiment and a data set of $198 \times 10^6~B\bar{B}$ pairs collected by the Belle~II experiment, both in electron-positron
  collisions at the $\Upsilon(4S)$ resonance. 
  We measure the \CP asymmetries to be $\mathcal{ A}_{\CP +} =~(+12.5 \pm 5.8 \pm 1.4)\% $ and $\mathcal{ A}_{\CP -} =~(-16.7 \pm 5.7 \pm 0.6)\%$,  
  and the ratios of branching fractions to be $\mathcal{ R}_{\CP+}=~1.164 \pm 0.081 \pm 0.036 $ and $\mathcal{ R}_{\CP-} =~1.151 \pm 0.074 \pm 0.019$. 
  The first contribution to the uncertainties is statistical, and the second is systematic. 
  The asymmetries $\mathcal{A}_{\CP +}$ and $\mathcal{A}_{\CP -}$ have similar magnitudes and opposite signs; their difference corresponds to 3.5~standard deviations. 
  From these values we calculate 68.3\% confidence intervals of ($8.5^{\circ}<\phi_{3}<16.5^{\circ}$) or ($84.5^{\circ}<\phi_{3}<95.5^{\circ}$) or ($163.3^{\circ}<\phi_{3}<171.5^{\circ}$) and $0.321<r_{B}<0.465$.
}

\keywords{$B$ physics, \CP-eigenstates, CKM angle $\phi_{3}$ ({$\gamma$}),
  $e^{+}e^{-} $ experiments}

\begin{document}

\vspace*{-5\baselineskip}
\begin{flushright}
Belle II Preprint 	2023-011		\\	
KEK Preprint 		2023-9 \\
\end{flushright}

\maketitle

\section{Introduction}

The Cabibbo-Kobayashi-Maskawa~(CKM) matrix parameterizes quark mixing in the standard model~\cite{Cabibbo,KM}. 
The angle $\phi_3$, also called $\gamma$, is the phase of a product of its elements
$-V_{\it ud} V^*_{\it ub}/ V_{\it cd}^* V_{\it cb}$.
Theoretical relationships connecting the angle $\phi_3$ with rates and \CP asymmetries of the decays $\Bpm\to\PD \Kpm$, where $D$ indicates a $D^0$ or \Dzb meson, are reliable and can be used for precise direct measurements of $\phi_3$.
Any inconsistency between direct measurements of $\phi_3$ and the value inferred from global CKM fits performed without this information would show that the CKM mechanism is not a complete description of \CP violation and reveal effects of physics beyond the standard model.
Gronau, London, and Wyler~(GLW) proposed a method to extract $\phi_{3}$ using decays 
in which the neutral \PD, $\PD_{\CP\pm}$, is reconstructed as a \CP~eigenstate~\cite{GLW1,GLW2}. 
We use this method to determine $\phi_{3}$ using combined data sets of the Belle and Belle~II experiments.

We measure \CP~asymmetries,
\begin{equation}
  \mathcal{A}_{\CP\pm} \equiv \frac{
    \mathcal{B}(\Bm \to \PD_{\CP\pm} \Km) - \mathcal{B}(\Bp \to \PD_{\CP\pm} \Kp)
  }{
    \mathcal{B}(\Bm \to \PD_{\CP\pm} \Km) + \mathcal{B}(\Bp \to \PD_{\CP\pm} \Kp)
  },
\end{equation}
and the ratio of branching fractions for decays in which the $D$ is reconstructed as a \CP eigenstate and decays in which the $D$ is reconstructed in a flavor-specific state:
\begin{equation}
  \mathcal{R}_{\CP\pm} \equiv \frac{
    \mathcal{B}(\Bm \to \PD_{\CP\pm} \Km) + \mathcal{B}(\Bp \to \PD_{\CP\pm} \Kp)
  }{
    (\mathcal{B}(\Bm \to \PD_{\rm flav} \Km) + \mathcal{B}(\Bp \to \Dbar_{\rm flav} \Kp))/2
  }.
\end{equation}
This ratio can be expressed as
\begin{equation}
  \mathcal{R}_{\CP\pm}
  \approx \frac{R_{\CP\pm}}{R_{\text{flav}}},
\label{eq:r_cp}
\end{equation}
where 
\begin{equation}
  R_{\CP\pm} \equiv \frac{
    \mathcal{B}(\Bm \to \PD_{\CP\pm} \Km) + \mathcal{B}(\Bp \to \PD_{\CP\pm} \Kp)
  }{
    \mathcal{B}(\Bm \to \PD_{\CP\pm} \pim) + \mathcal{B}(\Bp \to \PD_{\CP\pm} \pip)
  },
\end{equation}
and 
\begin{equation}
  R_{\rm flav} \equiv \frac{
    \mathcal{B}(\Bm \to \PD_{\rm flav} \Km) + \mathcal{B}(\Bp \to \Dbar_{\rm flav} \Kp)
  }{
    \mathcal{B}(\Bm \to \PD_{\rm flav} \pim) + \mathcal{B}(\Bp \to \Dbar_{\rm flav} \pip)
  }.
\end{equation}
In these ratios of branching fractions, most systematic uncertainties, such as those from reconstruction efficiencies and the known \PD branching fractions, cancel. 
The approximation in equation~\ref{eq:r_cp} is an equality if \CP is conserved in the $\Bpm \to \PD \pipm$ decay. 
Neglecting the small effects of \PD mixing and \CP violation in the \Dz decay~\cite{Rama:2013voa},
we relate $\mathcal{R}_{\CP\pm}$ and $\mathcal{A}_{\CP\pm}$ to $\phi_{3}$, the ratio $r_{B}$ of the magnitudes of the suppressed to favored $\Bpm \to D\Kpm$ amplitudes, and the relative interaction phase $\delta_B$ between them~\cite{PDG}:
\begin{alignat}{1}
  \mathcal{R}_{\CP\pm} &= 1 + r_{\PB}^2 \pm 2 r_{\PB} \cos\delta_{\PB} \cos\phi_{3},
  \label{eq:r_glw} \\
  \mathcal{A}_{\CP\pm} &= \pm2 r_{\PB} \sin\delta_{\PB} \sin\phi_{3} \, / \, \mathcal{R}_{\CP\pm}.
  \label{eq:a_glw}
\end{alignat}

The current precision on $\phi_{3}$ is about $3.5^{\circ}$~\cite{PDG, HFLAV}, dominated by recent measurements from the LHCb experiment~\cite{gammaLHCb}.
The Belle experiment reported a $\phi_{3}$-related measurement using the ADS
method~\cite{ADS1,ADS2} for $\Bpm \to \PD \Kpm$ decays with $\PD \to K^\pm \pi^\mp$ using its full data set~\cite{ADS_Belle}. A measurement using the
BPGGSZ method~\cite{GGSZ1,GGSZ2} for $\Bpm \to \PD h^{\pm}$
decays with $\PD \to \KS h^{\pm}h^{\mp}$, where $h$ is
a pion or kaon, using the full Belle data set and \SI{128}{fb^{-1}} of
data from Belle~II was reported recently~\cite{GGSZB2}.
However, Belle reported results using the GLW $\Bpm \to \PD^{(*)} \Kpm$ decays based {only a fraction of its data}~\cite{GLW_Belle}.
Here we report results based on the full Belle data set and also a fraction of the available data from Belle II.
These results supersede those of Ref.~\cite{GLW_Belle}.


\section{Data samples and detectors}
\label{sec:data}

We analyze samples containing \num{772e6} and \num{198e6}~$\BBbar$~pairs collected in
electron-positron collisions at the $\Upsilon(4S)$ resonance with
the Belle and Belle~II detectors.
The integrated luminosities of the corresponding data sets are \SI{711}{fb^{-1}} and \SI{189}{fb^{-1}} for Belle and Belle II. 
Belle operated at the KEKB
asymmetric-energy collider with electron- and positron-beam energies
of \SI{8}{GeV} and \SI{3.5}{GeV}~\cite{KEKB,KEKB_achievement}, respectively. 
Belle~II operates at its successor, SuperKEKB, designed to deliver
thirty times higher instantaneous luminosity than KEKB, with electron- and positron-beam energies of \SI{7}{GeV} and \SI{4}{GeV}~\cite{superKEKB}, respectively.

The Belle detector~\cite{Belle_detector,Belle_detector_achivement} was a large-solid-angle magnetic
spectrometer that consisted of a silicon vertex detector,
a 50-layer central drift chamber, an array of aerogel threshold Cherenkov
counters, a barrel-like arrangement of time-of-flight
scintillation counters, and an electromagnetic
calorimeter, all located within a superconducting solenoid coil
that provided a uniform \SI{1.5}{T} magnetic field collinear with the
beams. 
An iron flux-return yoke located outside the coil was
instrumented to detect \KL and muons.

The Belle~II detector~\cite{Belle2det} is an upgrade with several new
subdetectors 
designed to handle the significantly larger beam-related backgrounds of the new collider. 
It consists of a silicon vertex detector comprising two inner layers of pixel detectors and four outer layers of double-sided silicon strip detectors, a 56-layer central drift chamber, a time-of-propagation detector in the central detector volume and an aerogel ring-imaging Cherenkov detector in the forward region~(with respect to the electron-beam's direction) for charged particle identification~(PID), and an electromagnetic
calorimeter, all located inside the same solenoid as used for Belle. {A flux return outside the solenoid is instrumented with resistive-plate chambers, plastic scintillator modules, and an upgraded read-out system to detect muons, \KL mesons, and neutrons}.

We use simulated data to optimize selection criteria, determine detection efficiencies, train multivariate discriminants, identify sources of background, and obtain our fit models. 
The \textsc{EvtGen} software package is used to simulate the $\epem \to \Upsilon(4S) \to \BBbar$ process and our signal decays~\cite{EVTGEN}. 
The KKMC~\cite{KKMC} and Pythia~\cite{PYTHIA} generators are used to simulate the $\epem \to \qqbar$ continuum, where $\quark$ indicates a $ \uquark, \dquark, \squark $ or $\cquark$ quark. 
For Belle, the \textsc{Geant}3 package~\cite{GSIM} was used to model the detector response, whereas for Belle II the \textsc{Geant}4 package~\cite{GEANT4} is used. To account for final-state radiation, the \textsc{Photos} package~\cite{PHOTOS} is used.


\section{Reconstruction and candidate selection}
\label{sec:selection}

We use the Belle~II analysis software framework to reconstruct both Belle and Belle~II data~\cite{BASF2, basf2-zenodo, B2BII}.
Owing to the different performance of the detectors, separate sets of selection criteria are used for each data set. 

Online data-selection criteria are based on requirements of a minimum
number of charged particles and observed energy in an event. They are
fully efficient for signal and strongly suppress low-multiplicity
events. 
In the offline analysis, reconstructed charged-particle trajectories (tracks) are required to have distances
from the $\epem$ interaction point~(IP) smaller than
\SI{0.2}{cm} in the plane transverse to the beams, and smaller than
\SI{1.0}{cm} along the beam direction.
Charged kaon and pion candidates are identified based on information from PID detectors and the specific ionisation measured in the drift chamber. 
We use the ratio
%
%
$\mathcal{L}(K/\pi) \equiv {\mathcal{L}(K)}/{[\mathcal{L}(K) + \mathcal{L}(\pi)]}$
to identify the type of charged particles, where $\mathcal{L}(h)$
is the likelihood for a kaon or pion to produce the signals observed
in the detectors.  Charged particles with $\mathcal{L}(K/\pi) > 0.6$
are identified as kaons, and those with $\mathcal{L}(K/\pi) < 0.6$ as pions. 
To mitigate pion misidentification in the Belle~II data, we remove tracks with a polar angle $\theta > 120^{\circ}$,
since no PID detector covers this region~\cite{b2charm_conf}. 
No such veto is necessary for Belle data because the larger KEKB boost results in essentially all tracks being within the acceptance of the PID detectors.

We reconstruct \KS candidates in their $\pip\pim$ final state by forming each from a pair of oppositely charged particles (assuming they are pions) with a common vertex and mass in the range $[486, 509]$~\si{MeV}/$c^{2}$ for Belle data and $[491, 504]$~\si{MeV}/$c^{2}$ for Belle~II data. 
These ranges correspond to $3 \sigma$ in resolution in either direction from the known \Kz mass. 
To improve the purity of the \KS sample, we reject combinatorial background using neural networks for Belle data and boosted decision trees for Belle~II data~\cite{nakano_thesis, nakano_prd,GGSZB2}. 
Five input variables are common to the Belle and Belle~II discriminators: 
the angle between the \KS momentum and the direction from the IP to the \KS decay vertex;
the distance-of-closest-approach to the IP of the pion tracks; 
the flight distance of the \KS in the plane transverse to the beams; 
and the difference between the measured and known \KS masses divided by the uncertainty of the measured mass. 
The Belle discriminator uses seven additional variables, including the \KS momentum and the shortest distance between the two track helices projected along the beam direction~\cite{nakano_thesis, nakano_prd}. 
Each \KS momentum is recalculated from a fit of the pion momenta that constrains them to a common origin.

We reconstruct \piz candidates via their decays to two photons.
In Belle data, each photon is required to have an energy above \SI{50}{MeV}; in Belle~II data, each photon is required to have an energy above \SI{80}{MeV} if detected in the forward endcap, \SI{30}{MeV} if detected in the barrel, and \SI{60}{MeV} if in the backward endcap. 
{Each photon} must also be unassociated with any track and have an energy-deposition distribution in the calorimeter consistent with an electromagnetic shower. Each \piz candidate must have a mass in the range $[120, 145]$~\si{MeV}/$c^{2}$, corresponding to $2.5 \sigma$ in resolution on either side of the known \piz mass, and momentum above \SI{0.6}{GeV}/$c$. 
Each \piz momentum is recalculated from a fit of the photon momenta that constrains them to a common origin and the diphoton mass to the known mass of the \piz.

A \PD candidate is formed from combinations of $\Km$ and $\pip$, $\Kp$ and $\Km$, and $\KS$ and $\piz$ candidates.
The mass of each \PD candidate is required to be consistent with the known \PD mass~\cite{PDG} within $[-20, +20]$~\si{MeV}/$c^{2}$ in Belle data and $[-12, +12]$~\si{MeV}/$c^{2}$ in Belle~II data for $\PD \to \Kpm h^\mp$ decays; and within $[-64, +47]$~\si{MeV}/$c^{2}$ in Belle data and $[-53, +36]$~\si{MeV}/$c^{2}$ in Belle~II data for the $\PD \to \KS \piz$ decays. 
These ranges are approximately $3\sigma$ in resolution on either side of the known mass. 
Each \PD momentum is recalculated from a fit of the momenta of its decay products that constrains them to a common origin and their invariant mass to the known mass of the~\PD meson.
 
A \PB candidate is formed from a \PD candidate and an $h^\pm$ candidate. 
To select signal candidates, we use the beam-energy-constrained mass,
\begin{equation}
M_{\text{bc}} \equiv c^{-2} \sqrt{ E_{\text{beam}}^{*2} - |\vec{p}_{\PB}c|^2},
\end{equation}
and the energy difference,
$\Delta E \equiv E_{\PB} - E_{\text{beam}}^{*}$,
calculated from the \PB energy $E_{\PB}$, momentum $\vec{p}_{\PB}$, and 
beam energy $E_{\text{beam}}^{*}$, all in the center-of-mass~(c.m.)\ frame. 
We require $M_{\text{bc}}$ to be in the range $[5.27, 5.29]$~\si{GeV}/$c^{2}$, which is $\pm 3\sigma$ in resolution around the known \PB mass~\cite{PDG}. 
We require $\Delta E$ to be in the range $[-0.13, 0.15]$~\si{GeV} to suppress partially reconstructed $\Bpm \to \Dstar h^\pm$ decays, which have negative $\Delta E$.

Most remaining backgrounds arise from continuum events, in which final-state particles are highly boosted into two jets that are approximately back-to-back in the c.m.\ frame. 
Since $\BBbar$ pairs are produced slightly {above kinematic threshold}, their final-state particles are isotropically distributed in the c.m.\ frame.
We use boosted decision trees~(BDTs) to suppress candidates from continuum events.
We train them on equal numbers of simulated signal and continuum events using variables that are uncorrelated with $\Delta E$.  
Those variables are well simulated, as verified by inspection of the flavor-specific channel.
The variables used are modified Fox-Wolfram moments~\protect\cite{KSFW1,KSFW2};
the cosine of the polar angle of the \PB momentum in the c.m.\ frame;
the absolute value of the cosine of the angle between the thrust axis of the \PB and the thrust axis of the rest of the charged particles and photons in the event~(ROE);
the longitudinal distance between the \PB vertex and the ROE vertex;
and the output of a \PB-flavor-tagging algorithm~\cite{flavor_tag1,flavor_tag2}. 
The thrust axis of a group of particles is the direction that maximizes the sum of the projections of the particle momenta onto it.
The BDT classifier output, \NB, is in the range $[0, 1]$, peaking at zero for
continuum background and at one for signal. 
We require $\NB > 0.15$, which retains 95\% of signal in Belle {data} and 97\% in Belle~II {data}, while rejecting 60\% and 63\% of background, respectively.

To suppress \PD decays from $\Dstar \to \PD \Ppi$ arising from $\epem \to c\overline{c} \to \PD^{*}\PD^{(*)}X$ processes, we use the observed difference between the mass of the \PD candidate and the mass of the \Dstar candidates reconstructed by associating to the \PD any \pipm or \piz in the ROE. 
We require that {the differences} all be outside $\pm 3\sigma$ in resolution from the known \Dstar-\PD mass difference~\cite{PDG};
the excluded ranges are [143.4, 147.5]~\si{MeV}/$c^{2}$ and [143.8, 147.0]~\si{MeV}/$c^{2}$ in Belle and Belle~II {data} for $\Dstarpm$, respectively, and [140.0, 145.0]~\si{MeV}/$c^{2}$ in both experiments for $\Dstarz$.
This retains 97\% of signal candidates and rejects 13\% of background
candidates in Belle {data} and 18\% in Belle~II {data}.
For $\Bpm \to \PD( \to \Kpm \pimp)\pipm$, we require that the dipion mass not be in the range $[3.08, 3.14]$~\si{GeV}/$c^{2}$ to veto candidates reconstructed from $\Bpm \to \jpsi (\to \ellell) \Kpm$ {decays} in which both leptons are misidentified. 
The $\Delta E$ distribution of such events peaks in the signal region.

In events with multiple \PB candidates, 2\% of events for the \CP-even mode and 7\% for the \CP-odd mode, we retain the candidate with the smallest $\chi^{2}$ calculated from the reconstructed \PD mass, $M_{\text{bc}}$ and their resolutions; for decays with $\PD_{\CP-}$, the reconstructed \piz mass and its resolution are also used in the $\chi^{2}$ calculation. 
This selects the correct signal candidate in 70\%--80\% of such events in simulation.

\section{Fits to data}
\label{sec:extraction}

{The final event sample consists of signal, cross-feed background that comes from mis-identifying the $h^\pm$ of a signal event, other $\BBbar$ background sources, and continuum background.} 
To determine the {numbers} of signal candidates, we fit to the distributions of $\Delta E$ and \NB, the variables that best discriminate between signal and the remaining background. 
To make \NB easier to model, we transform it to a new variable $\NB'$, such that signal is distributed uniformly in~$[0, 1]$ and background is exponentially distributed~\cite{mu_trans}. 
We perform an unbinned extended maximum-likelihood fit to candidates with $\Delta E \in [-0.13, 0.14]~\si{GeV}$ and $\NB'$ in its full range.

Simulation shows that $\Delta E$--$\NB'$ correlations in the distributions of candidates from all sample components are negligible, and thus we factorize the two-dimensional probability density function~(PDF) for each component in the fit. 
For each decay mode, we divide the data into 12 subsets defined by the product of the two possible electric charges of the \PB, the three possible \PD final states (two \CP-specific and one flavor-specific), and whether $h^\pm$ is identified as a kaon or pion.
The {fit models} are mostly common in all decay modes and data subsets, but the shape parameters are different in each.

For signal, the $\Delta E$ PDF is the sum of two Gaussian functions
and an asymmetric Gaussian function, with all parameters fixed from
simulated data except for the common mean of all three
$\PD$ decay modes and a common multiplier for all signal widths.
These parameters are determined by the fit and account for differences in resolution between the experimental and simulated data.
The $\NB'$ PDF is a straight line with {its slope} fixed to the value fitted
in the simulated data,
except for the PDF used for the Belle
$\PD\Ppi$ data, in which the slope is a free parameter.

The cross-feed $\Delta E$ PDF is same to the signal one, but with its own set of parameters. 
When determining the fixed parameters of the cross-feed PDF for $\PD\Ppi$, the simulated data are corrected for momentum-dependent differences in particle misidentification rates between the experimental and simulated data. 
The $\NB'$ PDF is the sum of a straight line and an exponential function, with parameters fixed from the simulated samples.

For the $\BBbar$ background component, the $\Delta E$ PDF is the sum of an exponential function and a uniform distribution for the \CP modes, and the sum of an exponential function and a Novosibirsk function~\cite{novo} for the flavor-specific mode. 
The $\NB'$ PDF is a straight line whose slope is fixed from simulated data. 
A peaking background originates from events in which a $B$ decays directly to $Xh$ without the production of an intermediate charmed meson in the decay chain.
This background is estimated from the \PD mass sidebands in data,
We find no peaking structure in the sideband of the $\PD_{\CP-}$ mode, while for the $\PD_{\CP+}$ mode we see a significant peaking structure and estimate its yield in the signal $\Delta E$ region to be $132\pm17$ events {in Belle data} and $24\pm 4$ {in Belle II data}.
These yields are estimated by linearly extrapolating the results obtained in eight \PD mass sidebands in data, as discussed in Appendix~\ref{appendixB}.
In the final fit, the PDF shape of peaking background is fixed from a simulated sample.

For the continuum component, the $\Delta E$ PDF is a straight line and the $\NB'$ PDF is the sum of two exponential functions. The larger {exponential} component has its parameter fixed to the value fit from simulated data, and the other free to vary, which accounts for differences between the distributions in experimental and simulated data.

We perform a simultaneous fit to all decay modes in both the Belle and Belle~II data, to determine the six charge asymmetries and three branching-fraction ratios. 
The yields of $\Bpm \to \PD h^{\pm}$ with the \PD decaying to the state $X$ and the charged hadron identified as $h'^\pm$,
denoted as $Y_{h'}(\Bpm \to \PD_{\!X}h^\pm)$, are related to these physical observables via
\begin{alignat}{1}
\label{eq:yields} 
Y_{\pi}(\Bpm \to \PD_{\!X}\Kpm)  &= \tfrac{1}{2} [1 \mp \mathcal{A}(\PB \to \PD_{\!X}\PK)]  \, N(\PB \to \PD_{\!X}\Ppi)  \, R_X \, \delta \, (1 - \varepsilon_\pm) ,\\
Y_{K}(\Bpm \to \PD_{\!X}\Kpm)  &= \tfrac{1}{2} [1 \mp \mathcal{A}(\PB \to \PD_{\!X}\PK)]  \, N(\PB \to \PD_{\!X}\Ppi)  \, R_X \, \delta \, \varepsilon_\pm  ,\\
Y_{\pi}(\Bpm \to \PD_{\!X}\pipm) &= \tfrac{1}{2} [1 \mp \mathcal{A}(\PB \to \PD_{\!X}\Ppi)] \, N(\PB \to \PD_{\!X}\Ppi) \, (1-\kappa_\pm)  , \\
Y_{K}(\Bpm \to \PD_{\!X}\pipm) &= \tfrac{1}{2} [1 \mp \mathcal{A}(\PB \to \PD_{\!X}\Ppi)] \, N(\PB \to \PD_{\!X}\Ppi) \, \kappa_\pm ,
\end{alignat}
where $\mathcal{A}$ is the charge asymmetry,
$N$ is the total number of events regardless of how the charged
hadron was identified and of its sign, $\varepsilon_\pm$ is the
efficiency to identify a kaon with $\pm 1$ charge, and $\kappa_\pm$ is the rate for misidentifying a pion as a kaon with $\pm 1$ charge. 
The efficiency $\delta$ for reconstructing $\Bpm \to \PD\Kpm$ relative to that for $\Bpm \to \PD\pipm$ is independent of the $D$ final state and equals 0.975 in Belle data and 1.000 in Belle~II data.
We measure PID efficiencies and misidentification rates using control samples.  
For Belle, we measure $\kappa_+ = 7.7\%$, $\kappa_- = 8.2\%$, $\varepsilon_+ = 83.4\%$, and $\varepsilon_- = 84.3\%$~\cite{belle_pid}. For Belle~II, we measure $\kappa_+ = 7.2\%$, $\kappa_- = 8.7\%$, $\varepsilon_+ = 79.6\%$, and $\varepsilon_- = 78.9\%$.
Uncertainties on those values are typically 0.5\%.

The signal yields $N$ are independent for the Belle and Belle II data.
For each background component, separate yields are
fitted for \Bp and \Bm to account for their possible charge
asymmetries.

To check for fit biases, we perform the fit on five independent sets of simulated data. We also repeat the analysis on 1000 data sets simulated according to the fit model for seven different values of $\mathcal{A}_{\CP\pm}$: 0, $\pm 0.1$, $\pm 0.2$, $\pm 0.3$. 
In all cases, the fit results are consistent with the input values.

\section{Systematic uncertainties}
\label{sec:sys_uncer}

We consider several sources of systematic uncertainties, which are summarized in Table~\ref{tab:syst_glw}.
In general, for parameters fixed in the fits, we repeat the fits with the parameter varied by its uncertainty and take the resulting change in our results as the {fit-model} systematic uncertainty. 
We do this for the fixed PDF parameters, PID efficiencies and mis-identification rates, peaking background yields, and the efficiency ratio. 
We ignore correlations between those uncertainties and combine them by adding them in quadrature.
{
We also assign systematic uncertainties ({included in the "PDF parameters" item} of Table~\ref{tab:syst_glw}) from the difference between correcting and not correcting for the momentum-dependent pion misidentification rates when modeling the cross-feed PDF for the $\PD\Ppi$ data, and having common and independent mode parameters for the $\Delta E$ PDF's for $\PD\PK$ and $\PD\Ppi$.
}
We use a common mean for the signal $\Delta E$ PDFs for all modes. 
The corresponding systematic uncertainty ({"Signal-$\Delta E$ common mean" item of} Table~\ref{tab:syst_glw}) is estimated from the variations resulting from assigning different mean values to the $\Delta E$ PDFs,
i.e., $\PB \to \PD_{\!X}\PK $ and $\PB \to \PD_{\!X}\Ppi $ with the same or independent means, $\Bm$ and $\Bp$ with the same or independent means.
{For the slopes of the $\NB'$ PDF of the $\BBbar$ component}, we calculate the systematic uncertainty from the maximum difference among fit results in which the slope is taken from simulation, as in the nominal fit, or taken from the signal $\NB'$ PDF's slope in data, or determined by the fit itself.

\begin{table}[!h]

  \caption{Systematic and statistical uncertainties.}

  \label{tab:syst_glw}

  \begin{center}

    \sisetup{round-mode=places,round-precision=3}
    
    \begin{tabular}{l*{4}{S}}

\hline
      
      & {$\mathcal{R}_{\CP+}$}
      & {$\mathcal{R}_{\CP-}$}
      & {$\mathcal{A}_{\CP+}$}
      & {$\mathcal{A}_{\CP-}$}
      \\
\hline
      PDF parameters
      & 0.012
      & 0.014
      & 0.002
      & 0.002
      \\

      PID parameters
      & 0.009
      & 0.010
      & 0.003
      & 0.005
      \\

      Peaking background yields
      & 0.033
      & 0.002
      & 0.013
      & {---}
      \\

      Efficiency ratio
      & 0.001
      & 0.001
      & <0.001 
      & <0.001 
      \\
      
      {Signal-$\Delta E$ common mean}
      & 0.005
      & 0.006
      & <0.001 
      & <0.001 
      \\
      
\hline
      Total systematic uncertainty
      & 0.036
      & 0.019
      & 0.014
      & 0.006
      \\
\hline
      Statistical uncertainty
      & 0.081
      & 0.074
      & 0.058
      & 0.057
      \\
\hline
    \end{tabular}
    
  \end{center}
  
\end{table}

\section{Results}
\label{sec:combination}

Figures~\ref{fig:Kpi_B}, \ref{fig:KK_B}, and~\ref{fig:K0pi0_B} show distributions and the fit results for candidates satisfying $|\Delta E|<0.05~\rm GeV$ and $0.65<\NB'<1.0$ for Belle data; figures~\ref{fig:Kpi_B2},
\ref{fig:KK_B2}, and~\ref{fig:K0pi0_B2} {show the corresponding plots for Belle II data}. 
The fit results agree with the data; the small shifts seen for $\PB \to \PD_{\!X}\Ppi $ signal in $\Delta E$ are accounted for in the systematic uncertainty estimation.
Table~\ref{tab:signal_yields} summarizes the signal yields.

\begin{table}[!t]
  
  \caption{Signal yields extracted from the simultaneous fit in data.}
  
  \label{tab:signal_yields}
  
  \begin{center}
    
    \begin{tabular}{llcc}

\hline
        $\PD_{\!X}$  mode
      &
      & {$N(\PB \to \PD_{\!X}\PK)$}
      & {$N(\PB \to \PD_{\!X}\Ppi)$}
      \\

\hline      
      
      $\PD \to \Kpm \pimp$
      & Belle
      & $4238 \pm 94$
      & $59481 \pm 267$
      \\

      & Belle~II
      & $1084 \pm 44$
      & $14229 \pm 126$
      \\
\hline      
      
      $\PD \to \Kp\Km$
      & Belle
      & $476 \pm 36$ 
      & $5559 \pm 85$
      \\

      & Belle~II
      & $107 \pm 15$ 
      & $1336 \pm 40$ 
      \\
\hline      
      
      $\PD \to \KS\piz$
      & Belle
      & $541 \pm 42$ 
      & $6484 \pm 95$
      \\

      & Belle~II
      & $145 \pm 16$ 
      & $1763 \pm 46$ 
      \\
\hline      
      
    \end{tabular}
    
  \end{center}
  
\end{table}

From the combined Belle and Belle~II data, the ratios of branching-fraction ratios and \CP asymmetries of $\Bpm \to \PD_{\CP} \Kpm$ are
\begin{alignat}{1}
  \mathcal{R}_{\CP+} &= 1.164 \pm 0.081 \pm 0.036, \label{eq:rcpp_exp} \\
  \mathcal{R}_{\CP-} &= 1.151 \pm 0.074 \pm 0.019, \label{eq:rcpm_exp} \\
  \mathcal{A}_{\CP+} &=(+12.5 \pm 5.8 \pm 1.4)\%,  \label{eq:acpp_exp} \\
  \mathcal{A}_{\CP-} &=(-16.7 \pm 5.7 \pm 0.6)\%,  \label{eq:acpm_exp}
\end{alignat}
where the first uncertainty is statistical and the second is systematic.

The significances of \CP violation for \CP-even and \CP-odd \PD final states are approximated using $\sqrt{-2\ln(\mathcal{L}_{0}/\mathcal{L}_{\text{max}})} \, \sigma_{\text{stat}} / \sqrt{\sigma^{2}_{\text{stat}} + \sigma^{2}_{\text{syst}}}$, where $\mathcal{L}_{\text{max}}$ is the maximum likelihood value, $\mathcal{L}_0$ is the likelihood value obtained assuming \CP symmetry, and $\sigma$ are the statistical and systematic uncertainties. 
{We found $2.0\sigma$ and $2.8\sigma$ significances for \CP violation in the $\PD_{\CP+}$ and $\PD_{\CP-}$ modes, respectively}. 
{This corresponds to $3.5\sigma$ evidence for the asymmetries being different,} i.e., $\mathcal{A}_{\CP+} \neq \mathcal{A}_{\CP-}$.
The measured $\mathcal{R}_{\CP+}$ {value} is $2.2\sigma$ away from its expectation as estimated
from the world-average values~\cite{PDG, HFLAV} of $\phi_{3}$, $r_{\PB}$, and $\delta_{\PB}$, while the measured $\mathcal{R}_{\CP-}$ {value} agrees well with its expected value.
An underestimation of the peaking-background yield for $D_{\CP+}K$ could be a possible explanation, but this estimation is carefully done using {eight} different sidebands in data as described in Section~\ref{sec:extraction}.
{Fit bias is also excluded here; we examine data from both realistic simulation and simulation based on the fit model and find no fit bias} (Section~\ref{sec:extraction}).  
The \CP asymmetries of $\Bpm \to \PD_{\!X} \pipm$ and $\Bpm \to \PD_{\text{flav}} \Kpm$ are
$\mathcal{A}_{\CP+}^{\Ppi} = (-2.0 \pm 1.4 \pm 0.2)\%$,
$\mathcal{A}_{\CP-}^{\Ppi} = (-0.3 \pm 1.2 \pm 0.2)\%$, 
$\mathcal{A}_{\text{flav}}^{\Ppi} =(-0.5 \pm 0.4 \pm 0.2)\%$, and $\mathcal{A}_{\text{flav}}^{\PK} =(-1.4 \pm 1.7 \pm 0.1)\%$, 
{consistent with the expected \CP symmetry in these modes.}

\newcommand{\ResultFigureCaption}[2]{Distributions of $\Delta E$ and $\NB'$ for $\Bpm \to \PD (\to #1) h^\pm$ candidates in the #2 data with fit projections overlaid. Differences between data and fit results normalized by the uncertainty in data are shown in the bottom panels.}
\begin{figure}[!t]
  \begin{center}
    
    \begin{overpic}[width=0.46\textwidth]{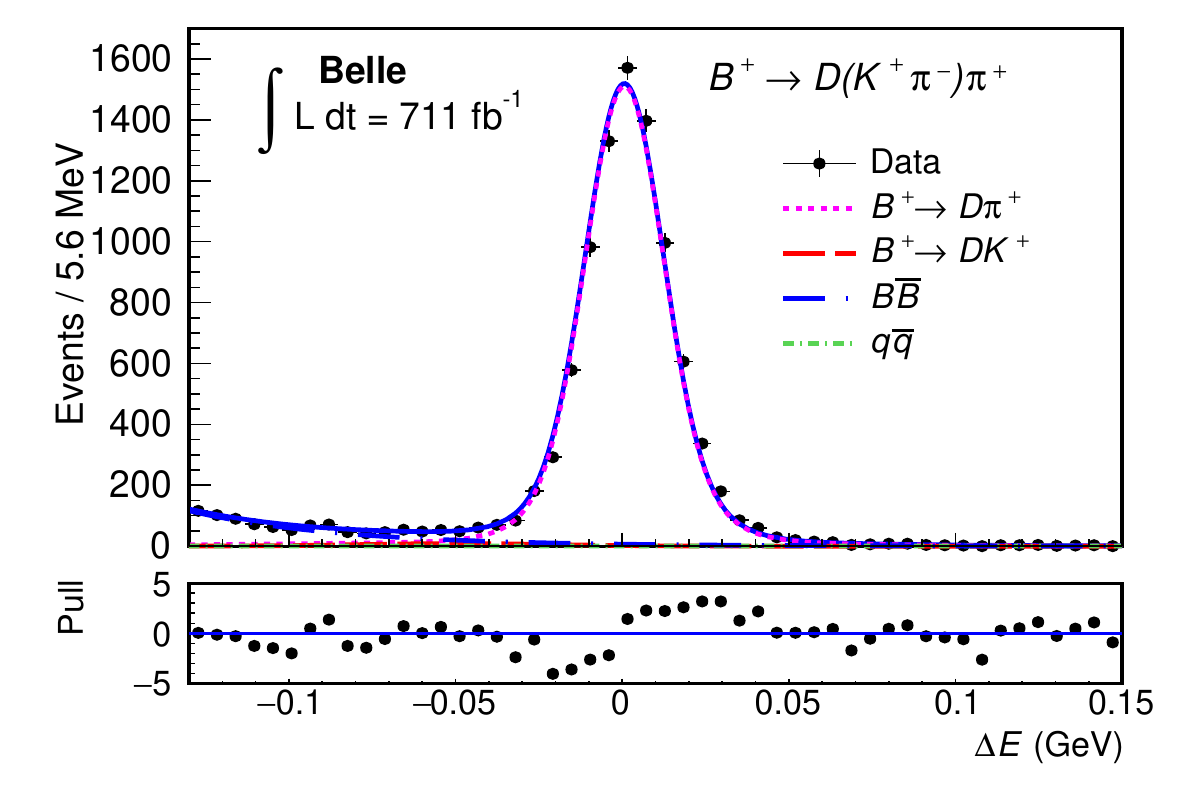}  \put(87,60){(a)} \end{overpic}
    \begin{overpic}[width=0.46\textwidth]{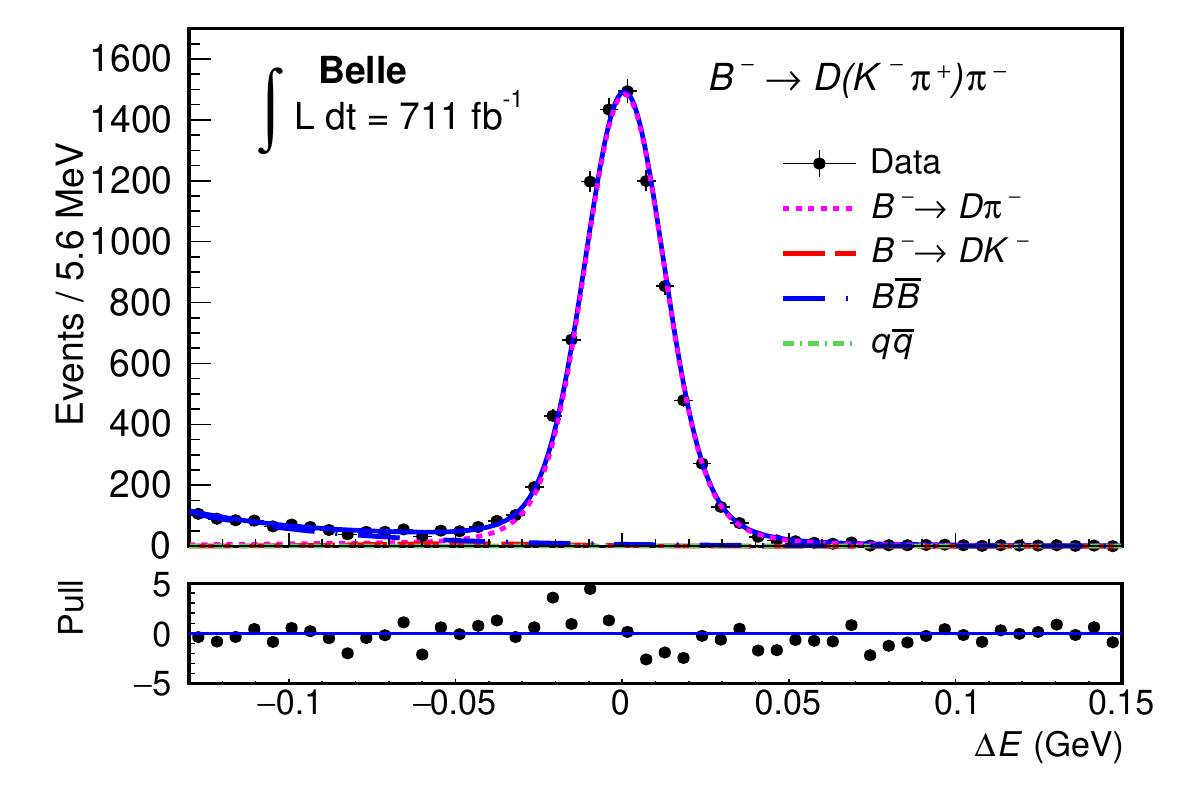} \put(87,60){(b)} \end{overpic}
    
    \begin{overpic}[width=0.46\textwidth]{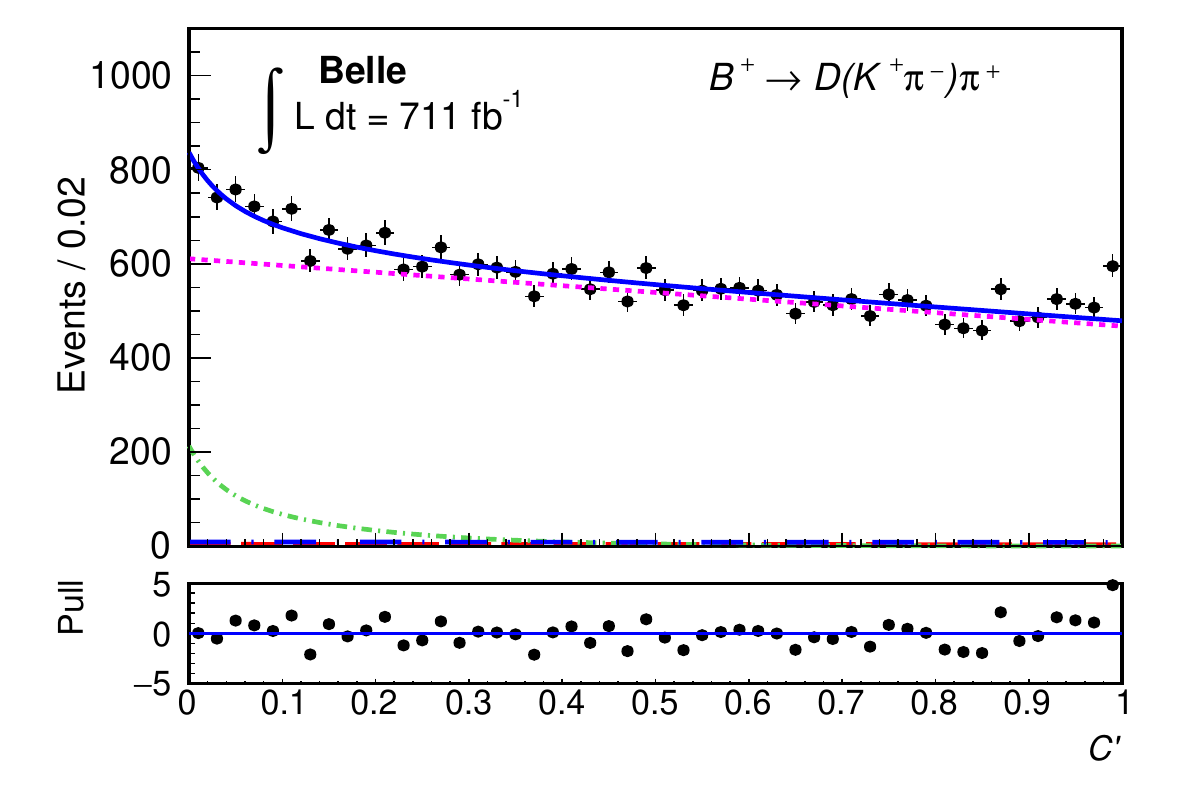}  \put(87,60){(c)} \end{overpic}
    \begin{overpic}[width=0.46\textwidth]{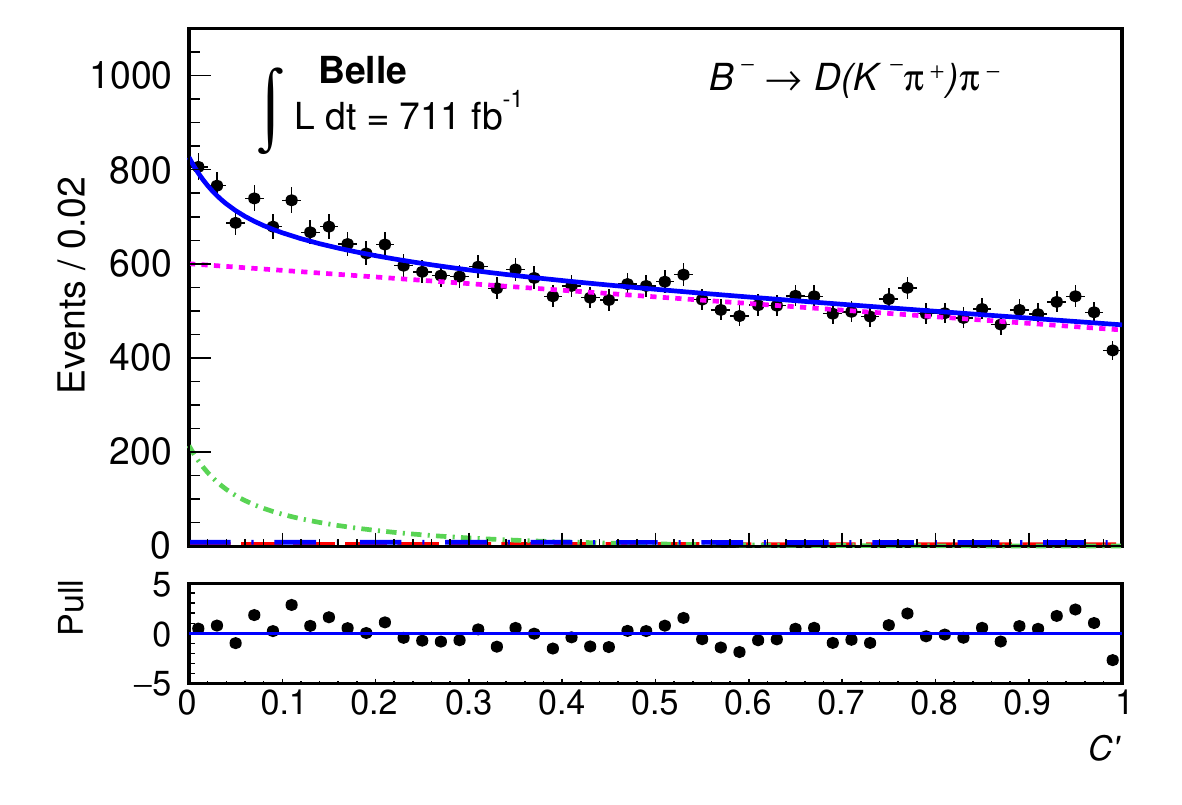} \put(87,60){(d)} \end{overpic}
    
    \begin{overpic}[width=0.46\textwidth]{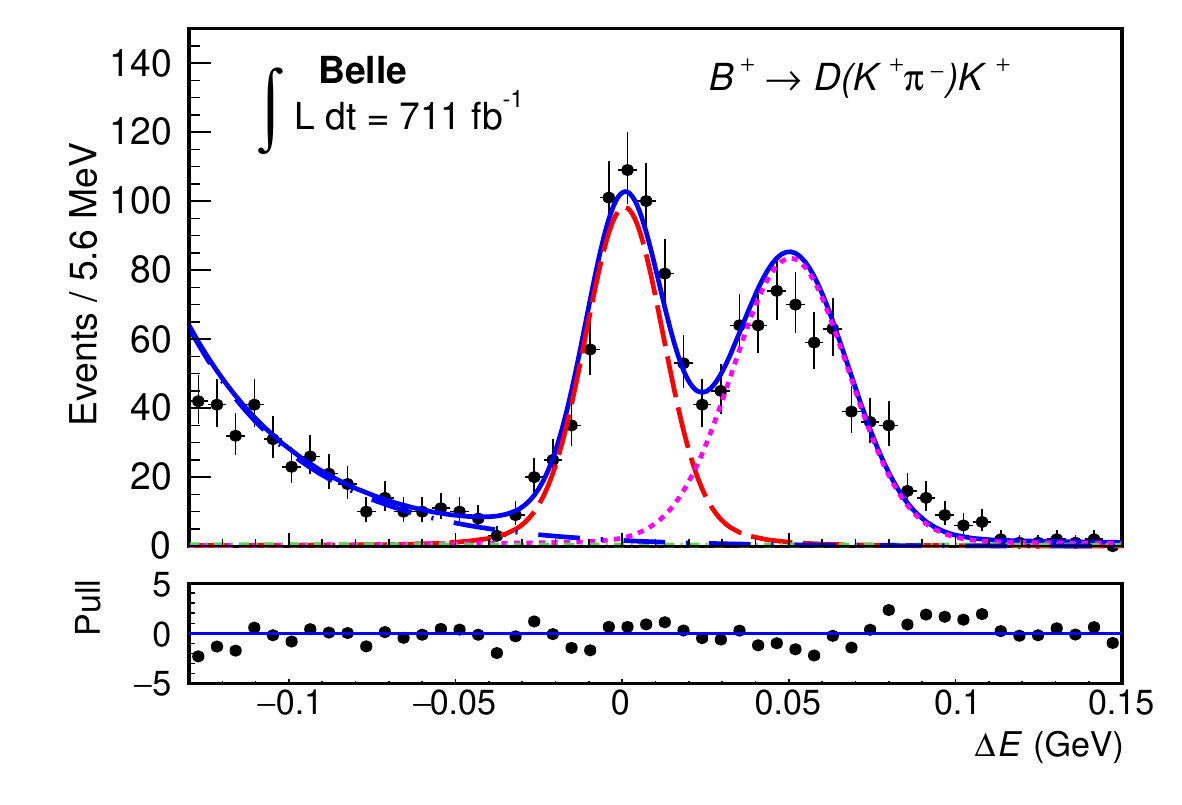}   \put(87,60){(e)} \end{overpic}
    \begin{overpic}[width=0.46\textwidth]{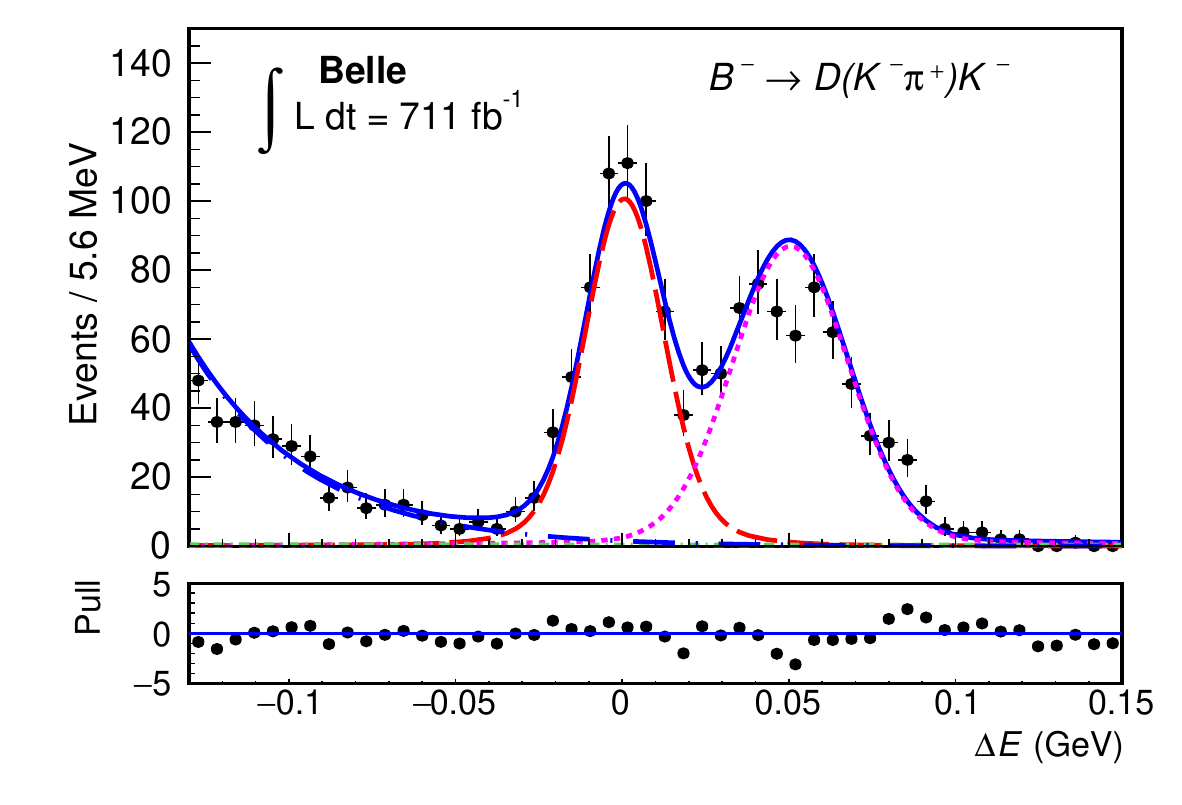}  \put(87,60){(f)} \end{overpic}
    
    \begin{overpic}[width=0.46\textwidth]{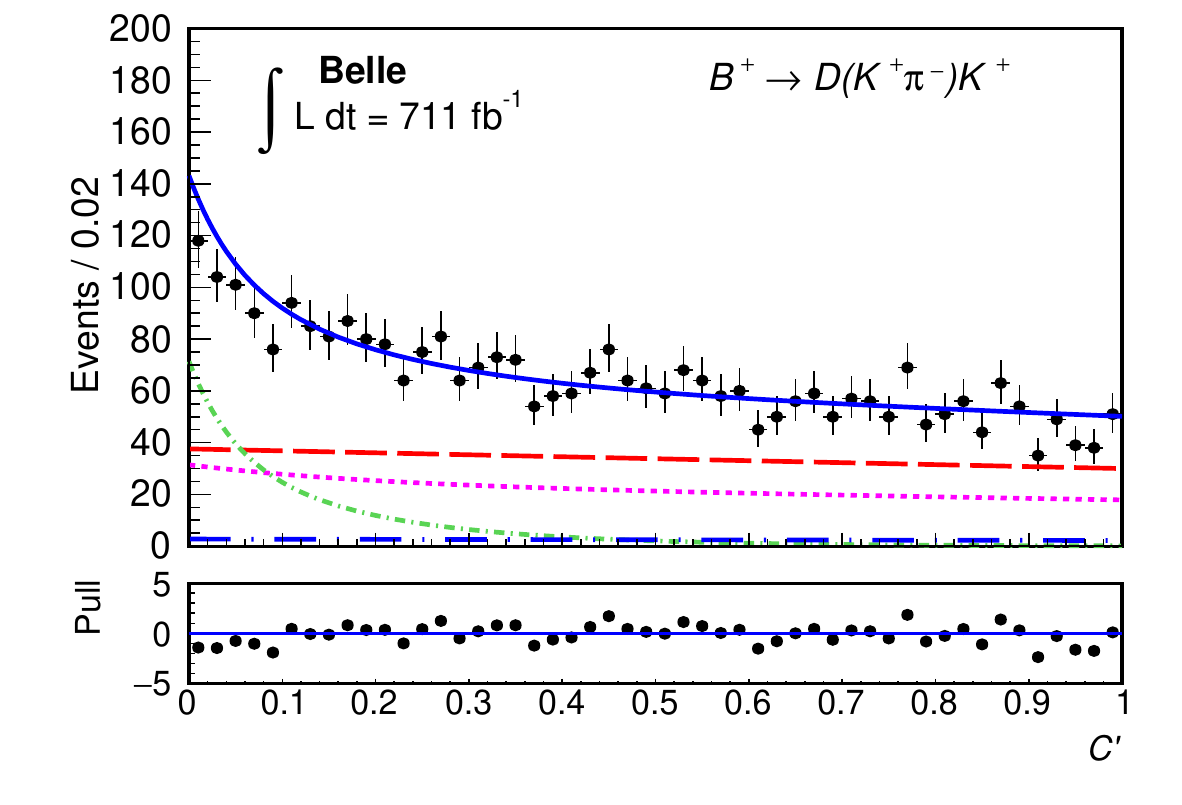}   \put(87,60){(g)} \end{overpic}
    \begin{overpic}[width=0.46\textwidth]{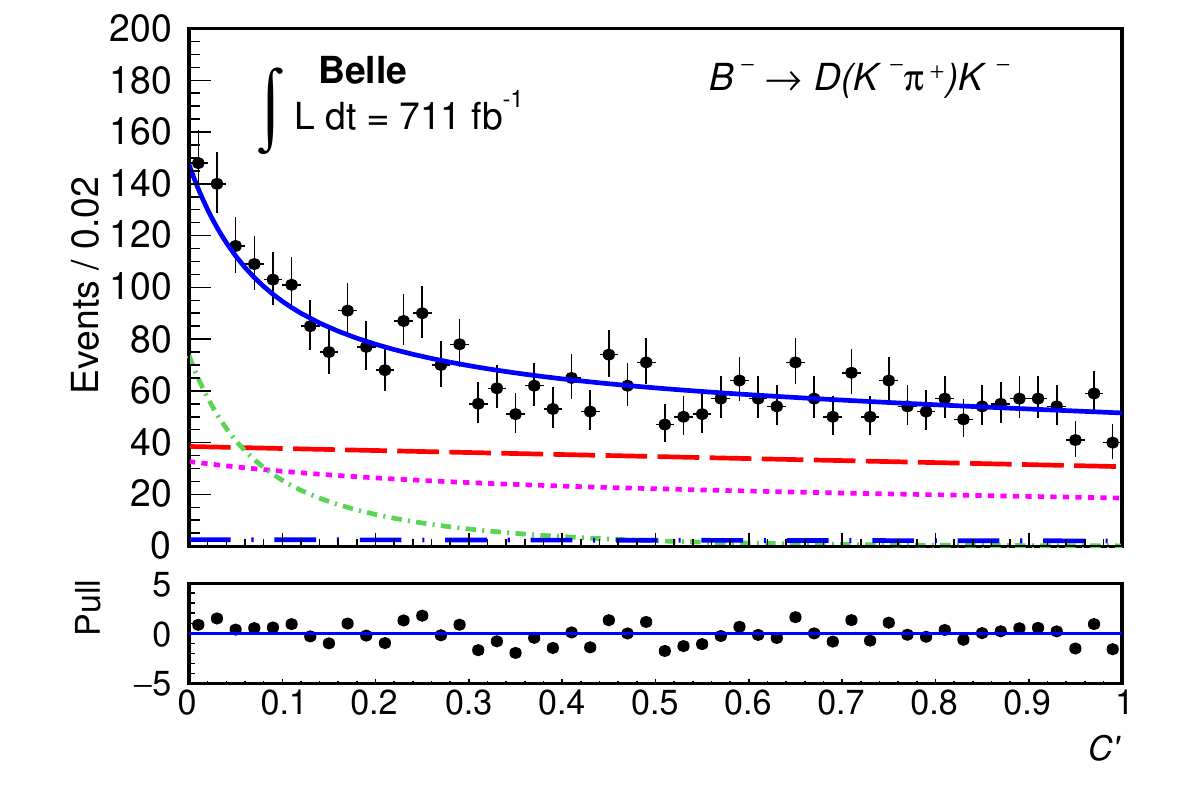}  \put(87,60){(h)} \end{overpic}
    
  \end{center}

  \caption{\ResultFigureCaption{\Kpm\pimp}{Belle}}

  \label{fig:Kpi_B}
\end{figure}

\begin{figure}[!t]
  \begin{center}

    \begin{overpic}[width=0.46\textwidth]{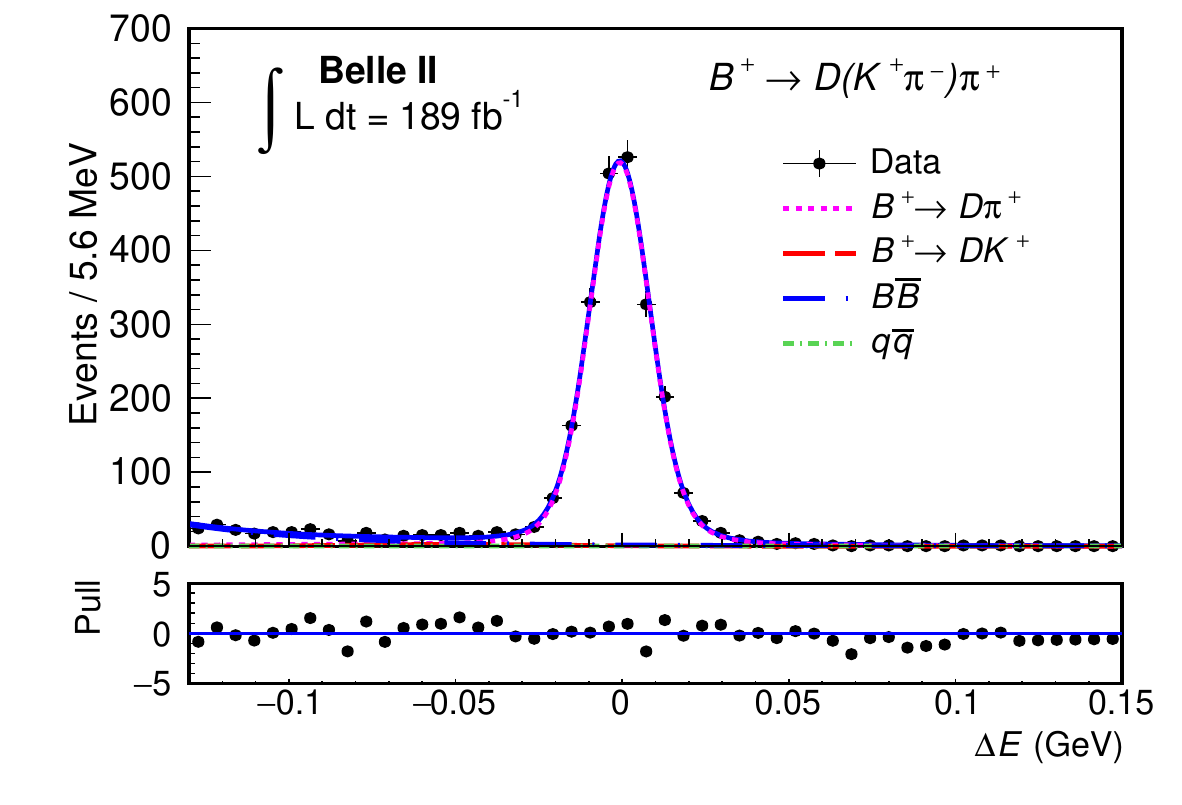}  \put(87,60){(a)} \end{overpic}
    \begin{overpic}[width=0.46\textwidth]{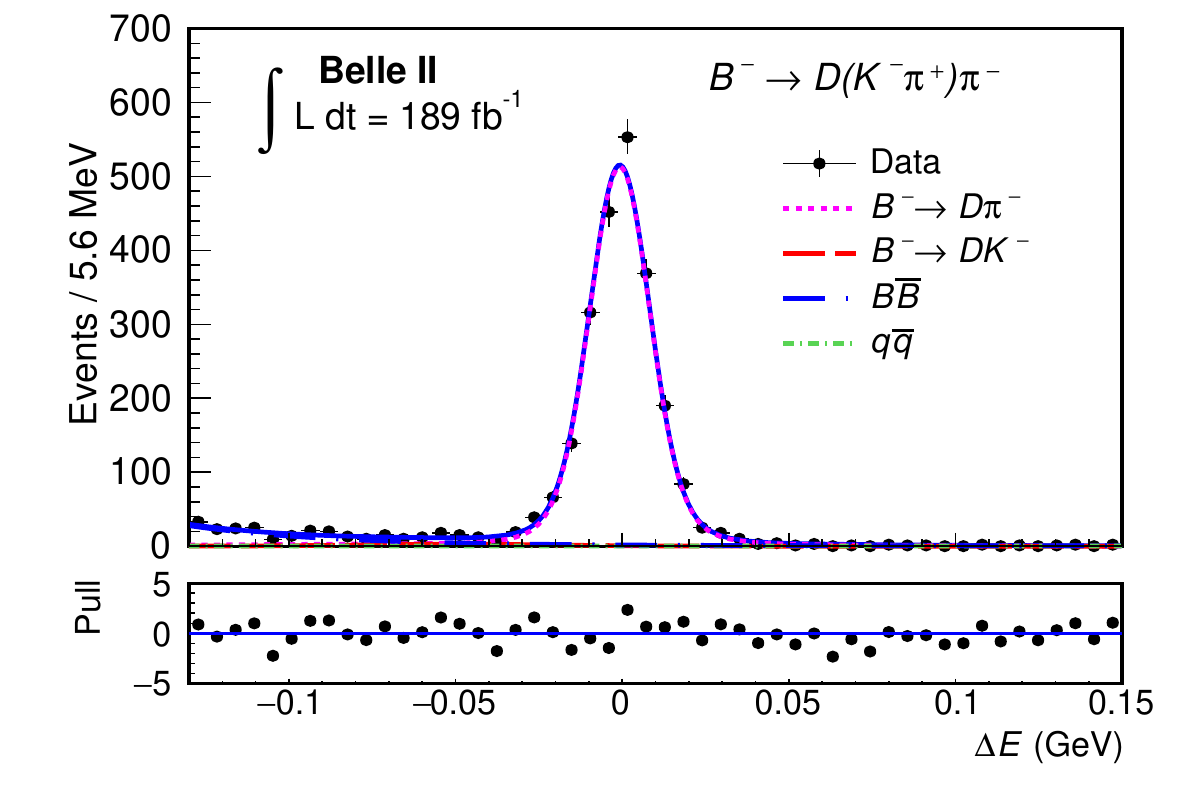} \put(87,60){(b)} \end{overpic}
    
    \begin{overpic}[width=0.46\textwidth]{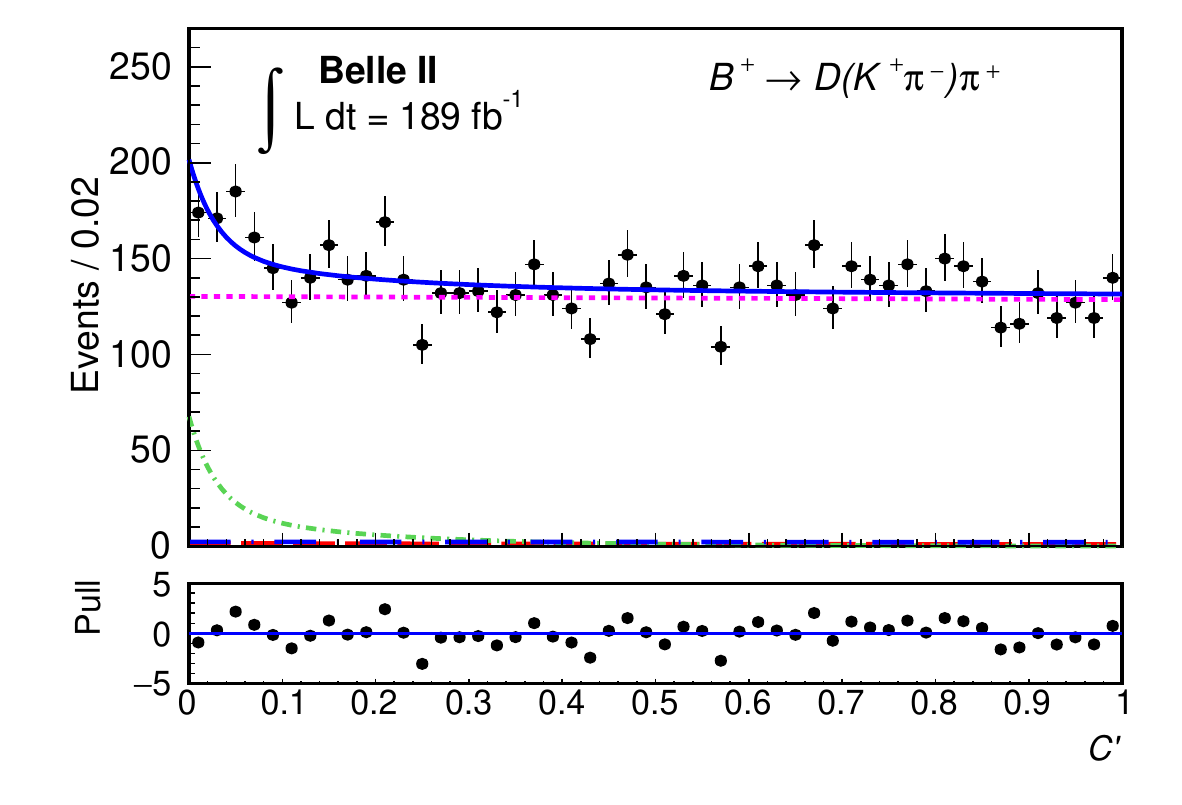}  \put(87,60){(c)} \end{overpic}
    \begin{overpic}[width=0.46\textwidth]{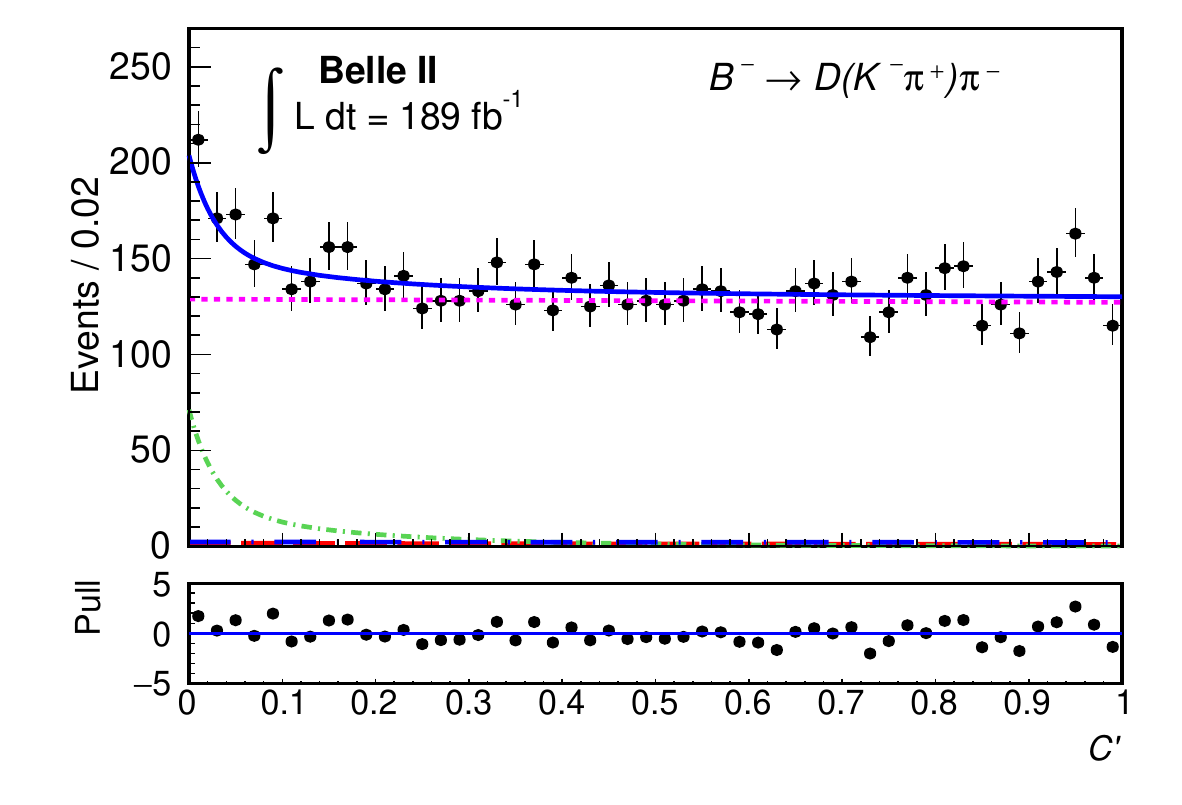} \put(87,60){(d)} \end{overpic}
    
    \begin{overpic}[width=0.46\textwidth]{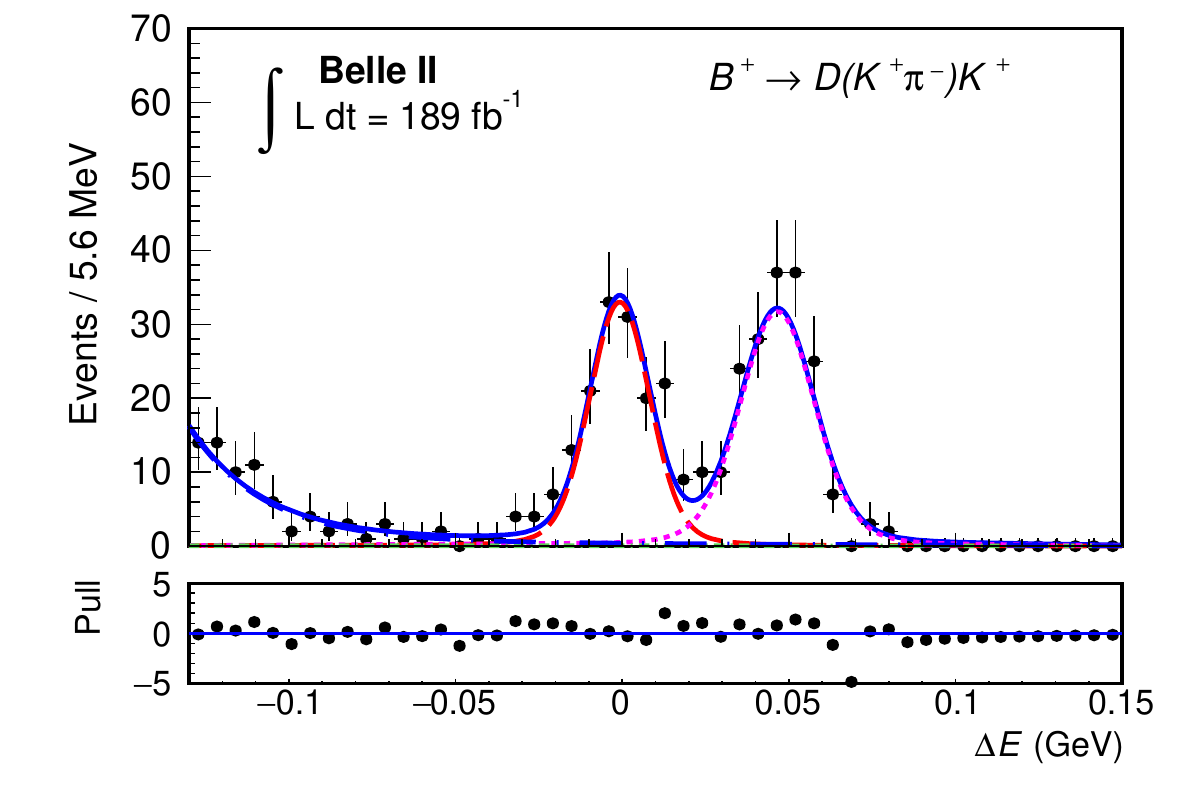}   \put(87,60){(e)} \end{overpic}
    \begin{overpic}[width=0.46\textwidth]{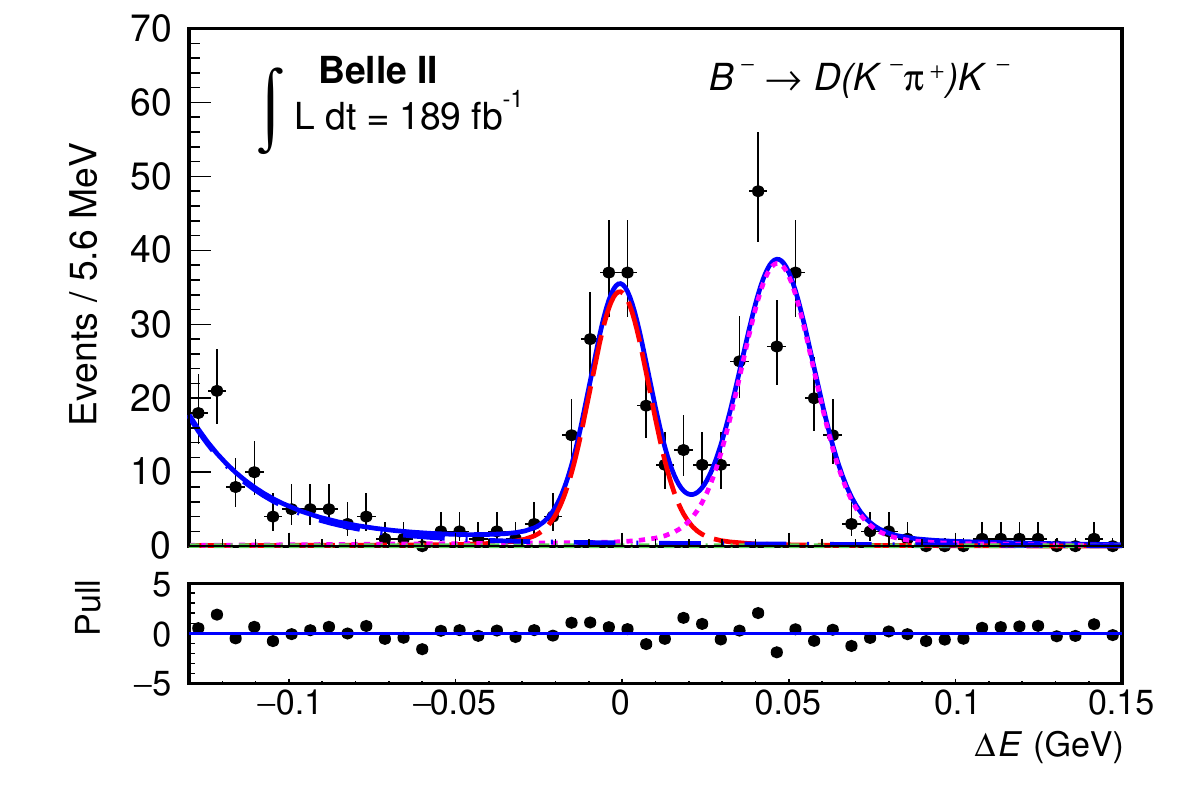}  \put(87,60){(f)} \end{overpic}
    
    \begin{overpic}[width=0.46\textwidth]{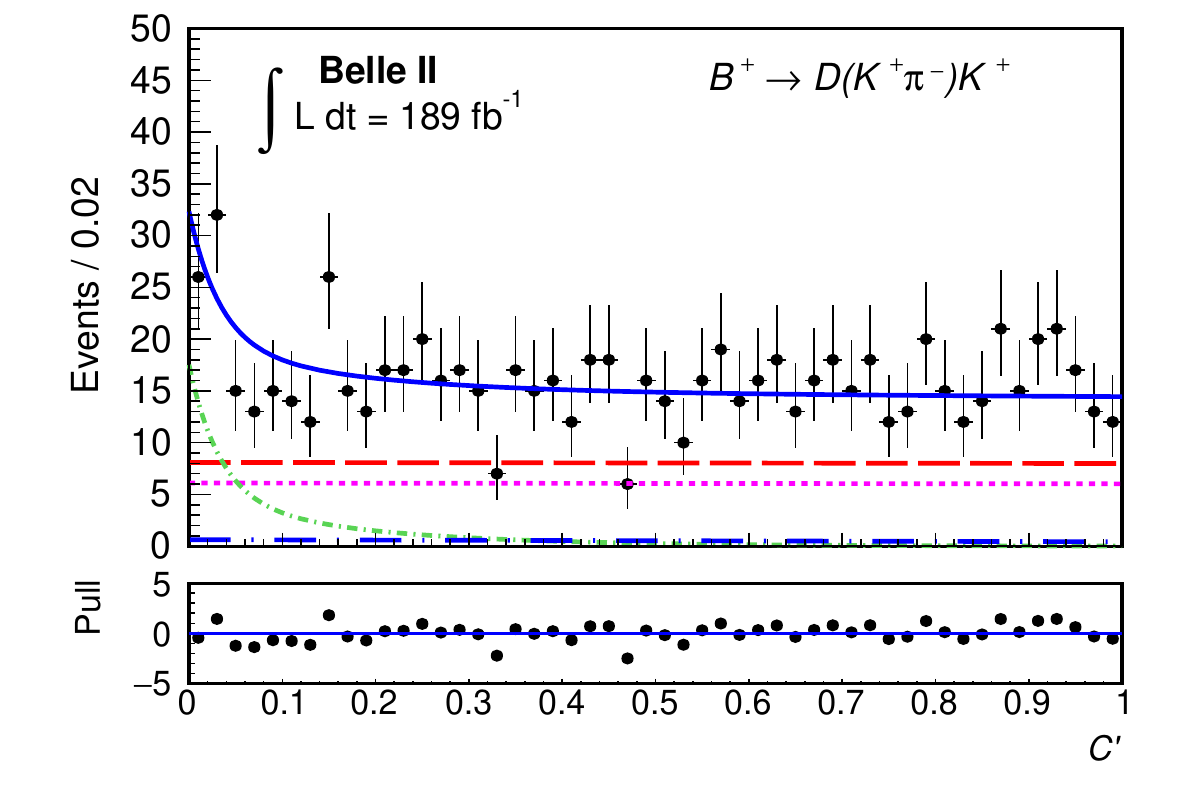}   \put(87,60){(g)} \end{overpic}
    \begin{overpic}[width=0.46\textwidth]{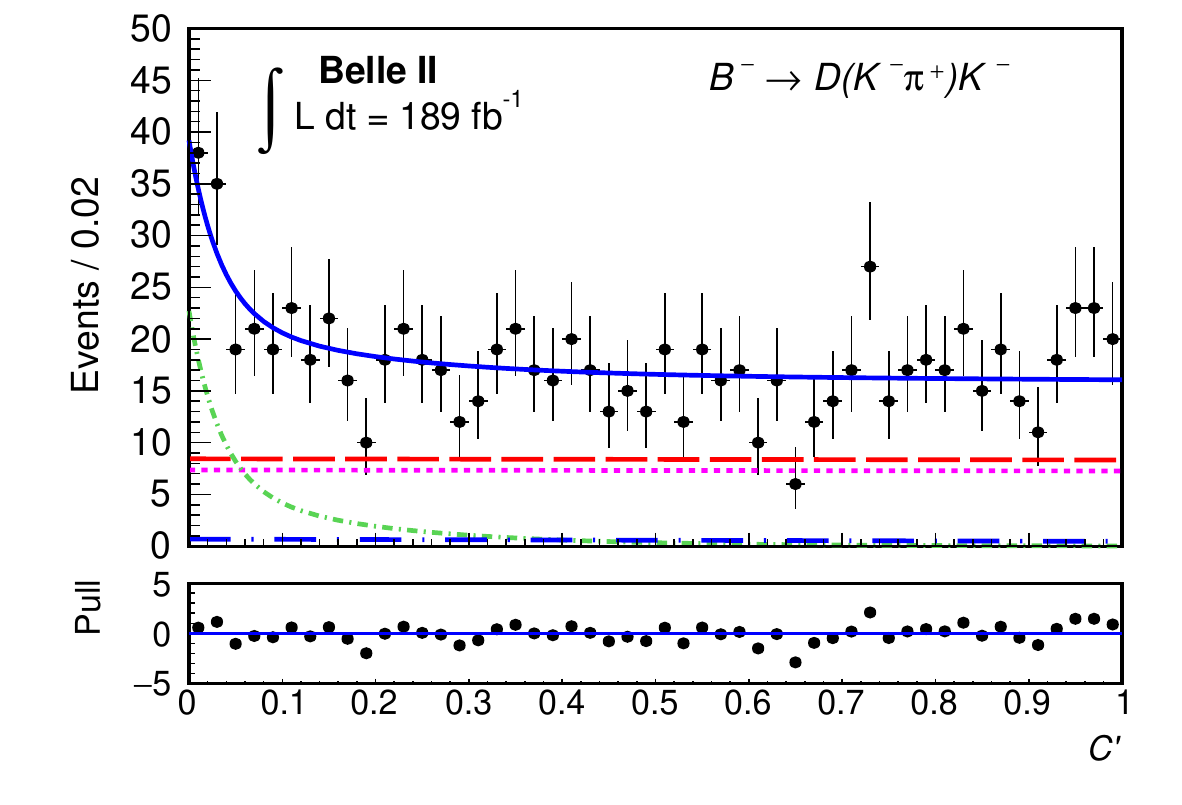}  \put(87,60){(h)} \end{overpic}

  \end{center}

  \caption{\ResultFigureCaption{\Kpm\pimp}{Belle~II}}

  \label{fig:Kpi_B2}
\end{figure}

\begin{figure}[!htb]
  \begin{center}

    \begin{overpic}[width=0.46\textwidth]{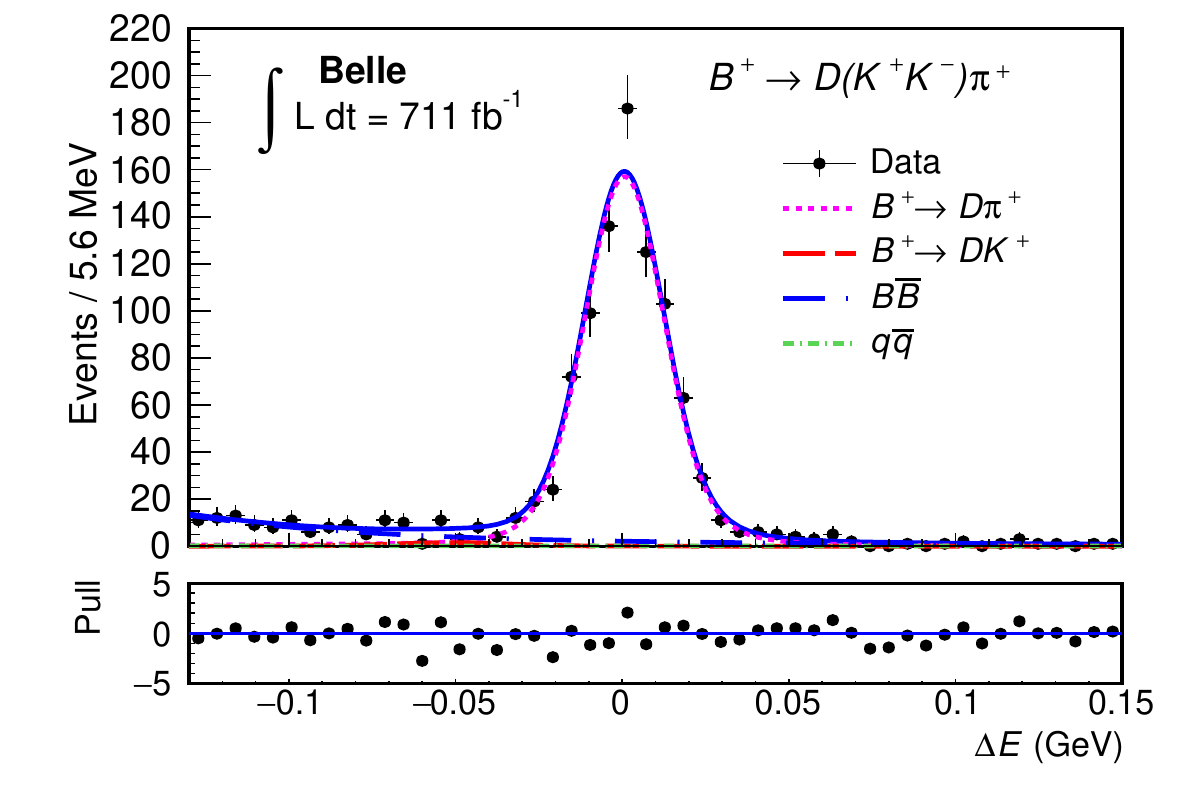}  \put(87,60){(a)} \end{overpic}
    \begin{overpic}[width=0.46\textwidth]{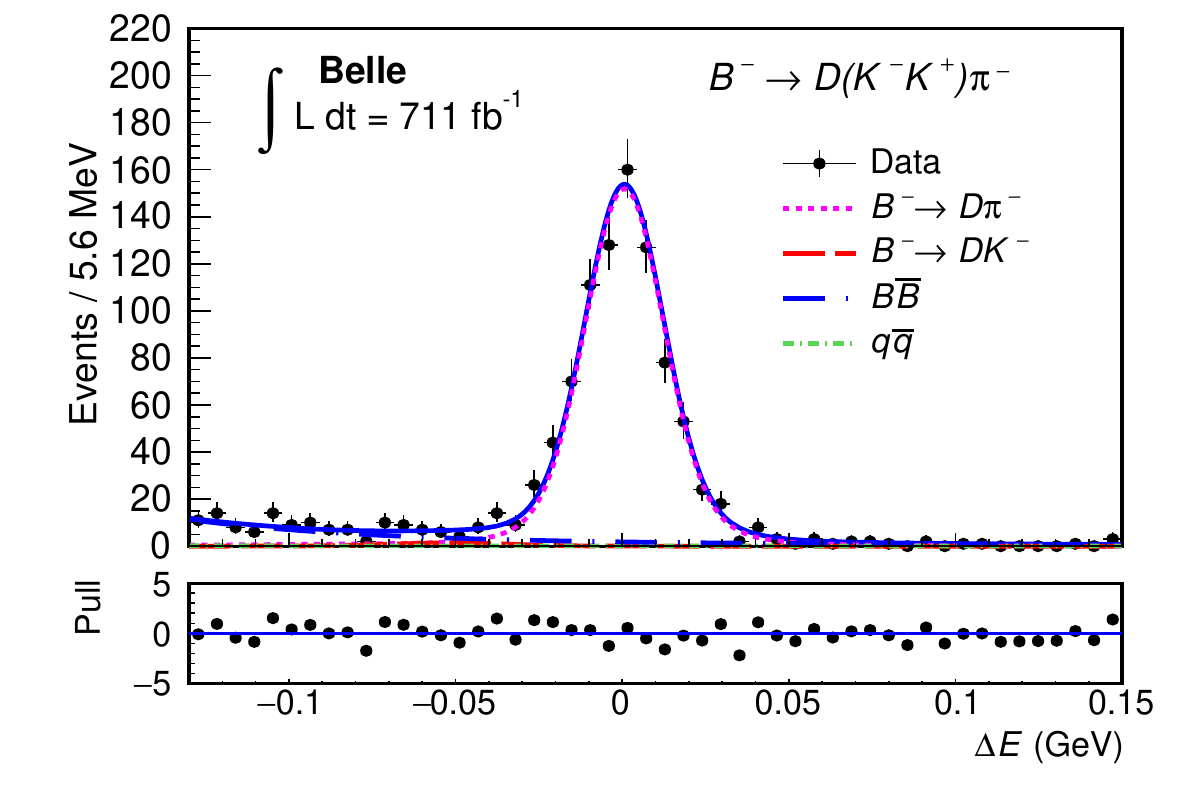} \put(87,60){(b)} \end{overpic}

    \begin{overpic}[width=0.46\textwidth]{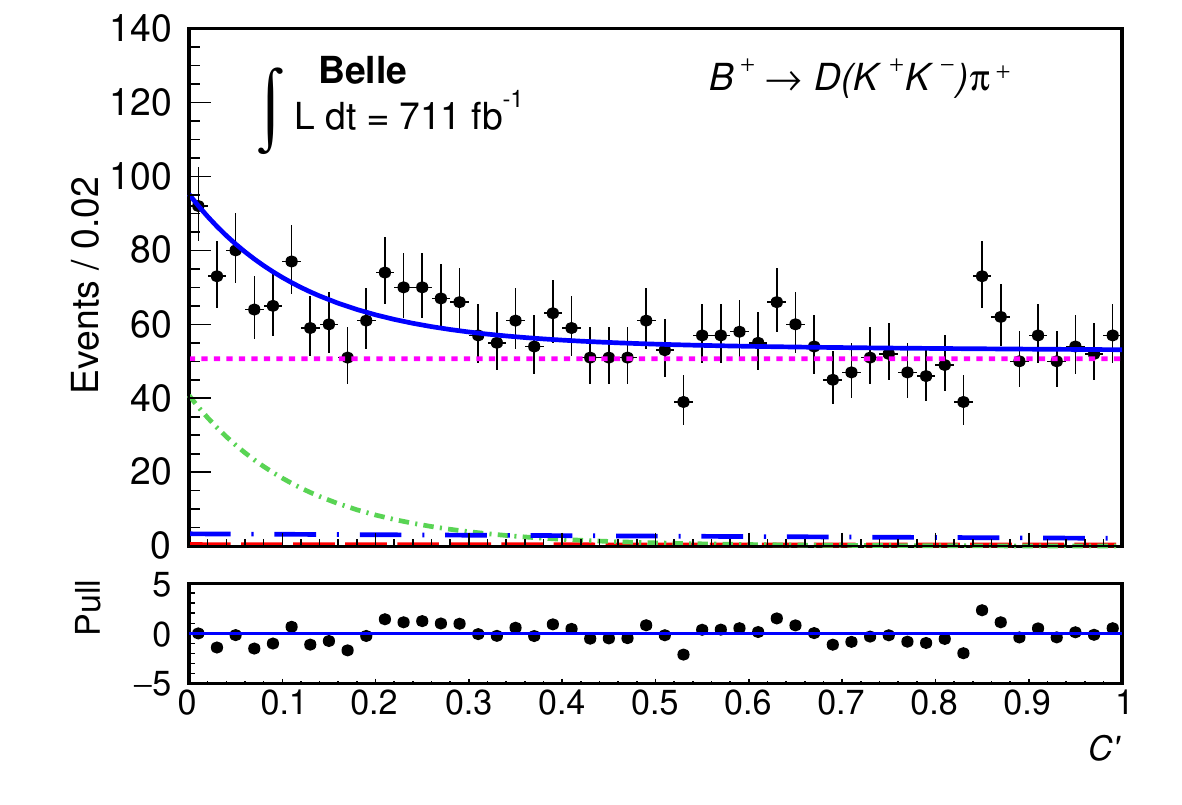}  \put(87,60){(c)} \end{overpic}
    \begin{overpic}[width=0.46\textwidth]{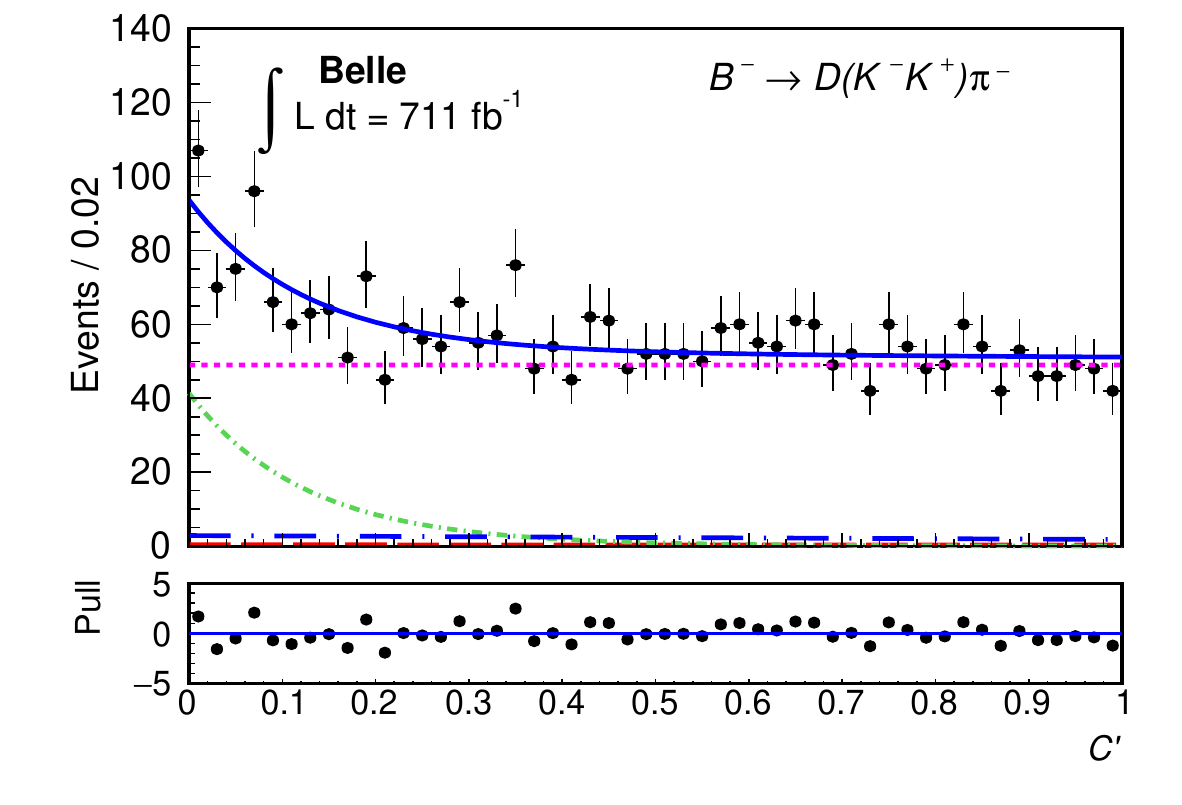} \put(87,60){(d)} \end{overpic}

    \begin{overpic}[width=0.46\textwidth]{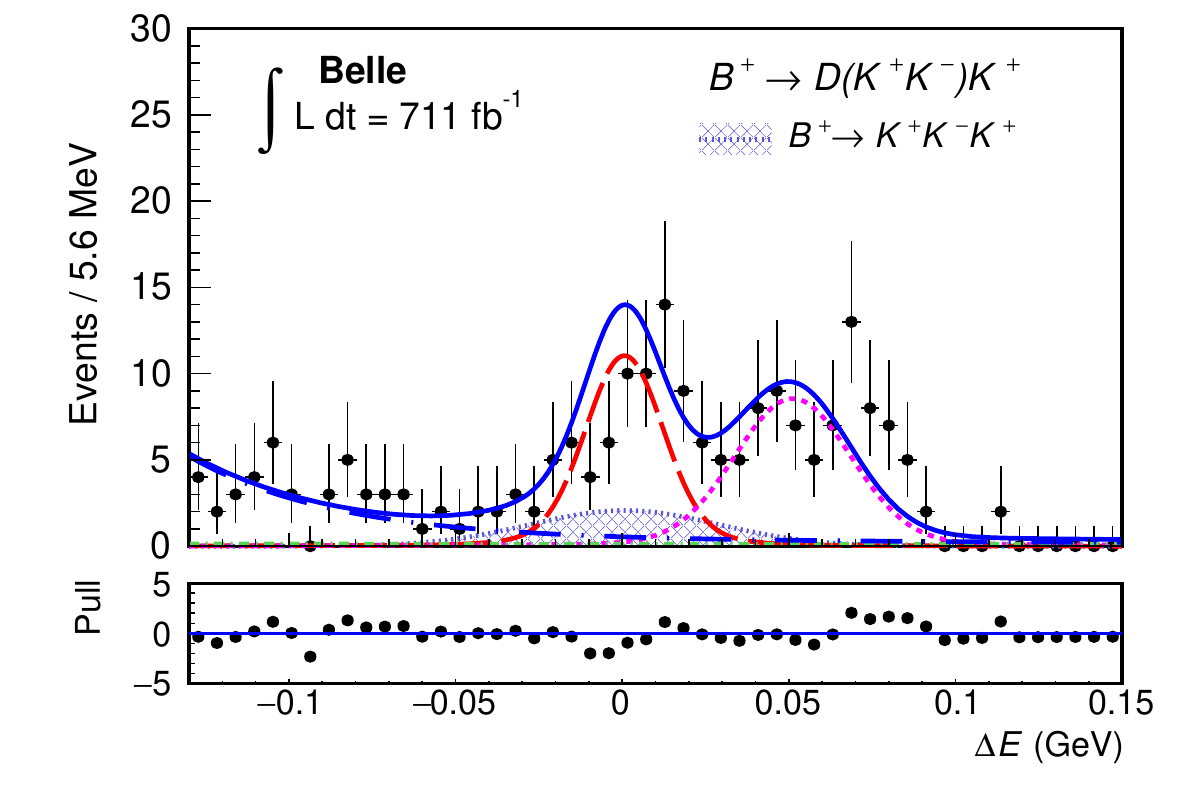}   \put(87,60){(e)} \end{overpic}
    \begin{overpic}[width=0.46\textwidth]{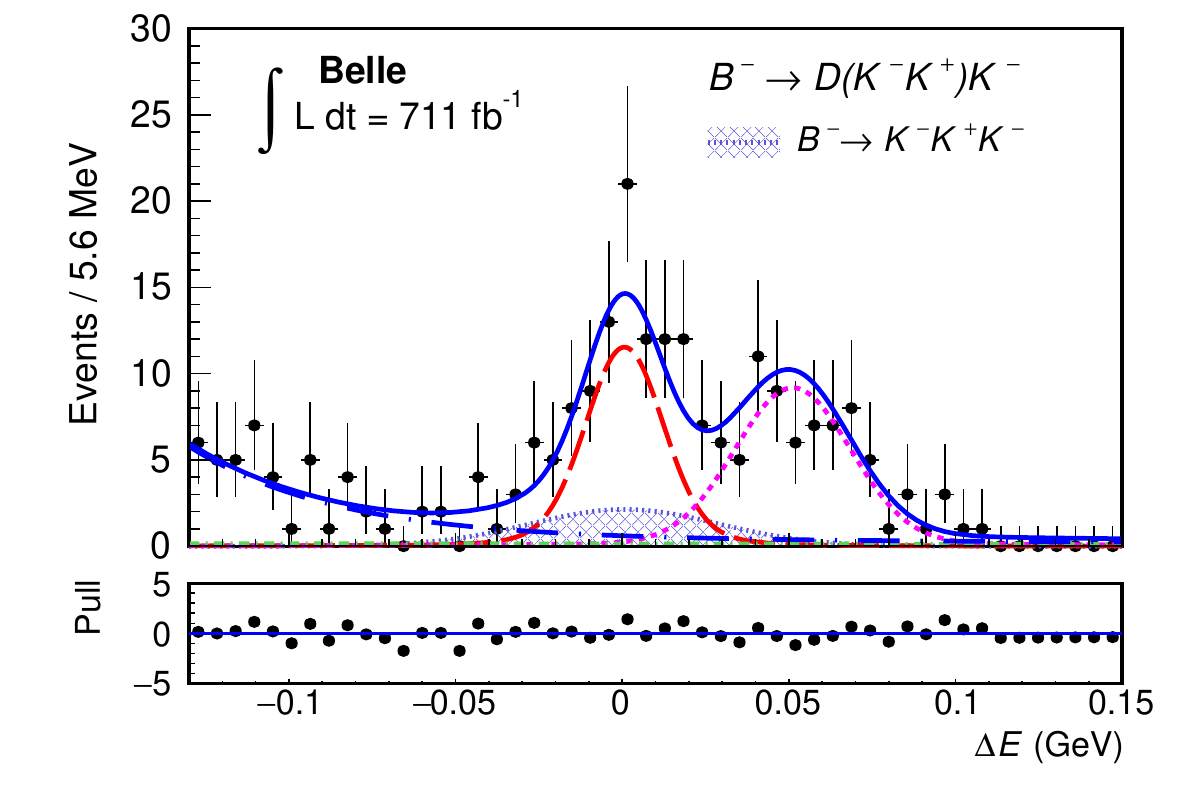}  \put(87,60){(f)} \end{overpic}

    \begin{overpic}[width=0.46\textwidth]{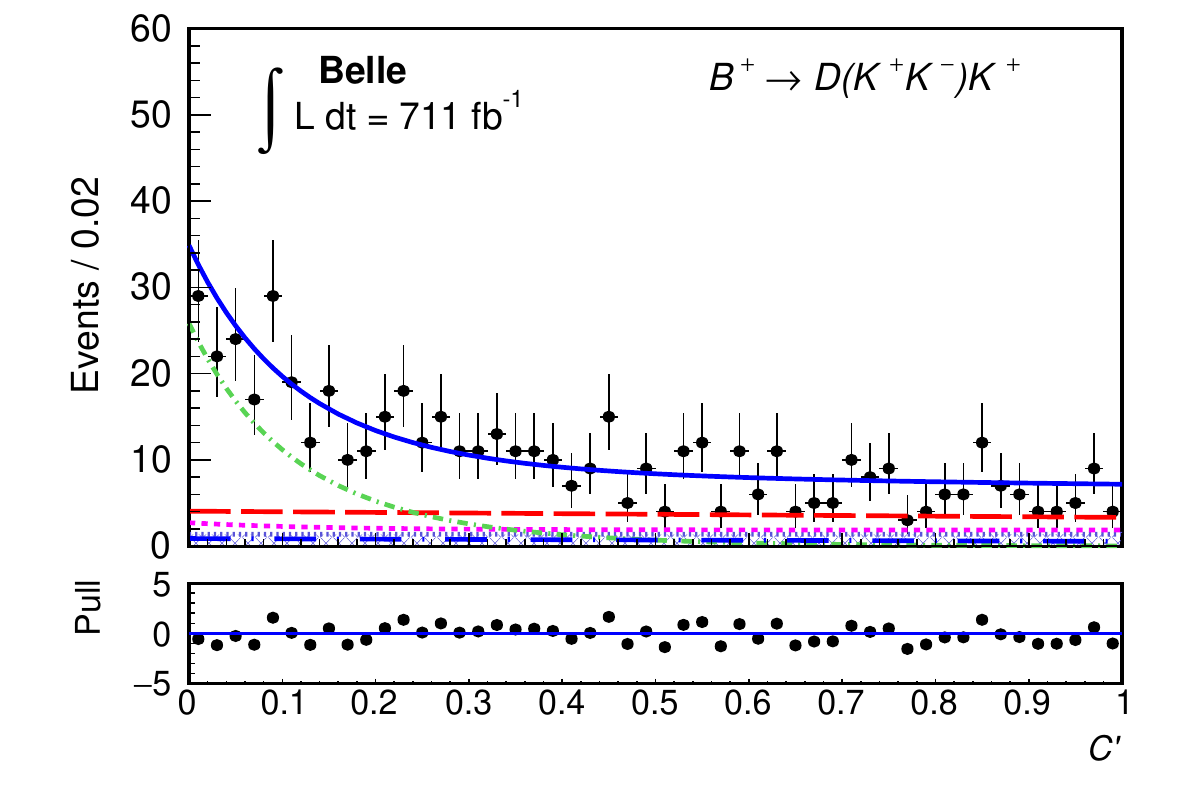}   \put(87,60){(g)} \end{overpic}
    \begin{overpic}[width=0.46\textwidth]{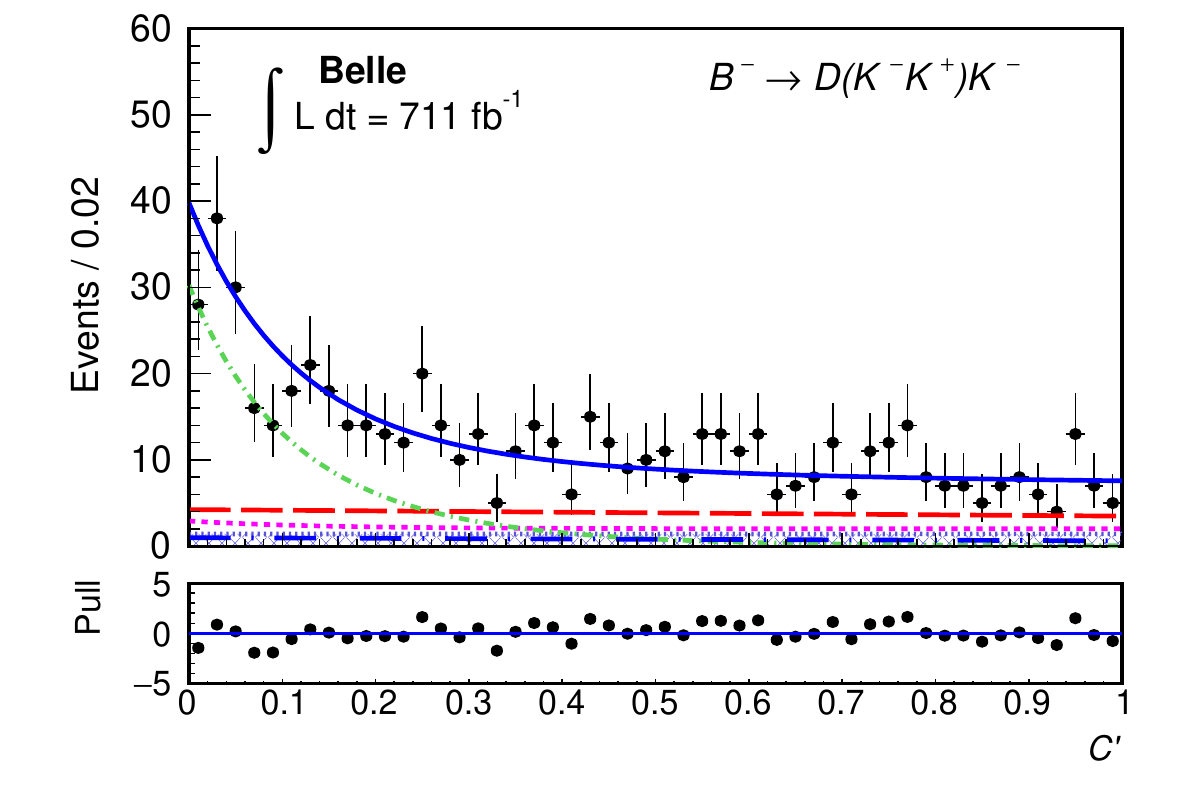}  \put(87,60){(h)} \end{overpic}
    
  \end{center}
  
  \caption{\ResultFigureCaption{\Kp\Km}{Belle}}
  
  \label{fig:KK_B}
\end{figure}

\begin{figure}[!t]
  \begin{center}

    \begin{overpic}[width=0.46\textwidth]{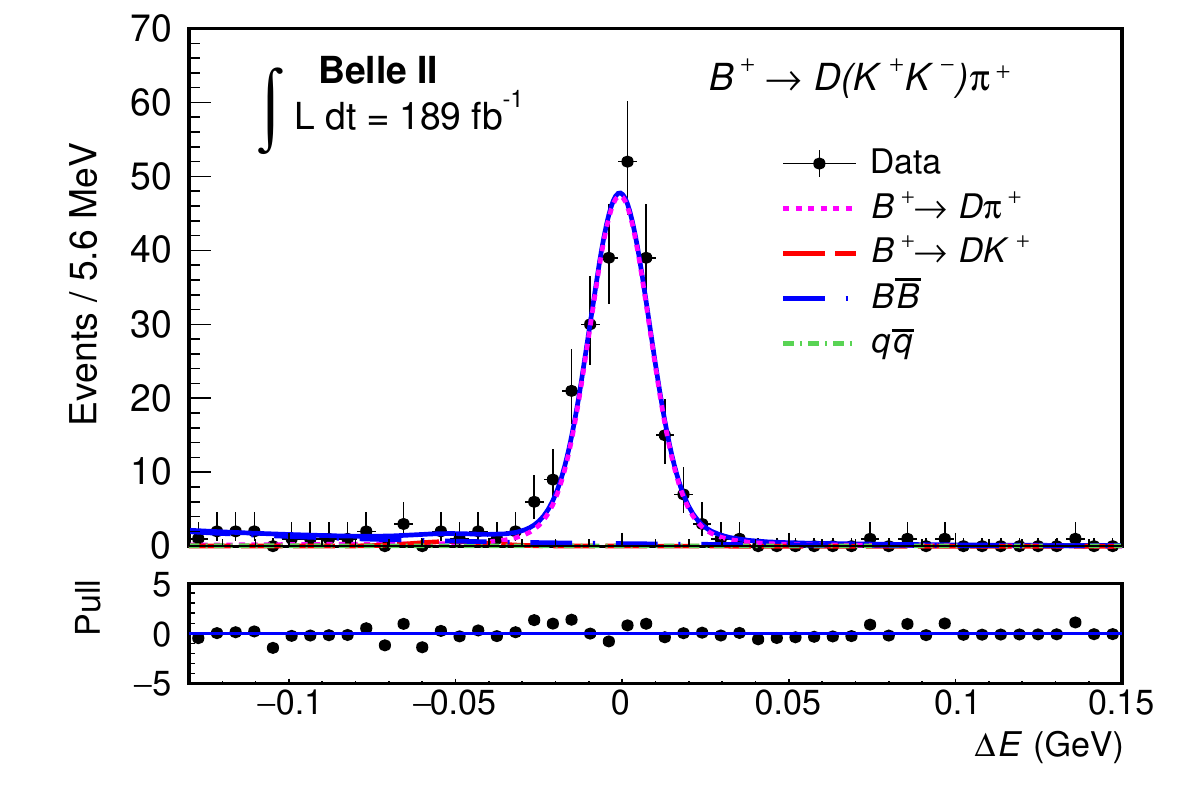}  \put(87,60){(a)} \end{overpic}
    \begin{overpic}[width=0.46\textwidth]{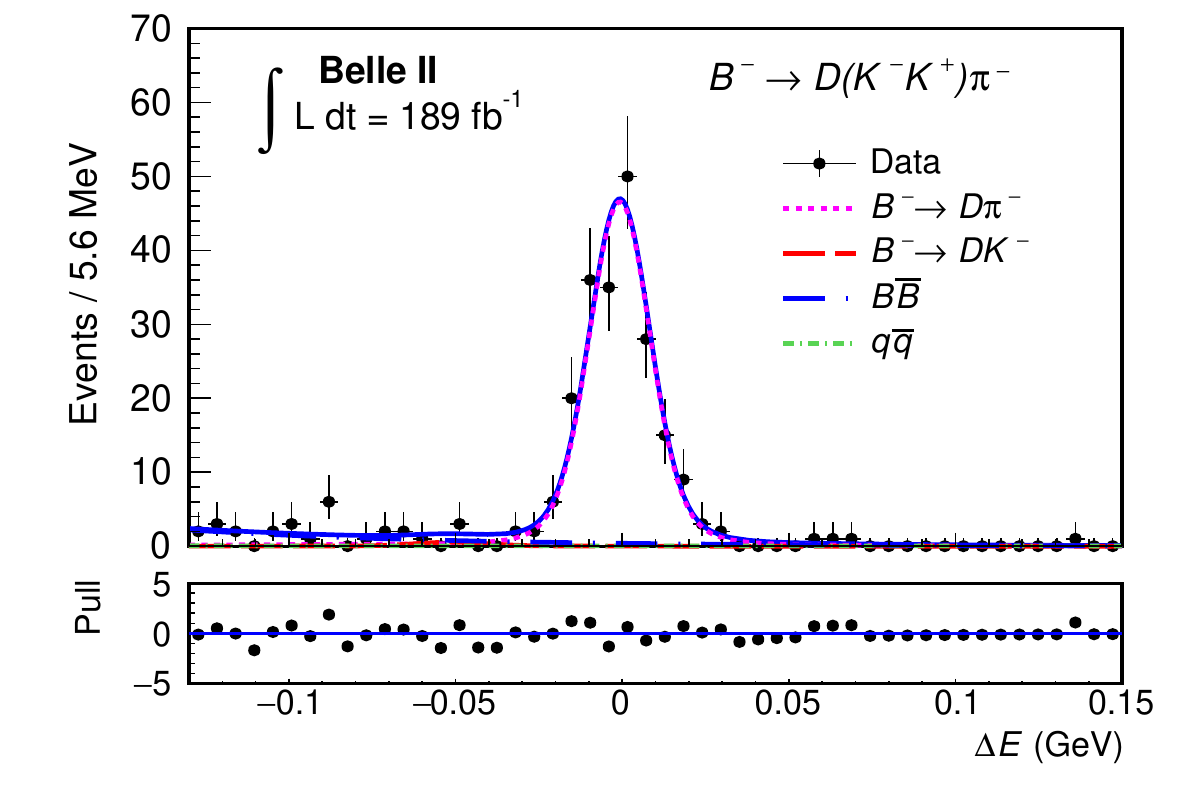} \put(87,60){(b)} \end{overpic}

    \begin{overpic}[width=0.46\textwidth]{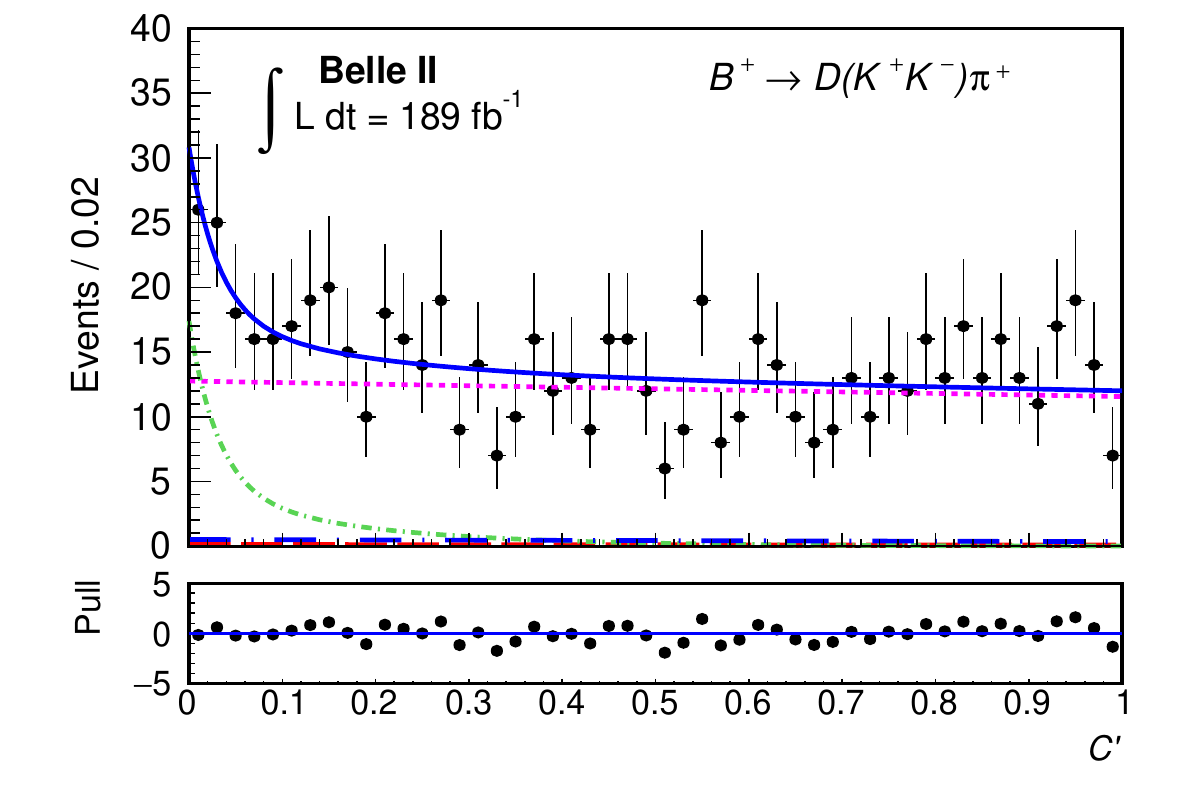}  \put(87,60){(c)} \end{overpic}
    \begin{overpic}[width=0.46\textwidth]{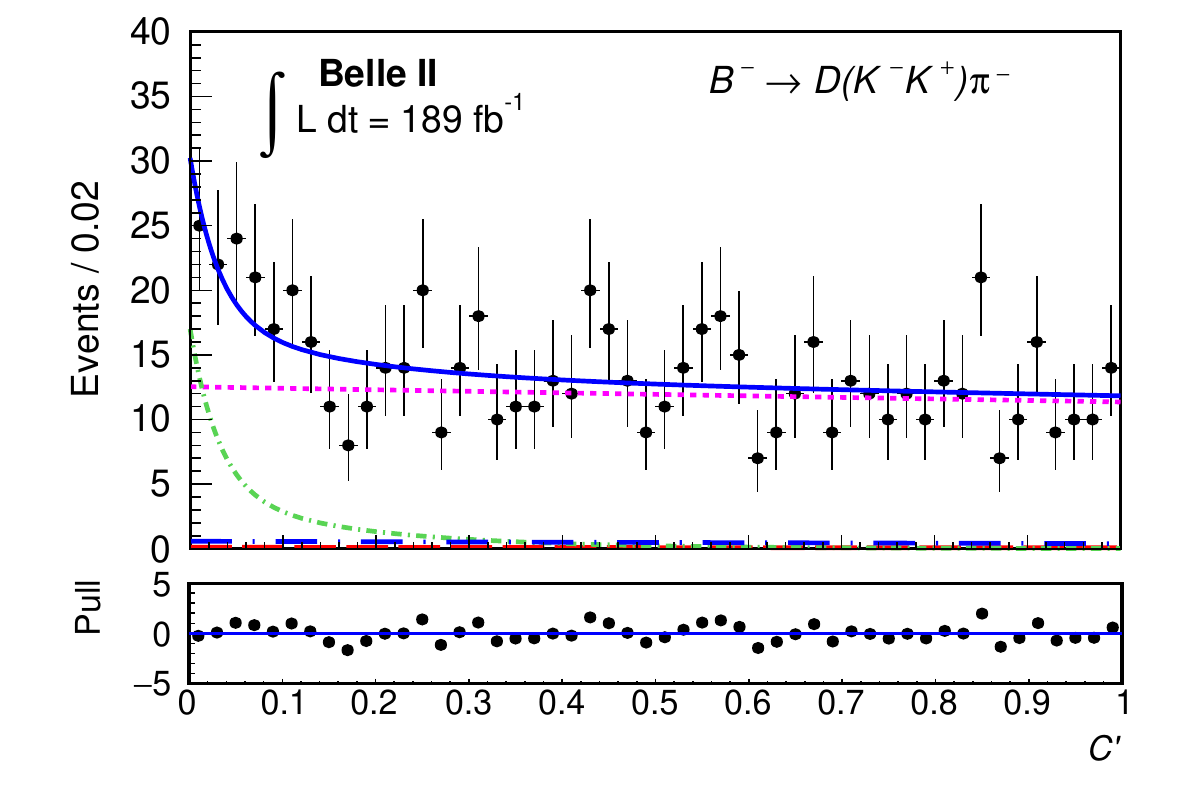} \put(87,60){(d)} \end{overpic}

    \begin{overpic}[width=0.46\textwidth]{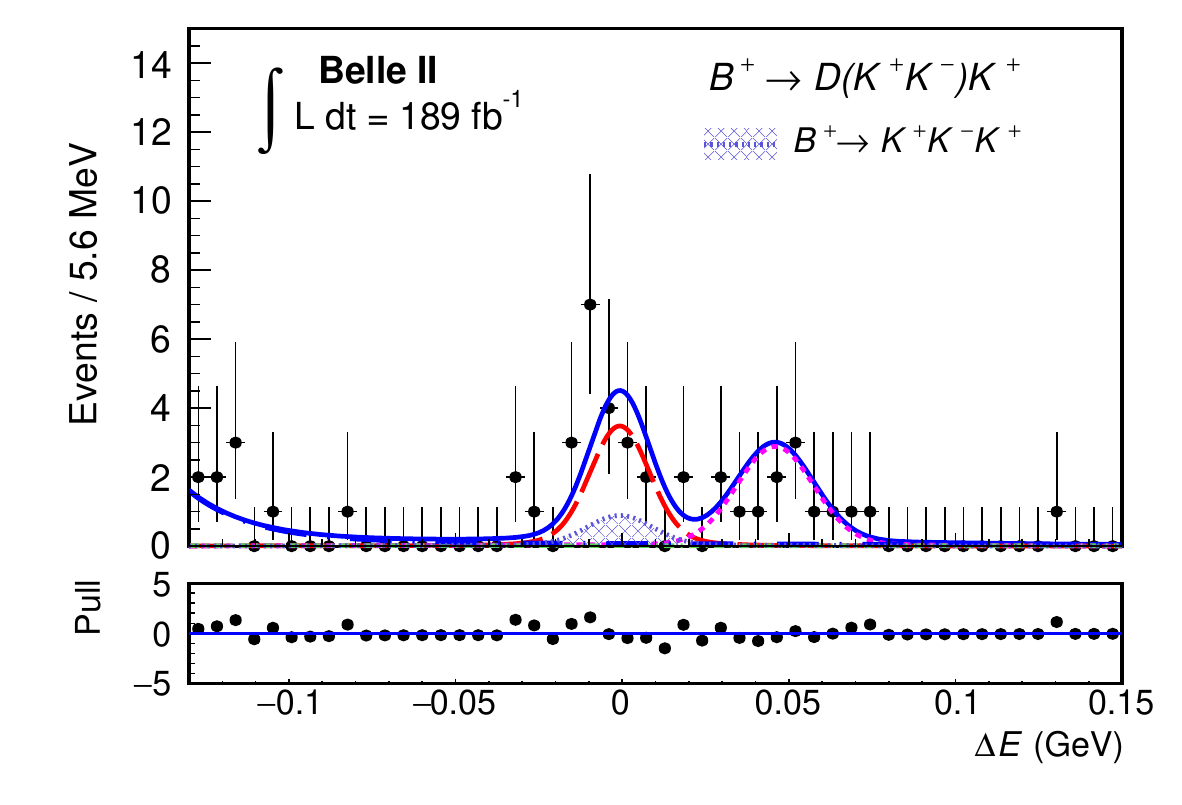}   \put(87,60){(e)} \end{overpic}
    \begin{overpic}[width=0.46\textwidth]{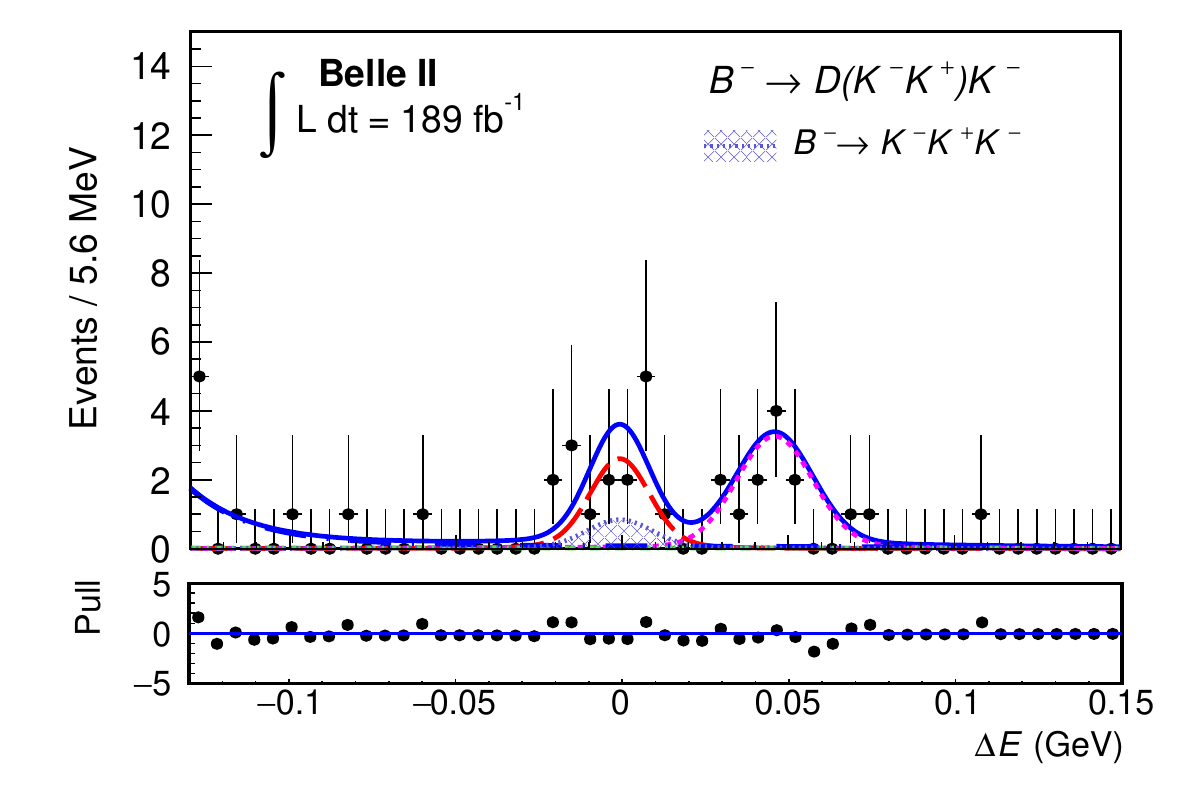}  \put(87,60){(f)} \end{overpic}

    \begin{overpic}[width=0.46\textwidth]{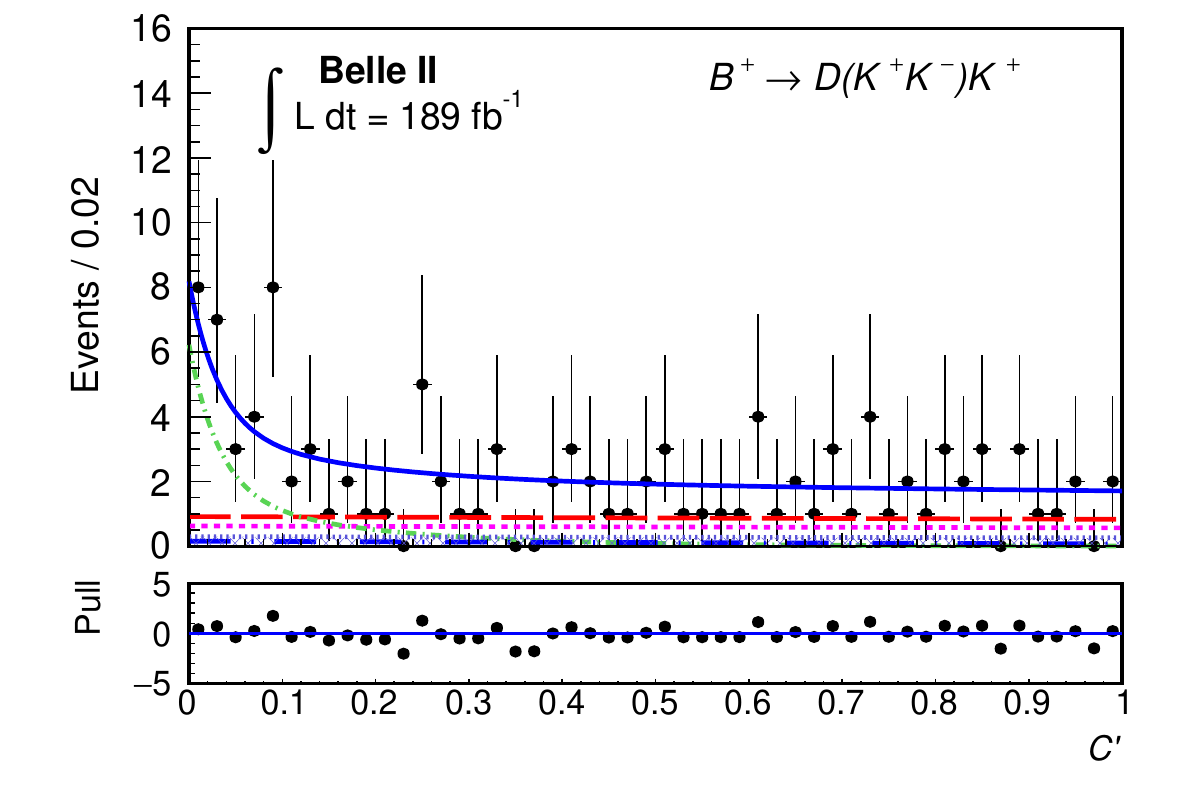}   \put(87,60){(g)} \end{overpic}
    \begin{overpic}[width=0.46\textwidth]{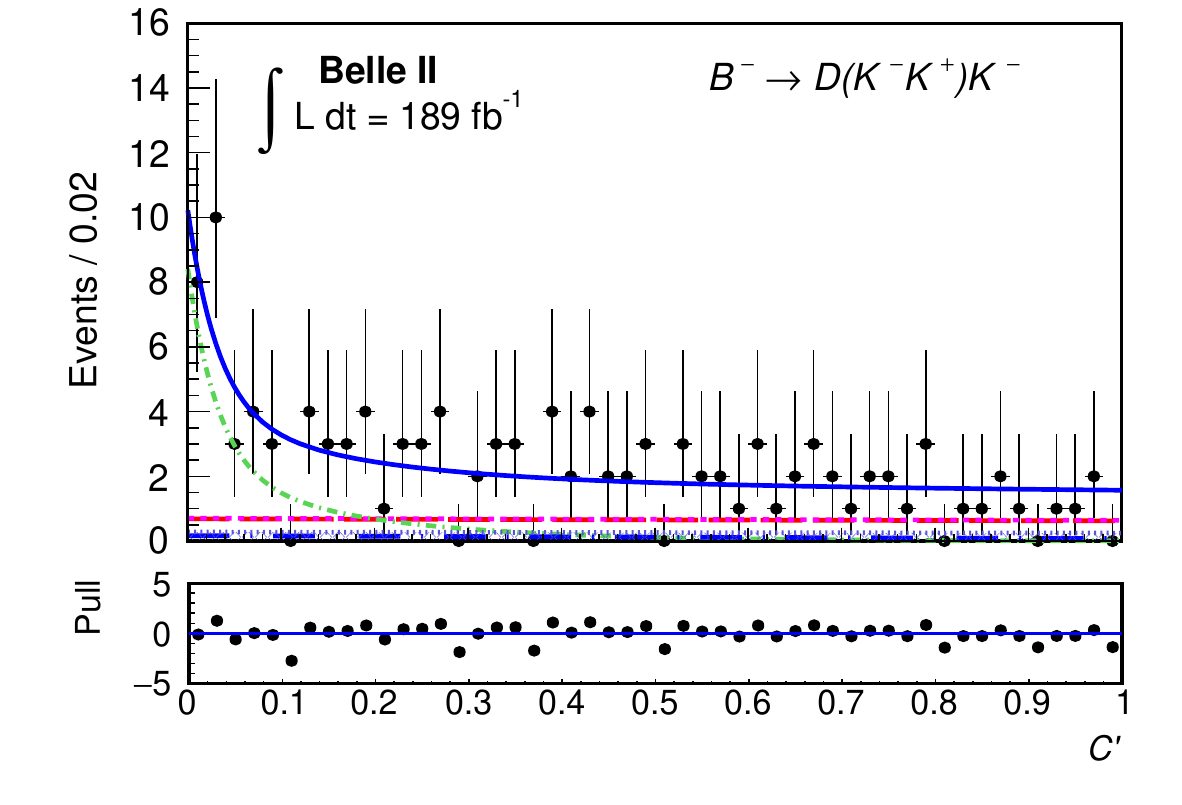}  \put(87,60){(h)} \end{overpic}

  \end{center}

  \caption{\ResultFigureCaption{\Kp\Km}{Belle~II}}

  \label{fig:KK_B2}
\end{figure}

\begin{figure}[!t]
  \begin{center}

    \begin{overpic}[width=0.46\textwidth]{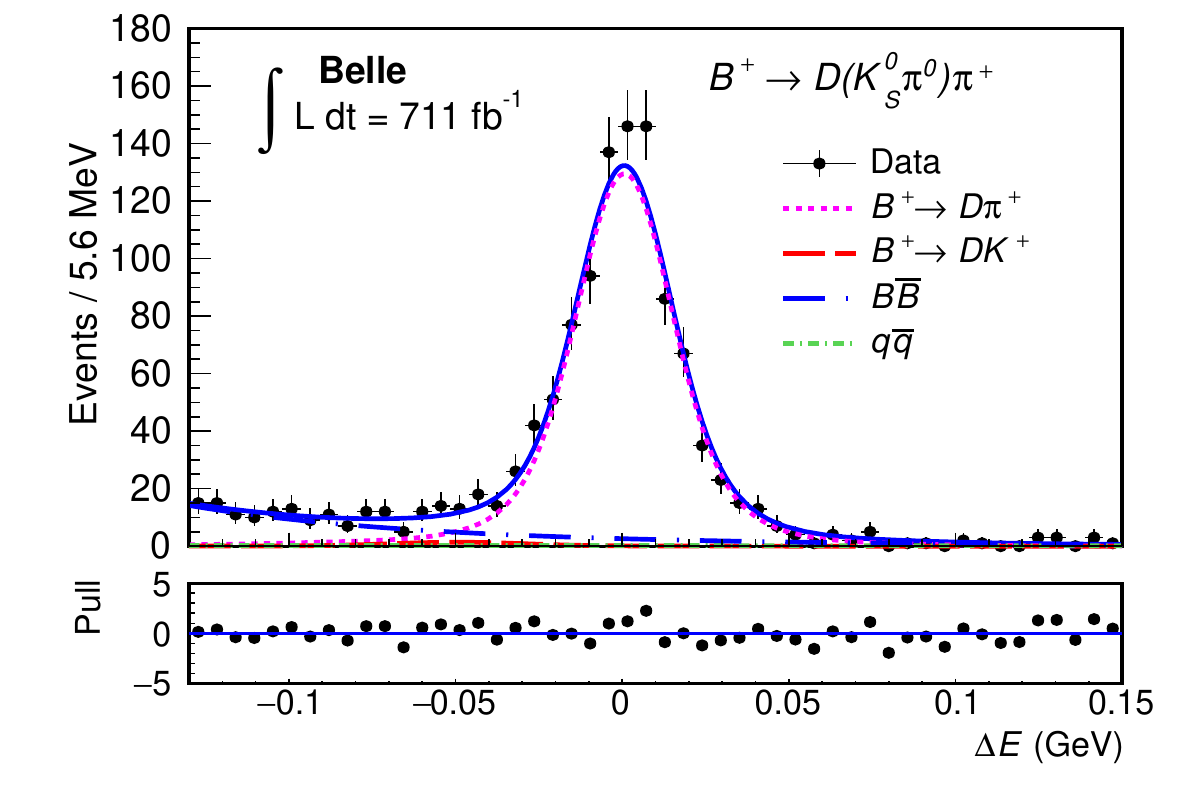}  \put(87,60){(a)} \end{overpic}
    \begin{overpic}[width=0.46\textwidth]{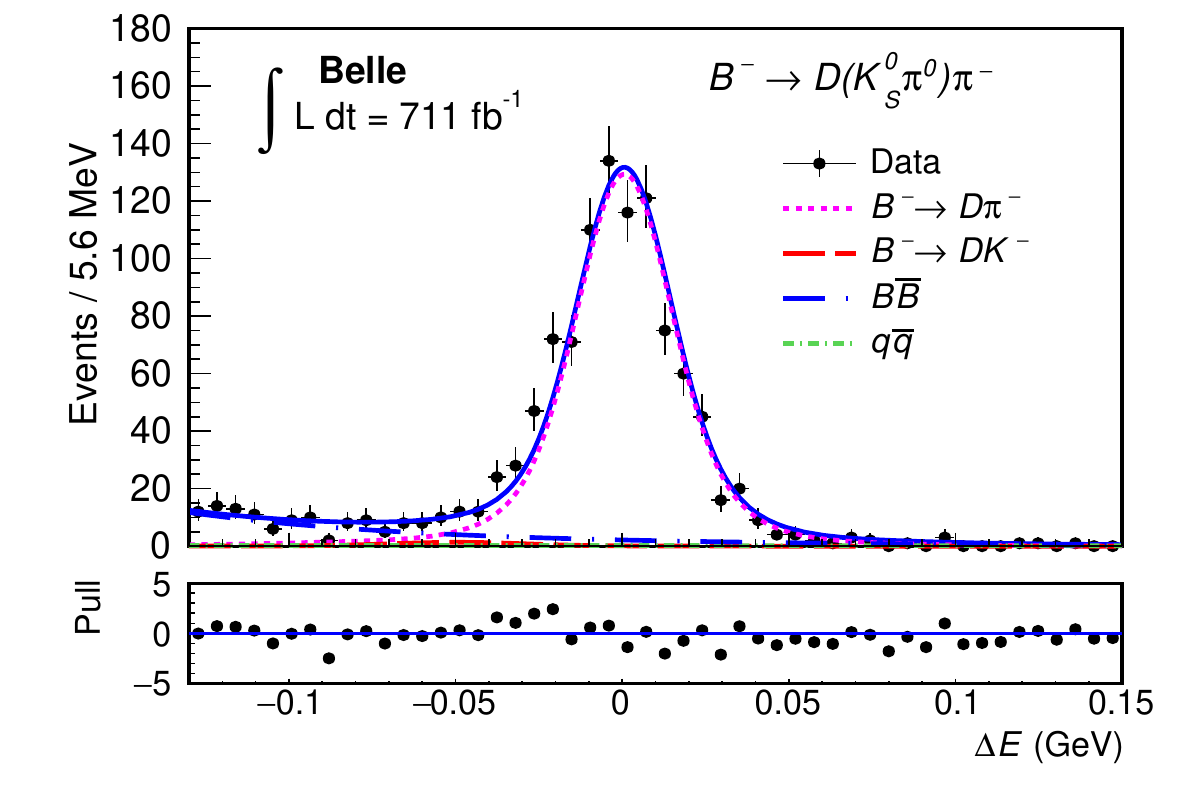} \put(87,60){(b)} \end{overpic}
    
    \begin{overpic}[width=0.46\textwidth]{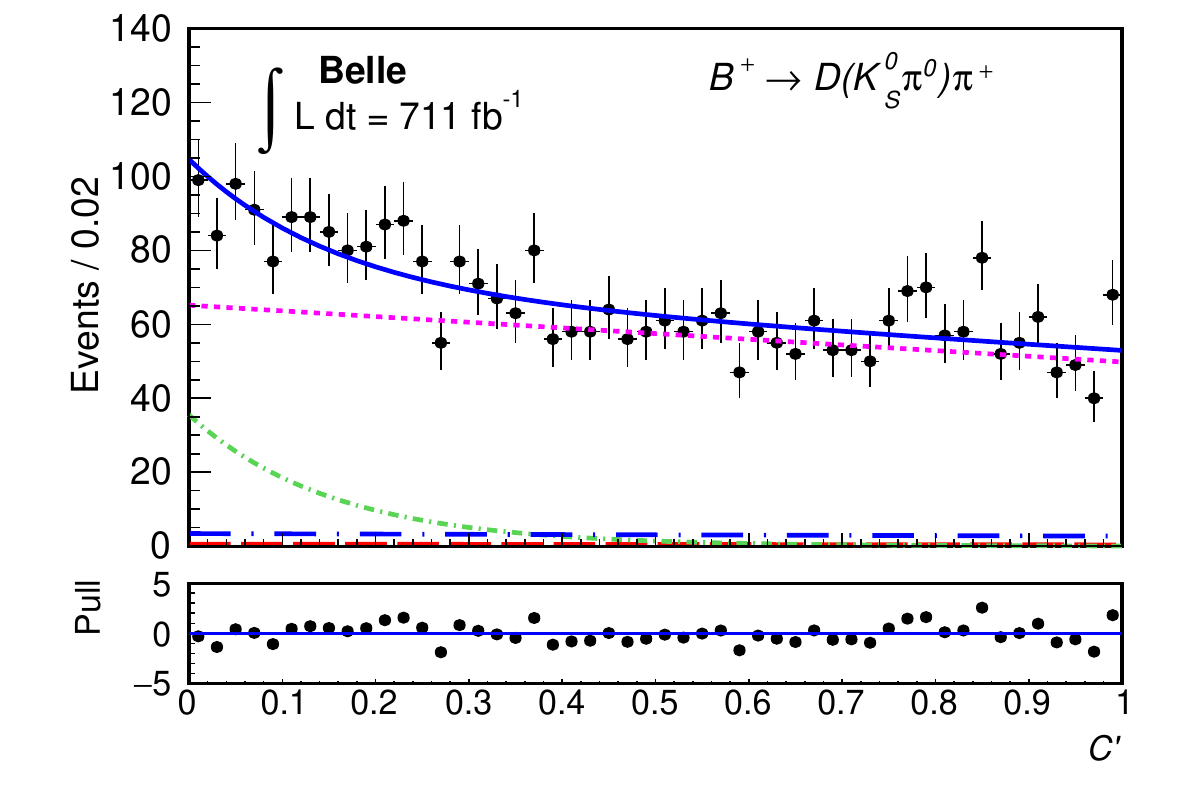}  \put(87,60){(c)} \end{overpic}
    \begin{overpic}[width=0.46\textwidth]{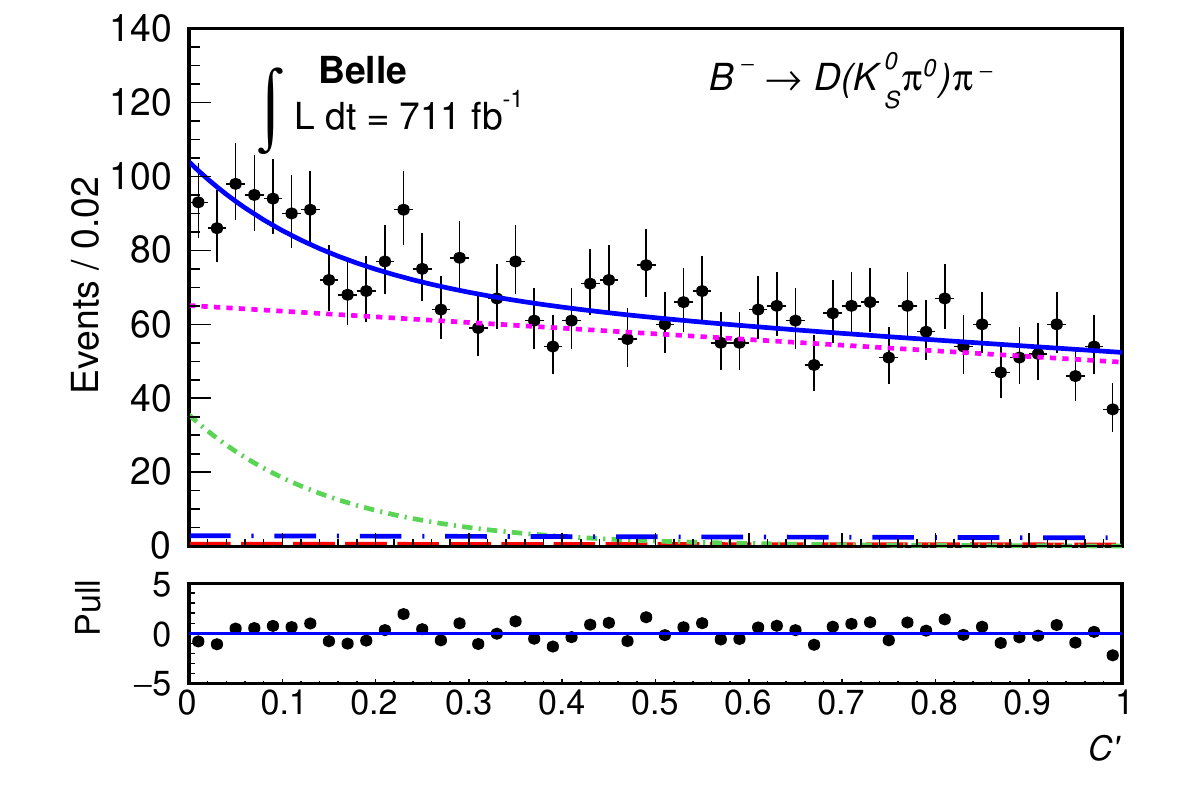} \put(87,60){(d)} \end{overpic}

    \begin{overpic}[width=0.46\textwidth]{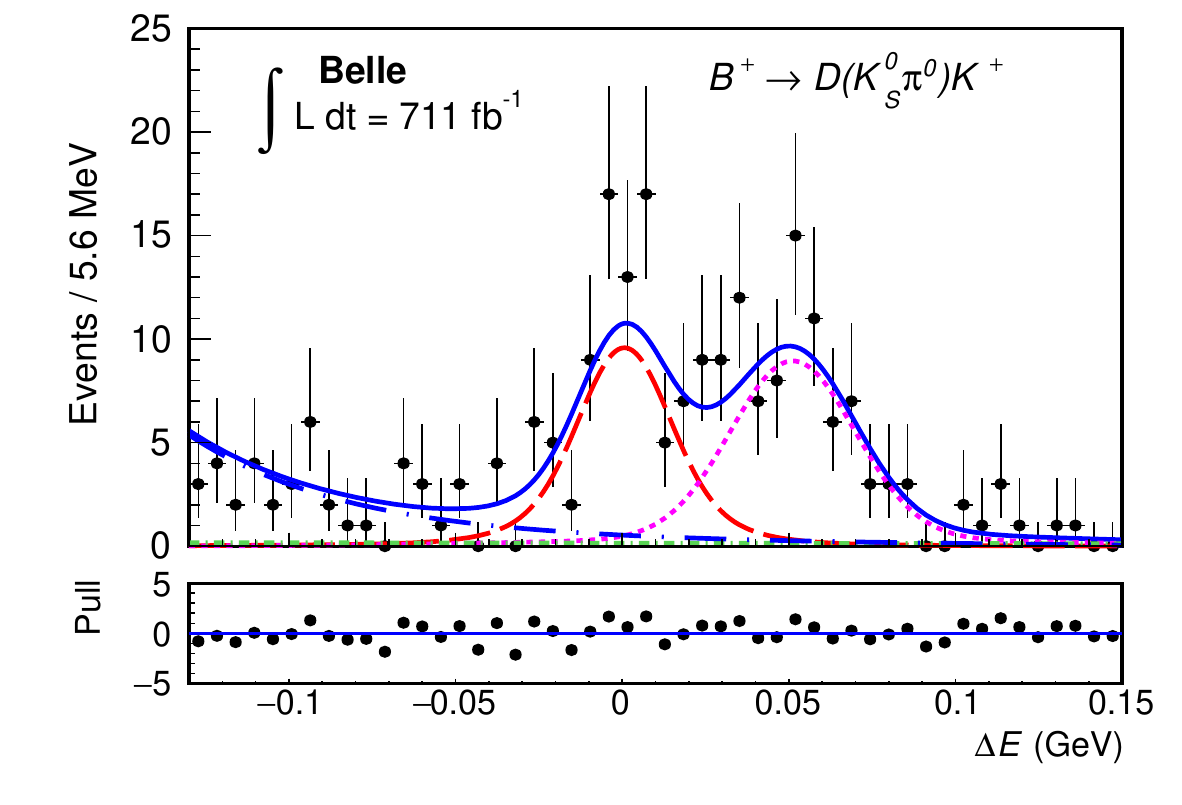}   \put(87,60){(e)} \end{overpic}
    \begin{overpic}[width=0.46\textwidth]{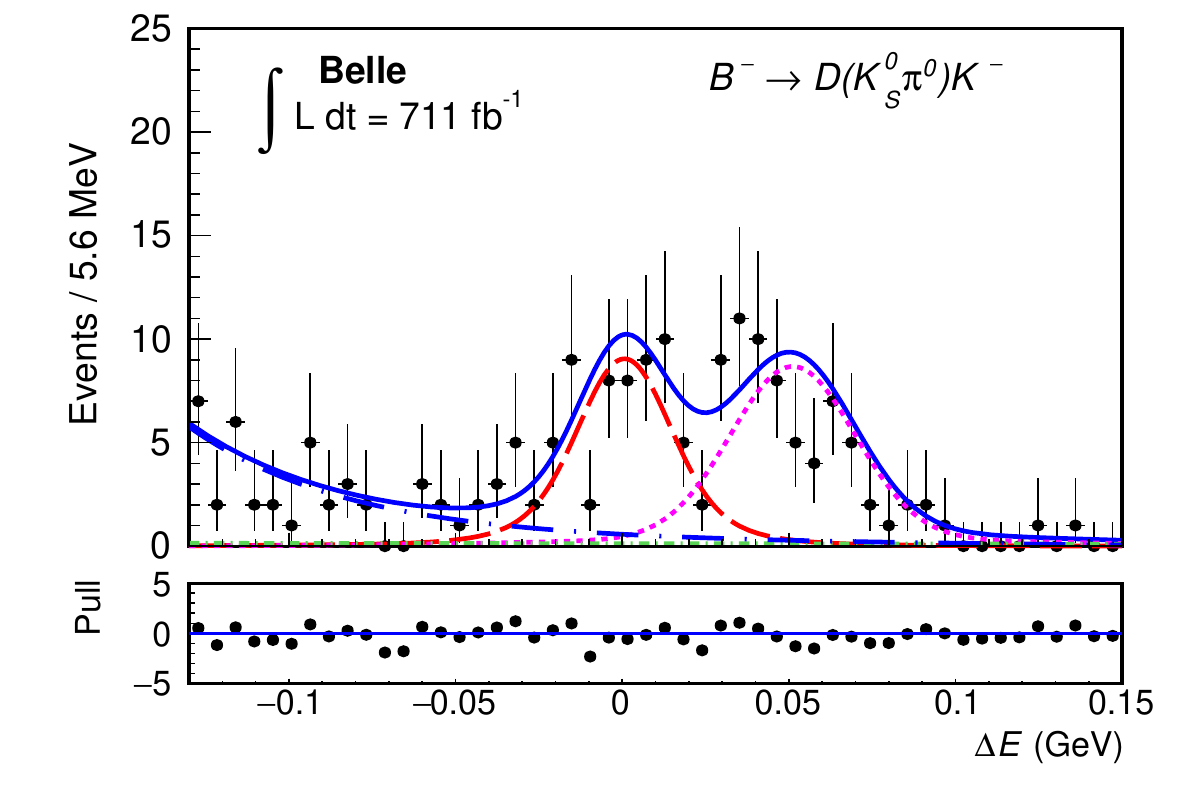}  \put(87,60){(f)} \end{overpic}

    \begin{overpic}[width=0.46\textwidth]{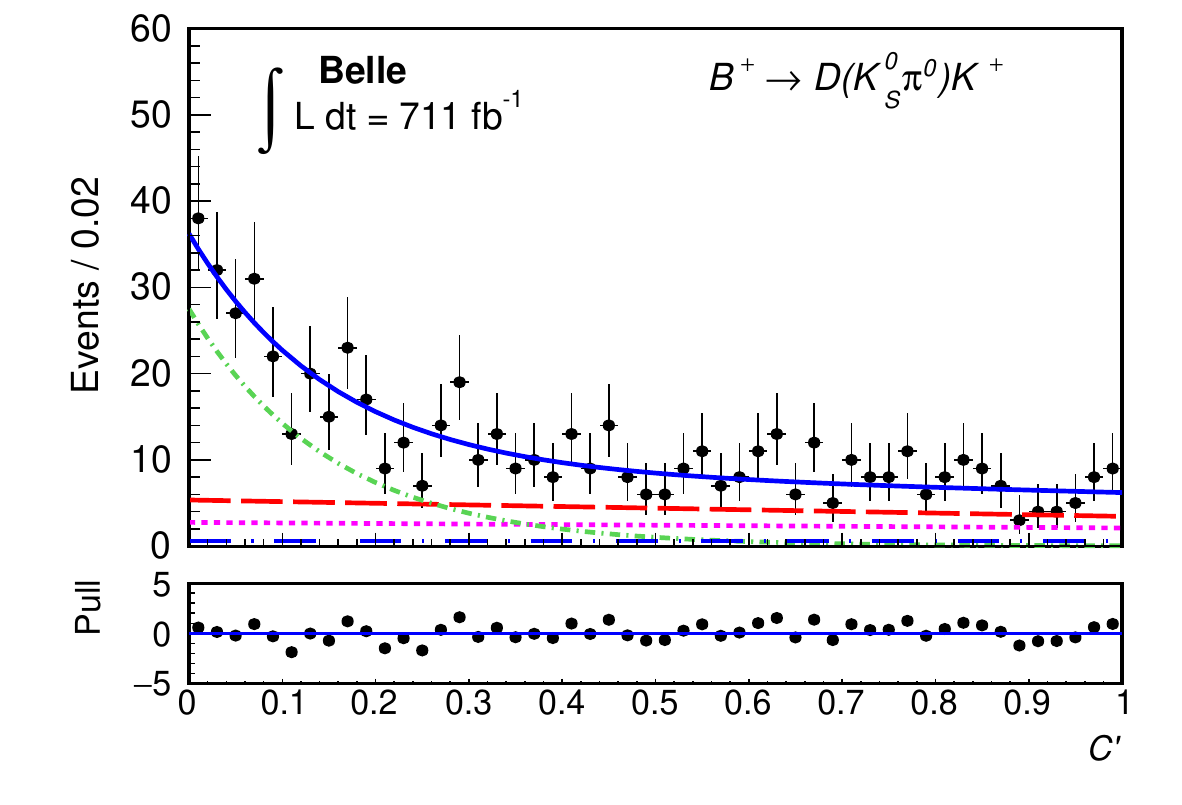}   \put(87,60){(g)} \end{overpic}
    \begin{overpic}[width=0.46\textwidth]{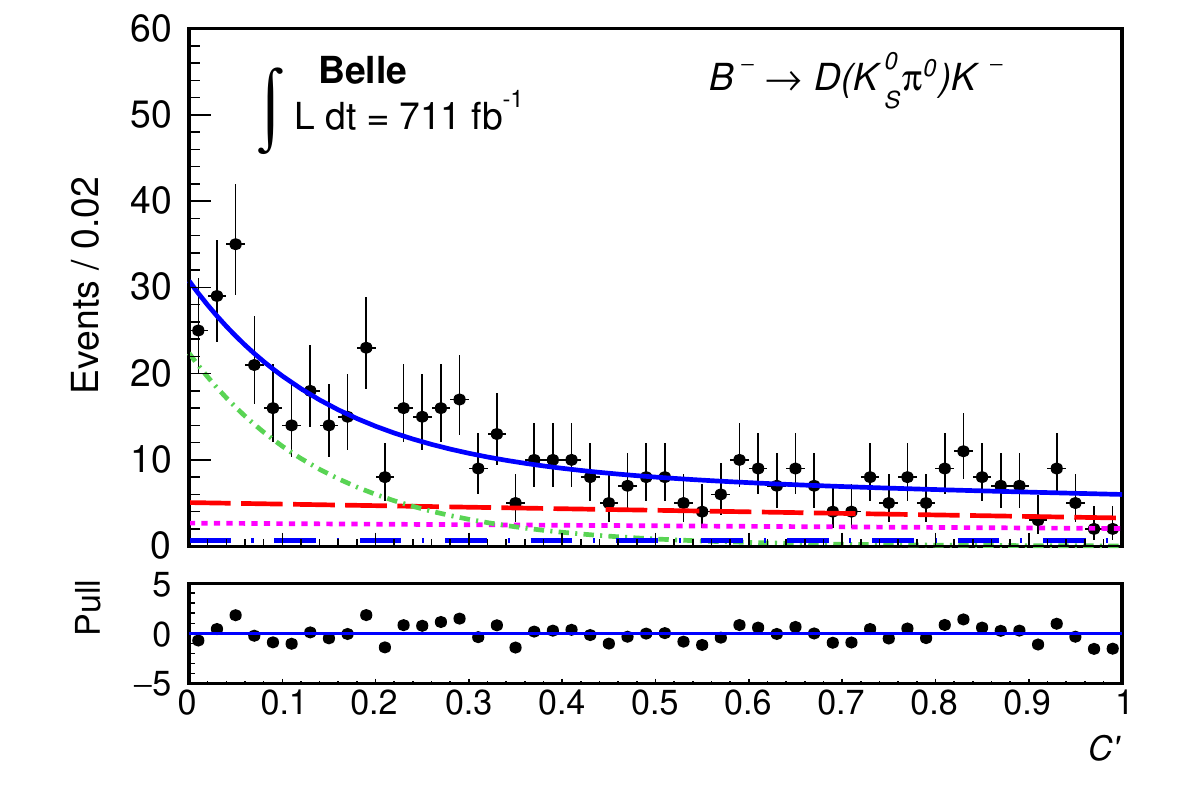}  \put(87,60){(h)} \end{overpic}

  \end{center}

  \caption{\ResultFigureCaption{\KS\piz}{Belle}}

  \label{fig:K0pi0_B}
\end{figure}

\begin{figure}[!t]
  \begin{center}

    \begin{overpic}[width=0.46\textwidth]{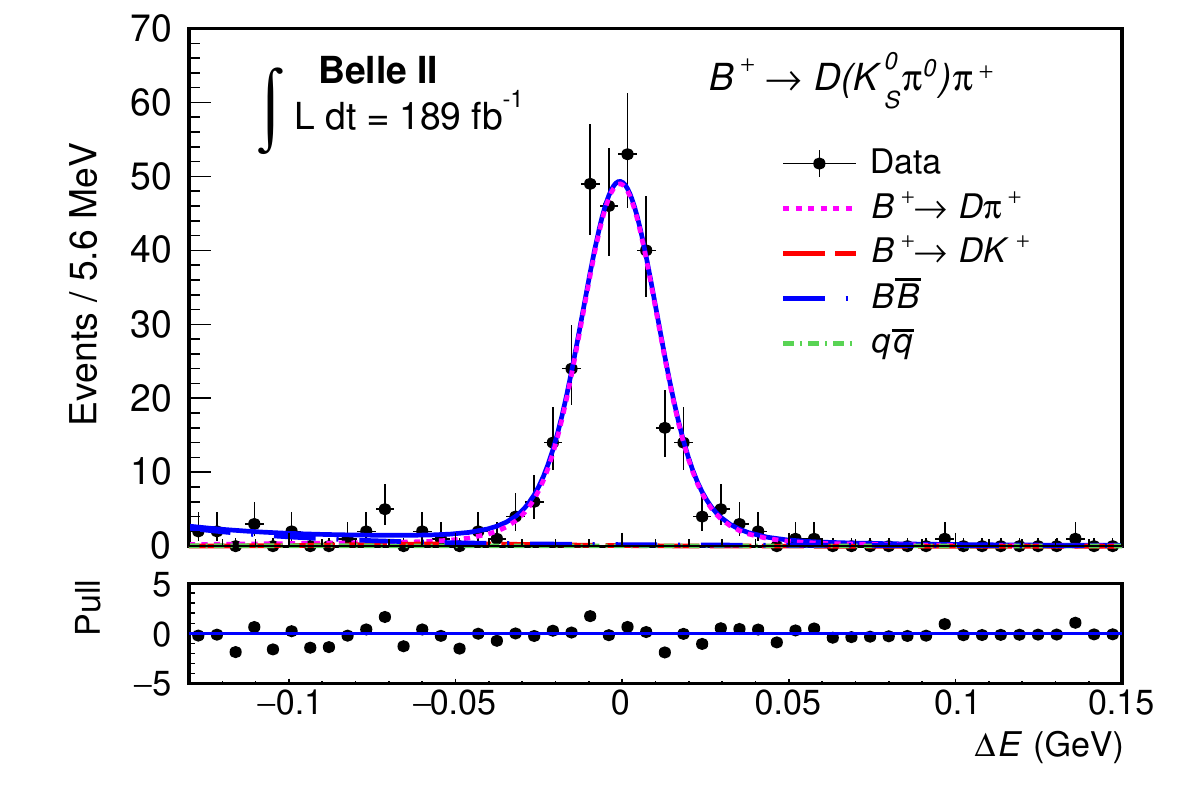}  \put(87,60){(a)} \end{overpic}
    \begin{overpic}[width=0.46\textwidth]{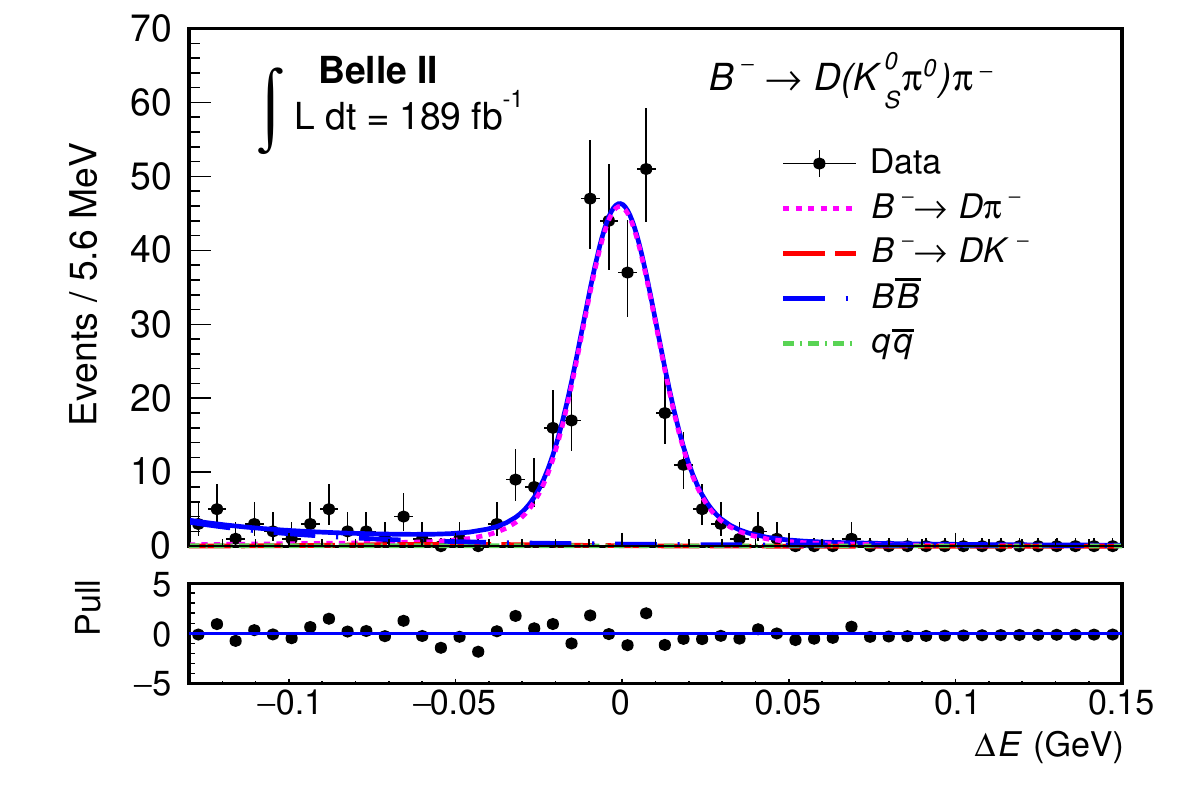} \put(87,60){(b)} \end{overpic}

    \begin{overpic}[width=0.46\textwidth]{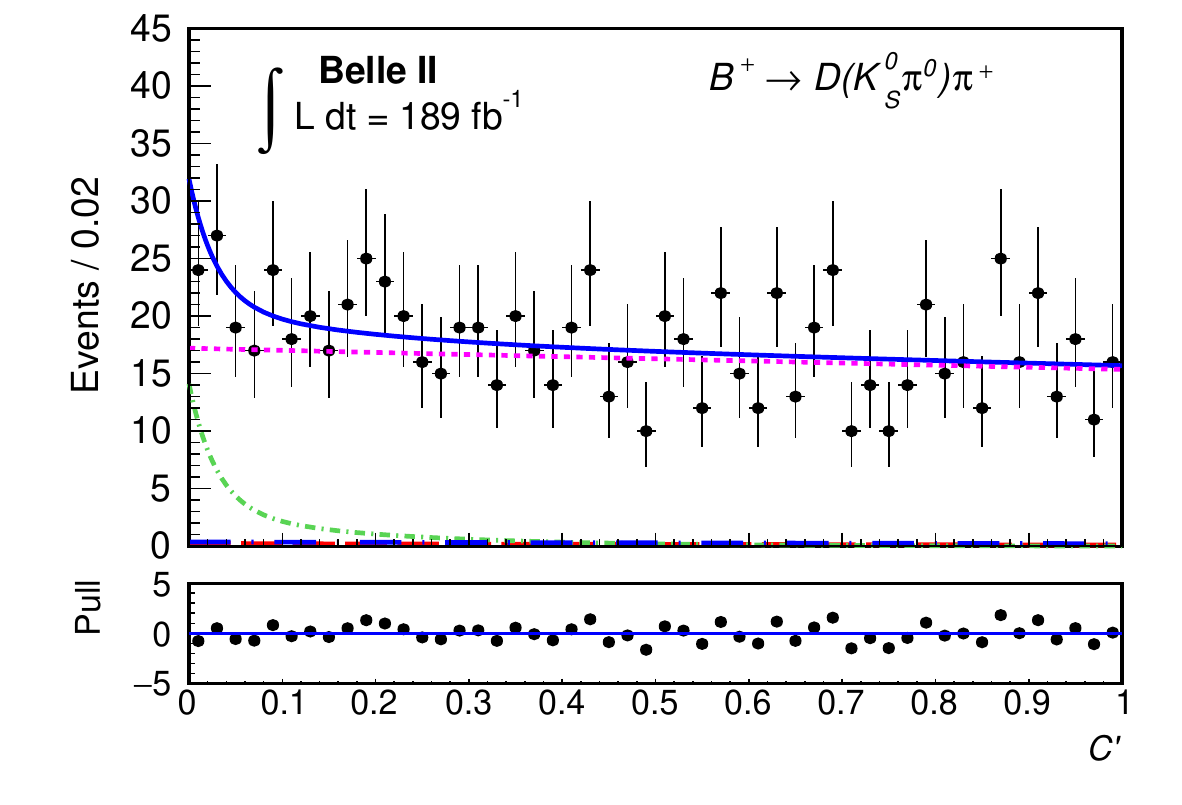}  \put(87,60){(c)} \end{overpic}
    \begin{overpic}[width=0.46\textwidth]{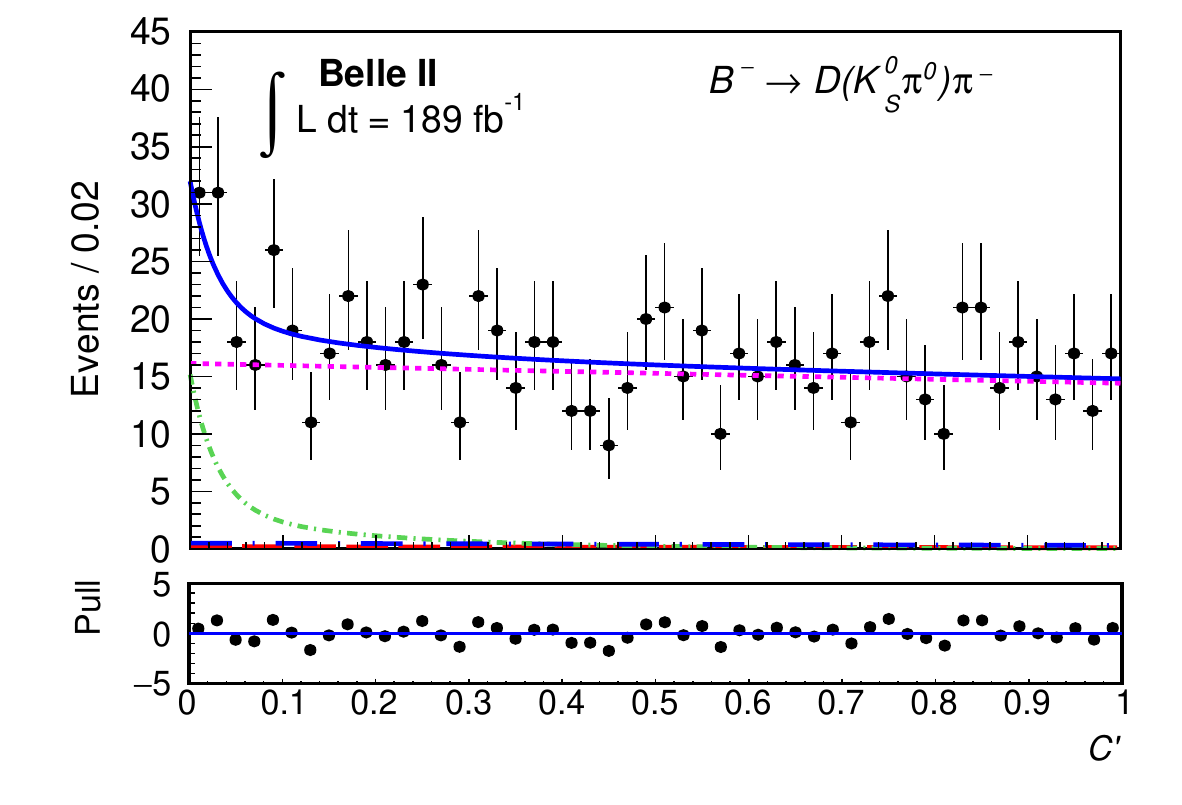} \put(87,60){(d)} \end{overpic}

    \begin{overpic}[width=0.46\textwidth]{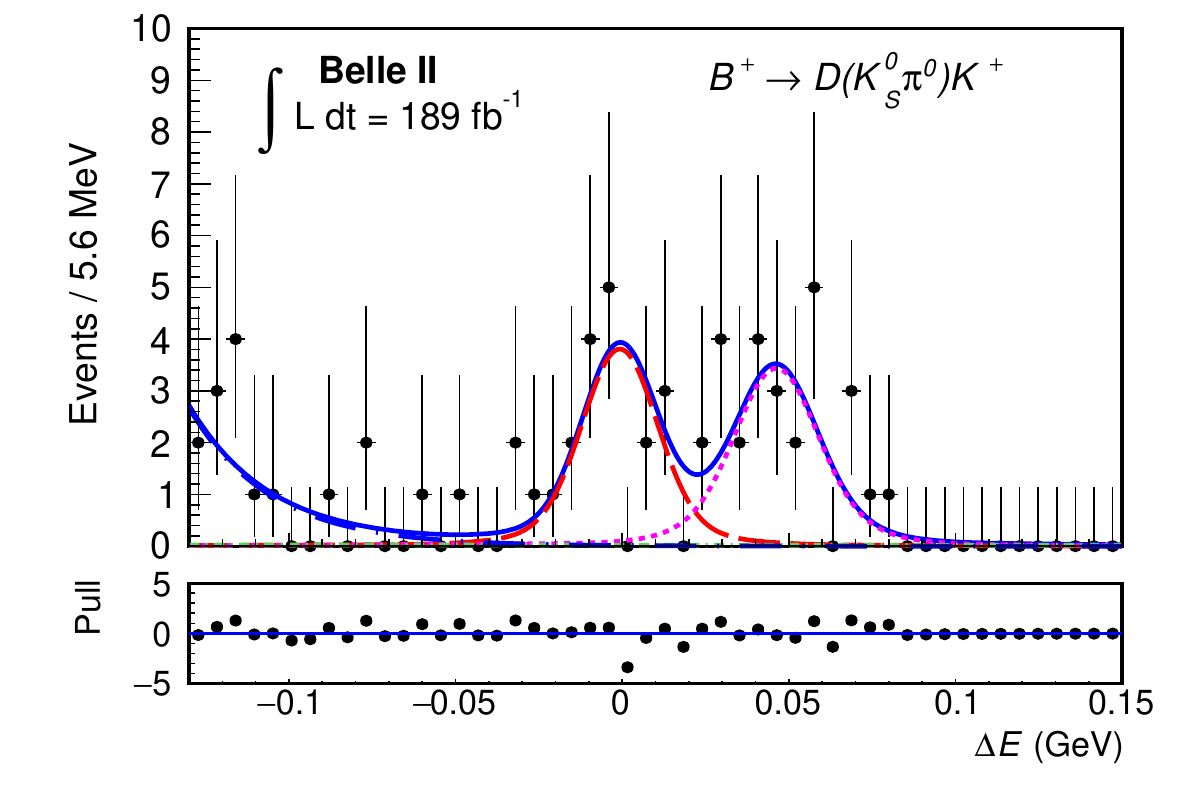}   \put(87,60){(e)} \end{overpic}
    \begin{overpic}[width=0.46\textwidth]{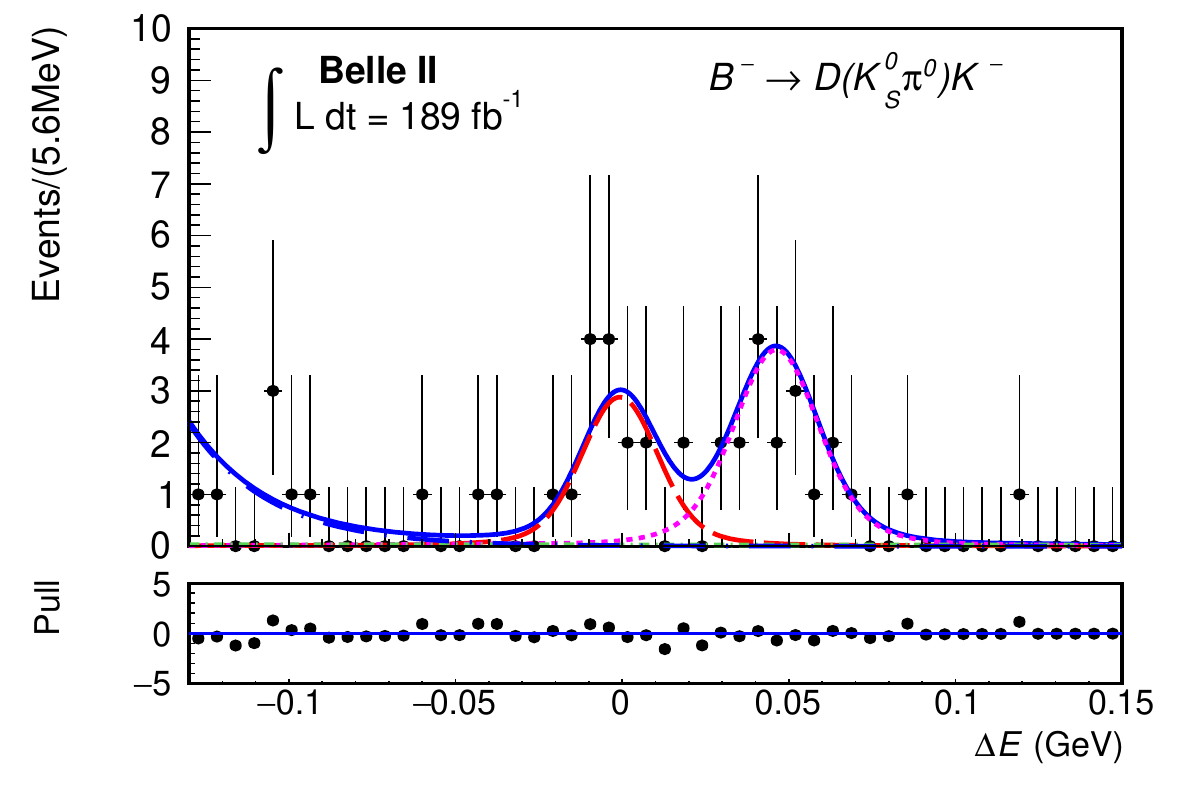}  \put(87,60){(f)} \end{overpic}

    \begin{overpic}[width=0.46\textwidth]{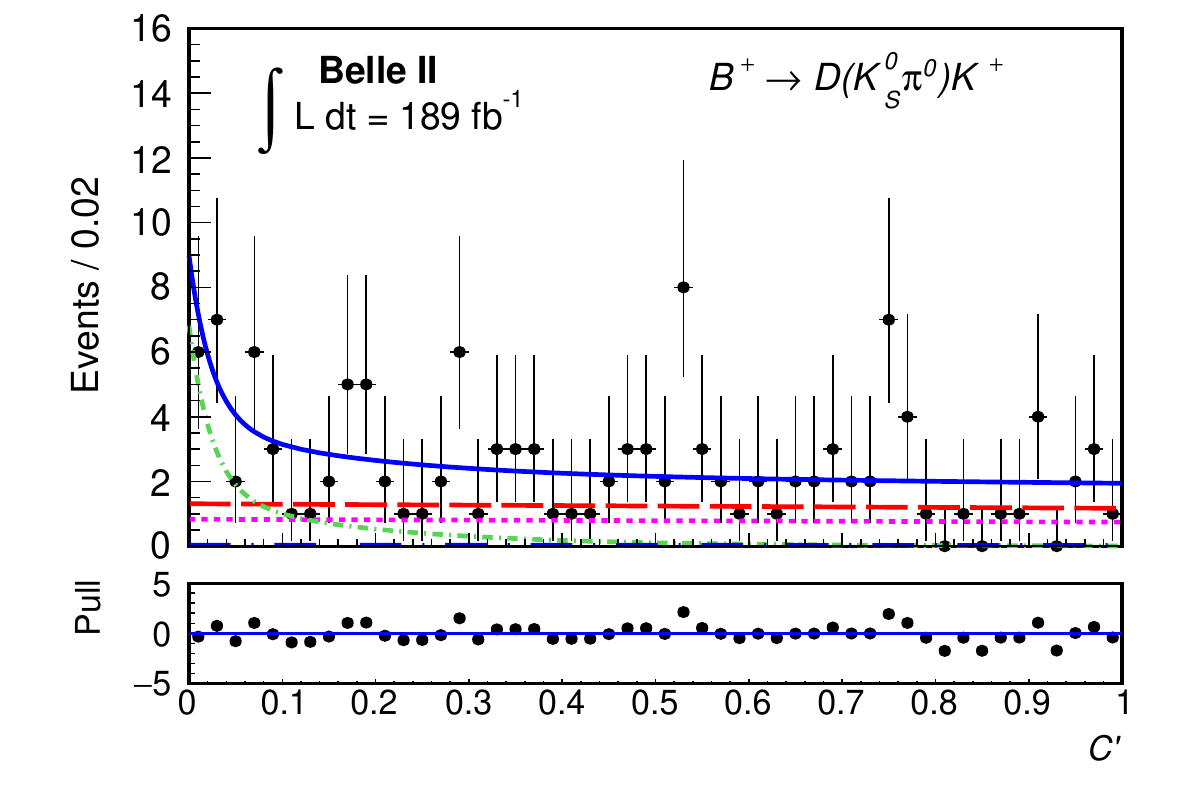}   \put(87,60){(g)} \end{overpic}
    \begin{overpic}[width=0.46\textwidth]{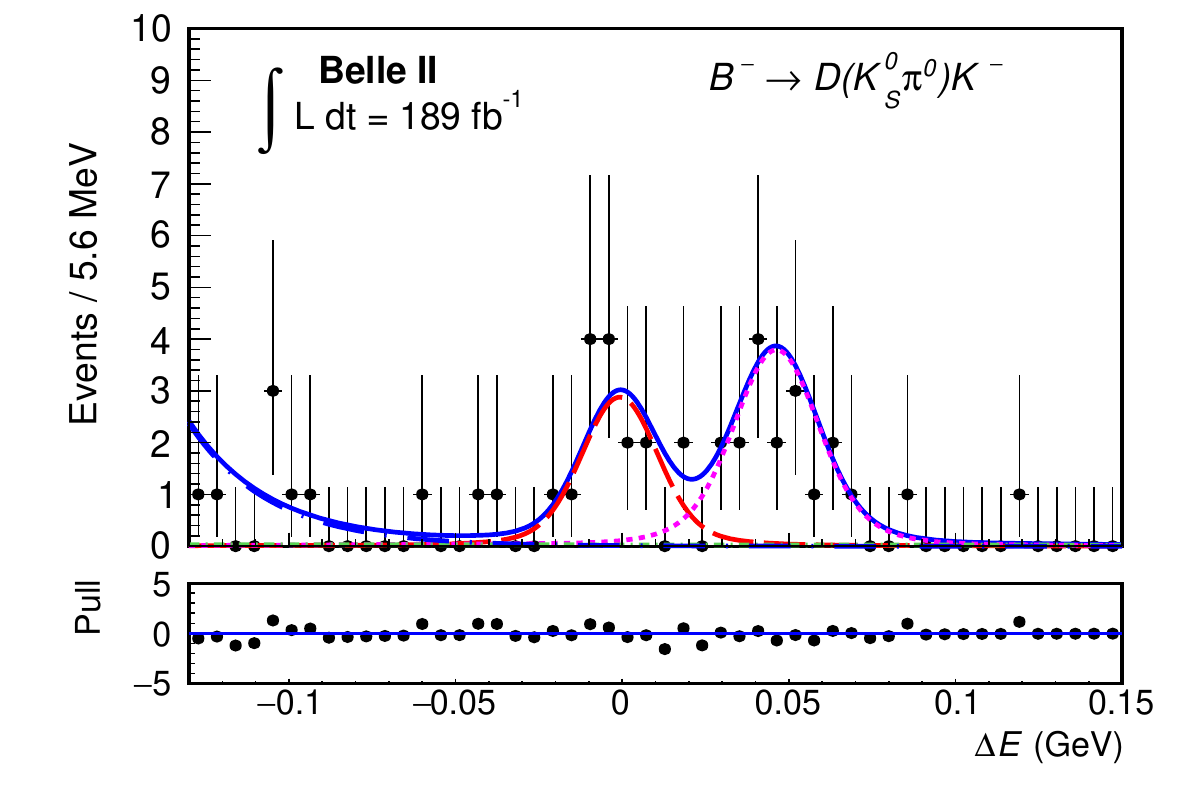}  \put(87,60){(h)} \end{overpic}

  \end{center}

  \caption{\ResultFigureCaption{\KS\piz}{Belle~II}}

  \label{fig:K0pi0_B2}
\end{figure}

\begin{figure}[!t]
  \begin{center}
    \includegraphics[width=0.46\textwidth]{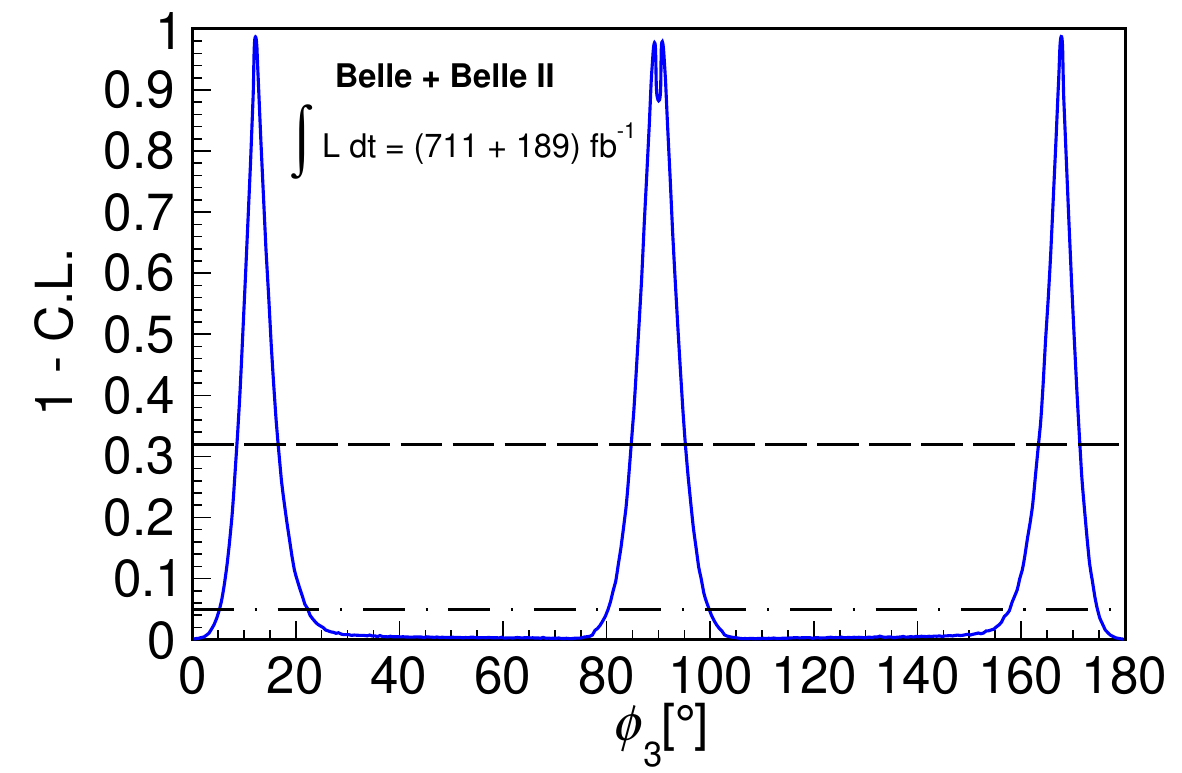}
    \includegraphics[width=0.46\textwidth]{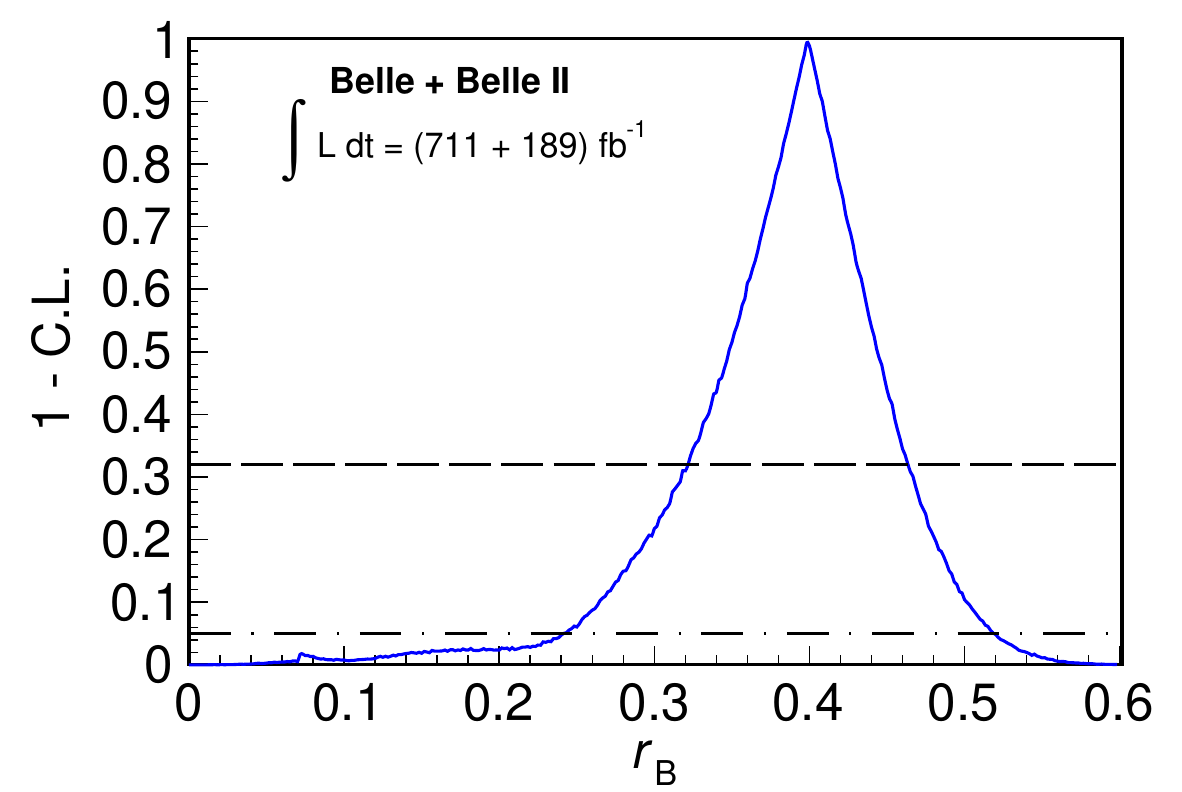}

    \caption{$p$-values ($1-\text{CL}$) as {functions} of $\phi_3$ (left) and $r_{\PB}$ (right). The dashed horizontal line shows the 68.3\% CL, and the dash-dotted line shows the 95.4\% CL.}

    \label{fig:gamma_results}
 \end{center}
\end{figure}

With these results for $R_{\CP\pm}$ and $\mathcal{A}_{\CP\pm}$, we constrain the angle $\phi_{3}$ using a frequentist approach as implemented in the \textsc{CkmFitter} package~\cite{CKMfitter}. 
Figure~\ref{fig:gamma_results} shows the resulting distributions of the $p$-value (the complement of the confidence level, $1-\text{CL}$). 
Given the $\delta_B \leftrightarrow \phi_3$ symmetry of equation~\ref{eq:r_glw} and~\ref{eq:a_glw}, the distribution for $\delta_{B}$ is identical to that for $\phi_3$.
Table~\ref{tab:summary_results} lists the 68.3\%- and 95.4\%-CL intervals for $\phi_{3}$ and $r_{B}$ for solutions with $\phi_3 \in [0^{\circ}, 180^{\circ}]$.
The large value measured for $\mathcal{R}_{\CP+}$ results in a relatively large $r_{\PB}$ which, in turn, gives a {stringent} constraint on $\phi_{3}$ due to the correlation between $r_{B}$ and $\phi_{3}$.

\begin{table}[!t]

  \caption{One-dimensional 68.3\% and 95.4\%  CL regions for $\phi_3$
    and $r_{\PB}$, for $\phi_3 \in [0^{\circ}, 180^{\circ}]$.}

  \label{tab:summary_results}

  \begin{center}
    \begin{tabular}{lll}
\hline      
      & 68.3\% CL
      & 95.4\% CL
      \\

\hline            
      & [8.5, 16.5]
      & [5.0, 22.0]
      \\

      $\phi_{3}$ (\si{\degree})
      & [84.5, 95.5]
      & [80.0, 100.0]
      \\

      & [163.3, 171.5]
      & [157.5, 175.0]
      \\
\hline            

      $r_B$
      &
      [0.321, 0.465]
      & [0.241, 0.522]
      \\
\hline                  
    \end{tabular}
  \end{center}
\end{table}

\section{Conclusion}
\label{sec:conclusion}

We measure the \CP asymmetries and ratios of branching-fraction ratios for $\Bpm \to \PD_{\CP\pm}\Kpm$ for the \CP-odd \PD final state $\Kp\Km$ and the \CP-even final state $\KS\piz$ with a combined analysis of the full Belle data set of \num{772e6} $\BBbar$ pairs and a Belle II data set containing \num{198e6} $\BBbar$ pairs. 
As expected, the asymmetries have opposite signs, 
{showing prominent \CP violation in $\Bpm \to \PD_{\CP} \Kpm$.}
The statistical and systematic precision of our results, based on a data set almost four times larger than the previous Belle measurement~\cite{GLW_Belle}, {is significantly improved}.
The results are consistent with those of the \babar~and LHCb experiments~\cite{babar_paper, lhcb_paper}. 
We obtain 68.3\%-CL intervals for the CKM angle $\phi_3$ and the amplitude ratio $r_{\PB}$:
\begin{alignat}{1}
  \phi_3 &\in [8.5^{\circ}, 16.5^{\circ}] \cup [84.5^{\circ}, 95.5^{\circ}] \cup [163.3^{\circ}, 171.5^{\circ}],\\
  r_{\PB} &\in [0.321, 0.465].
\end{alignat}

\newpage
\appendix

\section{ Correlation matrices }

Table~\ref{tab:stat_matrix} and~\ref{tab:syst_matrix} list the statistical and systematic correlation matrices {for} $\mathcal{A}_{\CP\pm}$ and $\mathcal{R}_{\CP\pm}$.
{We vary every fixed parameter randomly by Gaussian distribution for thousand times.}
{We repeat the fit with the varied values for every fixed parameter, which can result in Gaussian-like distributions of the measured observables.}
{The correlations are calculated by using those Gaussian-like distributions.}
These correlation matrices are used in the extraction of $\phi_{3}$, $\delta_{B}$ and $r_{B}$. 

\begin{table}[!hhh]

  \caption{Statistical correlation matrix of measured observables.}

  \label{tab:stat_matrix}

  \begin{center}
    \begin{tabular}{|c|cccc|}
\hline
      & $\mathcal{R}_{\CP+}$
      & $\mathcal{R}_{\CP-}$
      & $\mathcal{A}_{\CP+}$
      & $\mathcal{A}_{\CP-}$
      \\

\hline            
      $\mathcal{R}_{\CP+}$
      &  1
      &  $-0.081$
      &  0.060
      &  0.000
      \\

      $\mathcal{R}_{\CP-}$
      & 
      & 1
      & 0.000
      & 0.056
      \\

      $\mathcal{A}_{\CP+}$
      & 
      &
      & 1
      & 0.000
      \\
        
      $\mathcal{A}_{\CP-}$
      & 
      &
      &
      &  1
      \\
\hline                  
    \end{tabular}
  \end{center}
\end{table}

\begin{table}[!hhh]

  \caption{Systematic correlation matrix of measured observables.}

  \label{tab:syst_matrix}

  \begin{center}
    \begin{tabular}{|c|cccc|}
\hline
      & $\mathcal{R}_{\CP+}$
      & $\mathcal{R}_{\CP-}$
      & $\mathcal{A}_{\CP+}$
      & $\mathcal{A}_{\CP-}$
      \\

\hline            
      $\mathcal{R}_{\CP+}$
      &  1
      &  $-0.063$
      &  $0.342$
      &  $0.005$
      \\

      $\mathcal{R}_{\CP-}$
      & 
      & 1
      & $-0.128$
      & $-0.490$
      \\

      $\mathcal{A}_{\CP+}$
      & 
      &
      & 1
      & $0.542$
      \\
        
      $\mathcal{A}_{\CP-}$
      & 
      &
      &
      &  1
      \\
\hline                  
    \end{tabular}
  \end{center}
\end{table}

\section{\PD mass sidebands for the $\Bpm \to \PD_{\CP+}\Kpm$ mode}
\label{appendixB}

In Section~\ref{sec:extraction}, we use eight \PD mass sidebands of data to estimate the peaking background for the $\PD_{\CP+}$ mode. 
Table~\ref{tab:sidebands} lists the sideband mass ranges used in the Belle and Belle II analyses, respectively. 
{These sidebands are chosen to extend over the same range as the signal \PD region.}

\begin{table}[!h]

  \caption{\PD sideband mass regions for $\PD_{\CP+}$ mode, {in GeV/$c^{2}$ units}. }

  \label{tab:sidebands}

  \begin{center}
    
    \begin{tabular}{ccc}

\hline
      
       {Analysis}
      & {Lower sidebands}
      & {Upper sidebands}

      \\
\hline
      Belle
      & [1.67,1.71][1.71,1.75]
      & [1.90,1.94][1.94,1.98]
      \\

      & [1.75,1.79][1.79,1.83]
      & [1.98,2.02][2.02,2.06]
      \\

\hline

      Belle II
      & [1.706,1.732][1.732,1.758]
      & [1.758,1.784][1.784,1.810]
      \\

      & [1.920,1.946][1.946,1.972]
      & [1.972,1.998][1.998,2.024]
      \\

\hline
    \end{tabular}
    
  \end{center}
  
\end{table}

\begin{figure}[!t]
  \begin{center}

    \begin{overpic}[width=0.46\textwidth]{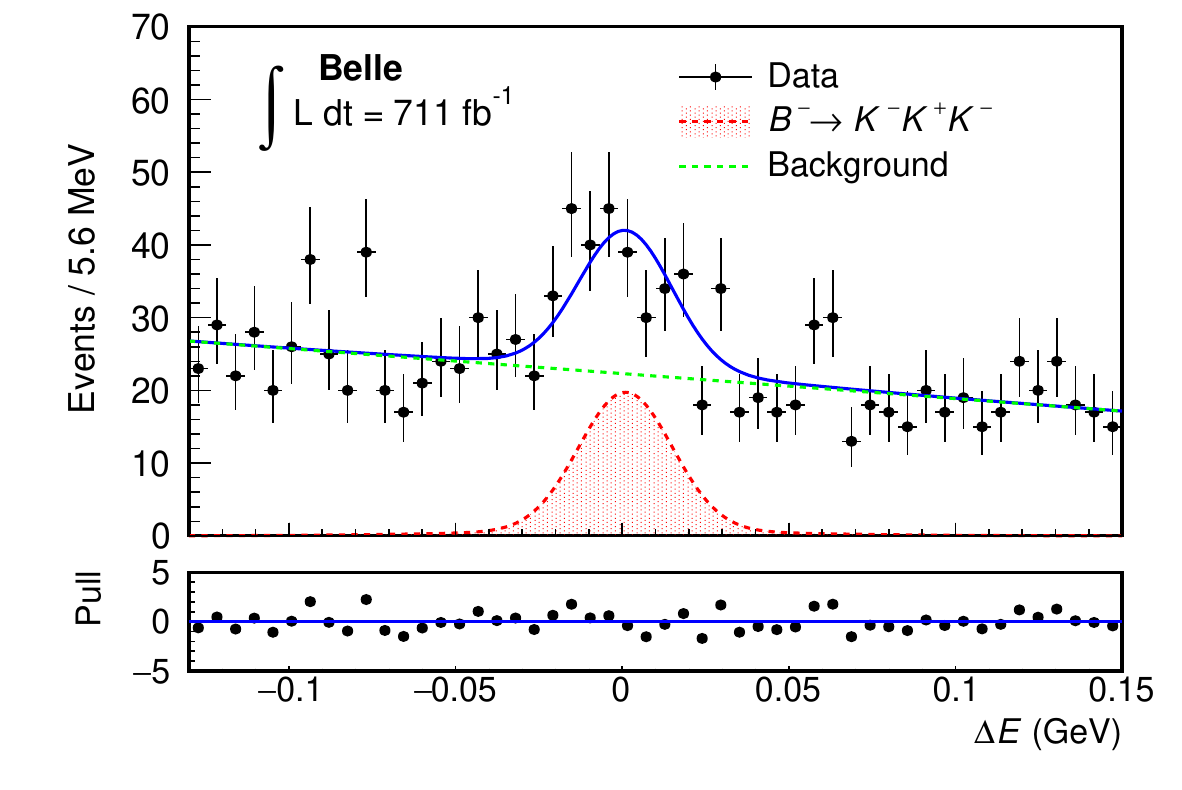}  \put(87,59){(a)} \end{overpic}
    \begin{overpic}[width=0.46\textwidth]{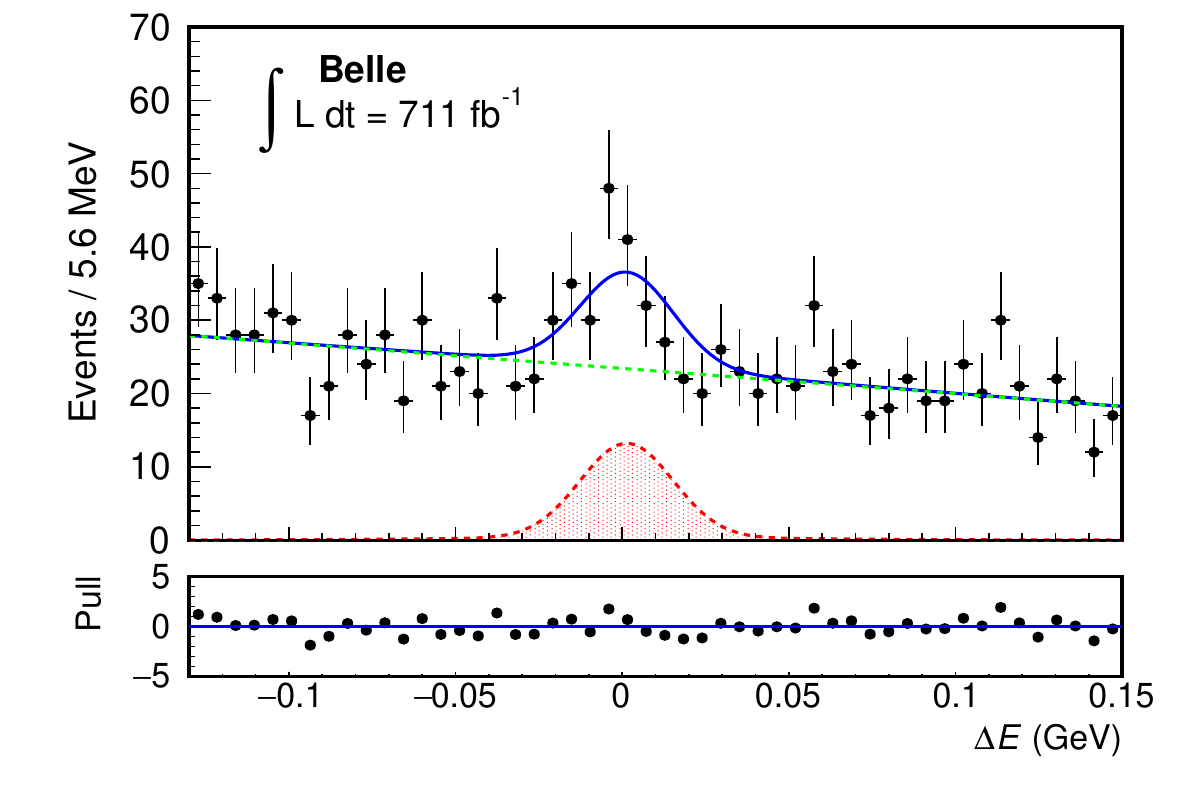} \put(87,59){(b)} \end{overpic}

    \begin{overpic}[width=0.46\textwidth]{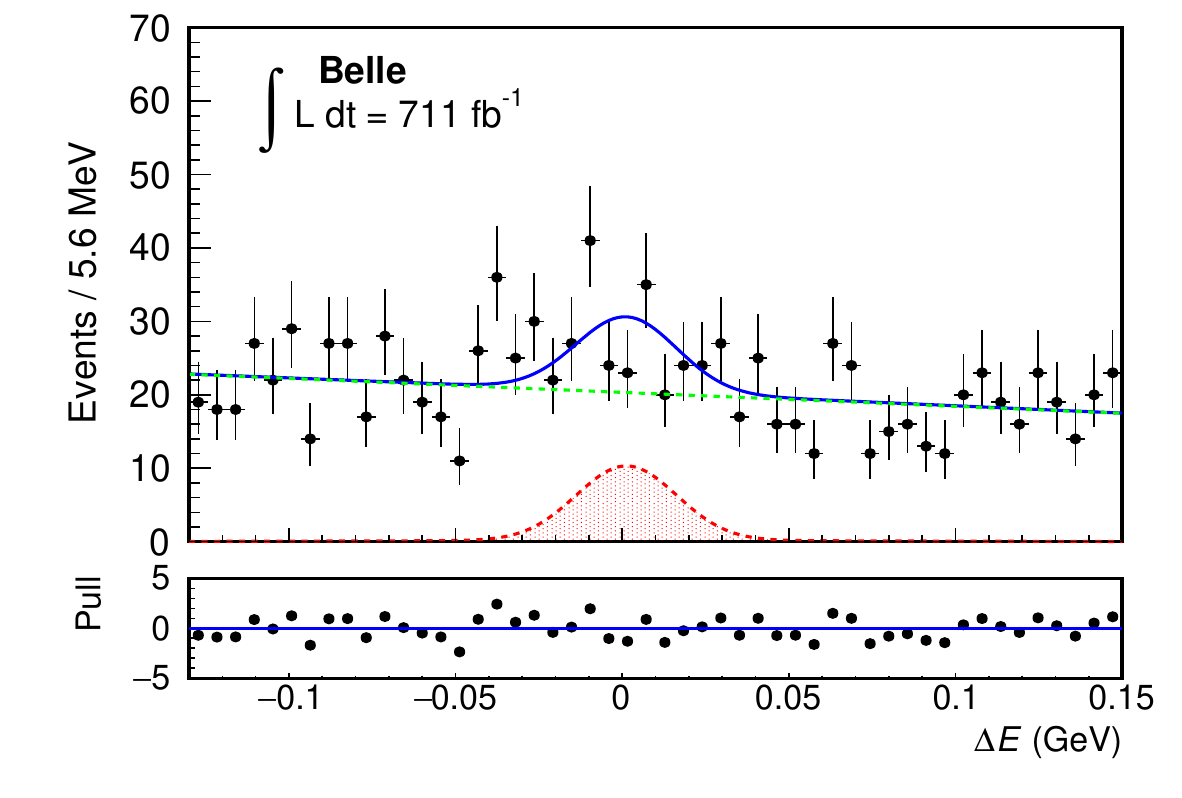}  \put(87,59){(c)} \end{overpic}
    \begin{overpic}[width=0.46\textwidth]{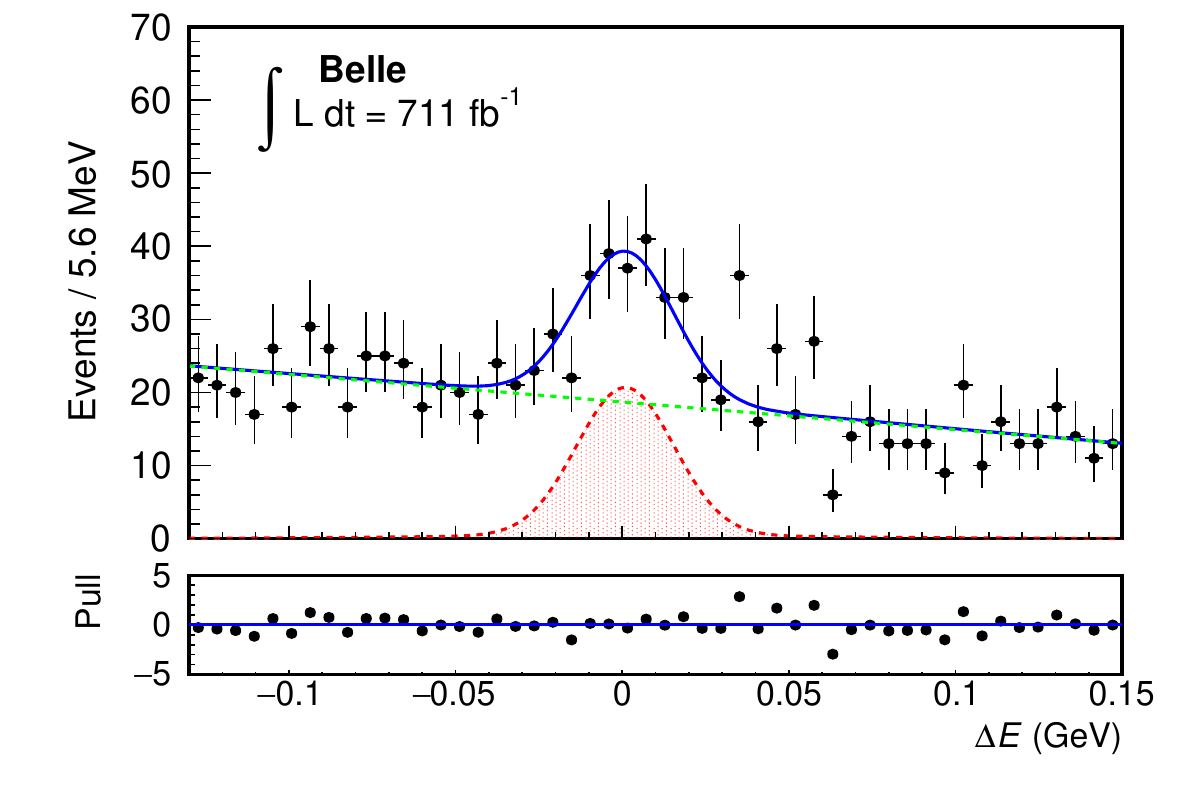} \put(87,59){(d)} \end{overpic}

    \begin{overpic}[width=0.46\textwidth]{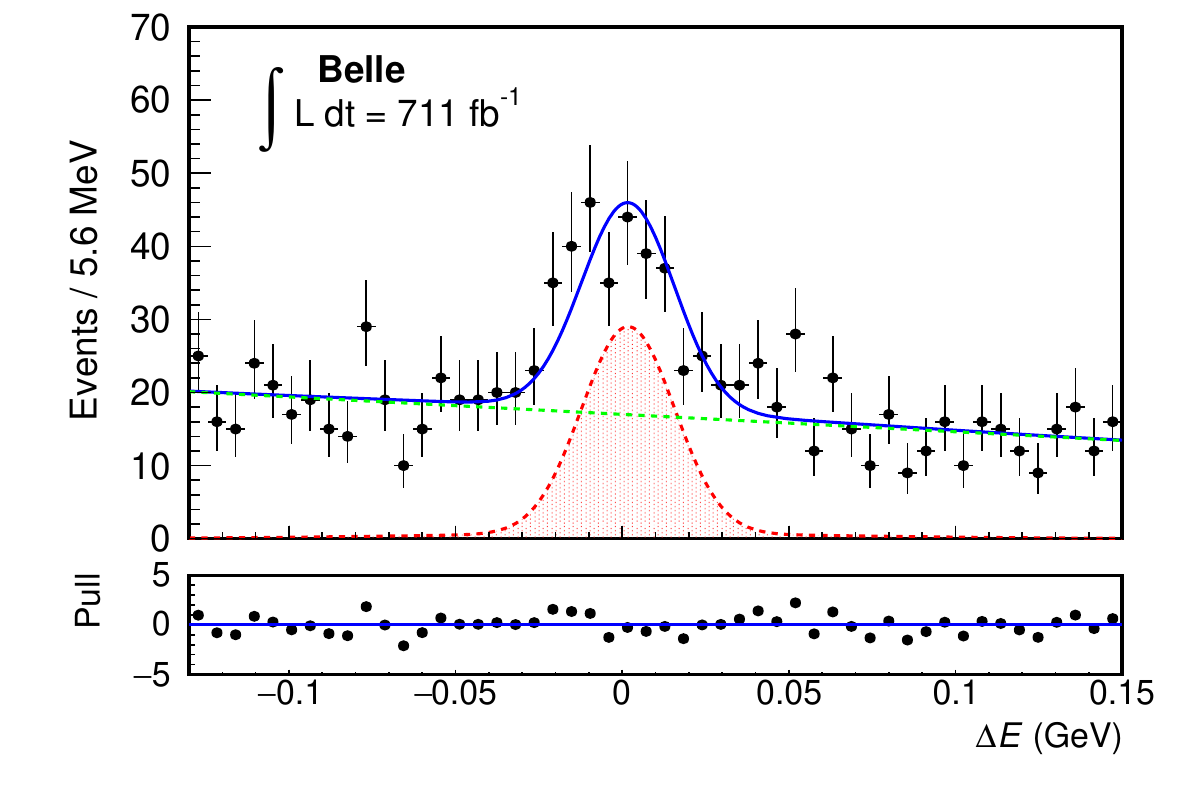}   \put(87,59){(e)} \end{overpic}
    \begin{overpic}[width=0.46\textwidth]{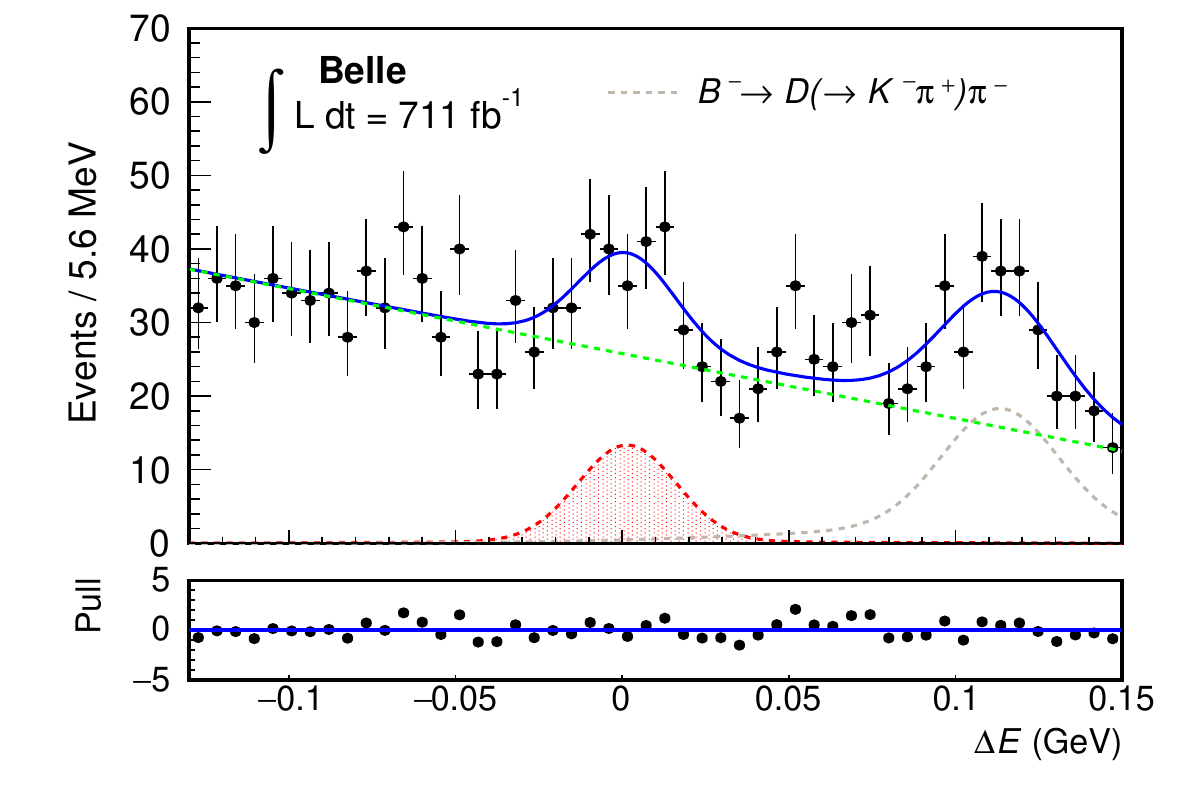}  \put(87,59){(f)} \end{overpic}

    \begin{overpic}[width=0.46\textwidth]{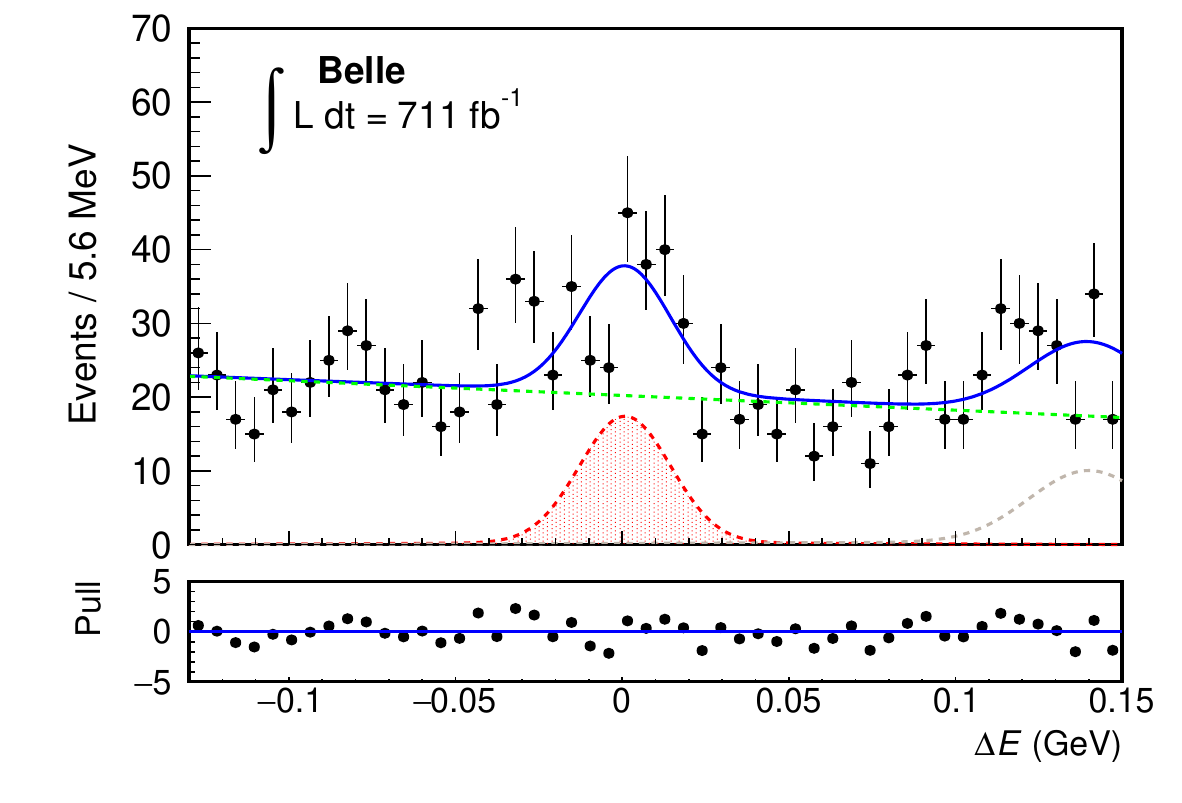}   \put(87,59){(g)} \end{overpic}
    \begin{overpic}[width=0.46\textwidth]{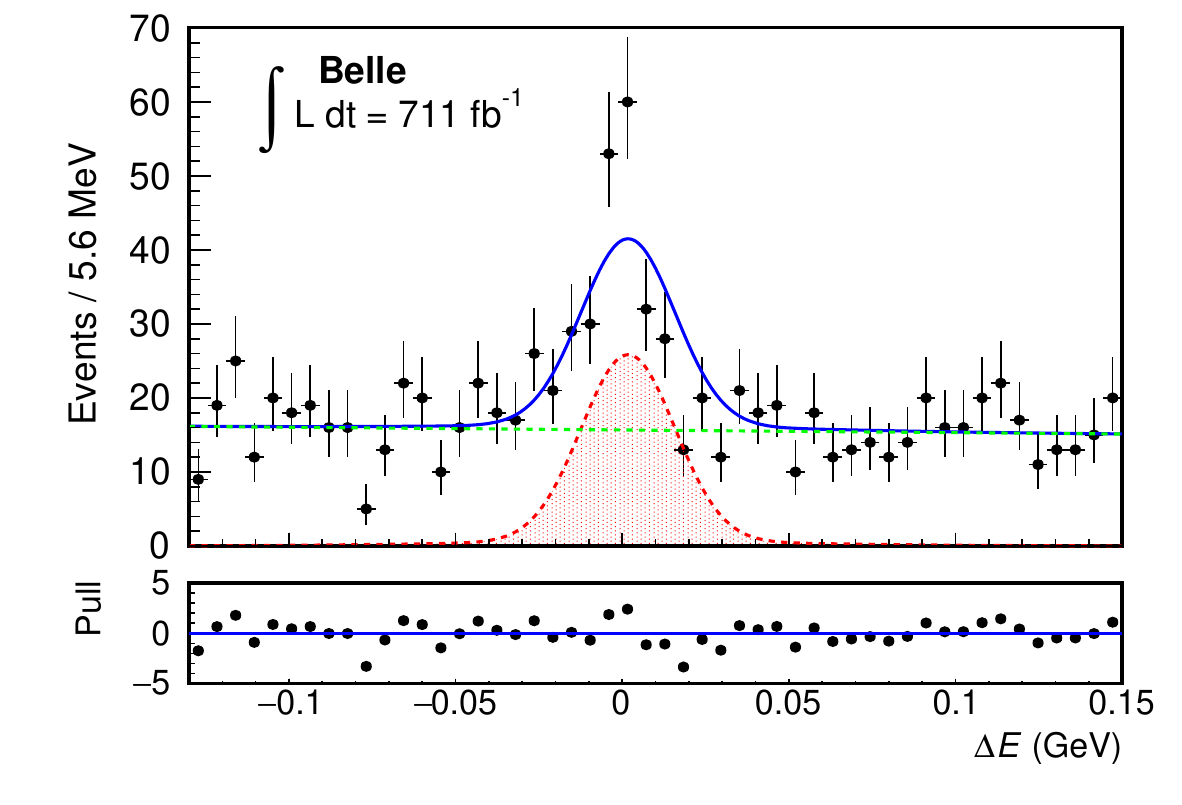}  \put(87,59){(h)} \end{overpic}

  \end{center}
  \caption{$\Delta E$ distributions of $\Bpm \to \PD(\to \Km \Kp) \Kpm$ in Belle data with the fit projections overlaid for the lower sidebands [1.67, 1.71] $\mathrm{GeV}/c^{2}$ (a), [1.71, 1.75] $\mathrm{GeV}/c^{2}$ (b), [1.75, 1.79] $\mathrm{GeV}/c^{2}$ (c), and [1.79, 1.83] $\mathrm{GeV}/c^{2}$ (d), and the upper sidebands [1.90, 1.94] $\mathrm{GeV}/c^{2}$ (e), [1.94, 1.98] $\mathrm{GeV}/c^{2}$ (f), [1.98, 2.02] $\mathrm{GeV}/c^{2}$ (g), and [2.02, 2.06] $\mathrm{GeV}/c^{2}$ (h).}


  \label{fig:sideband_fit}
\end{figure}

Fig~\ref{fig:sideband_fit} shows distributions and fit-result {projections} {in the data sidebands} for the Belle analysis. 
We obtain the peaking background yield for each sideband and interpolate those yields {linearly}. 

The Belle II data sample used in this analysis has only an integrated luminosity of \SI{189}{fb^{-1}}, {which is insufficient} to estimate the yield of peaking background. 
Instead we obtain the Belle II yield by scaling the Belle yield by the {reconstruction} efficiencies $\epsilon$ of $\Bm \to \Km \Kp \Km$ in simulated data and the luminosities~($L$) of Belle and Belle~II data samples:
\begin{equation}
    Y_{B2}(\Bm \to \Km \Kp \Km) = Y_{B}(\Bm \to \Km \Kp \Km) \frac{\epsilon_{B2}L_{B2}}{\epsilon_{B}L_{B}},
\end{equation}
where subscripts $B$ and $B2$ stand for Belle and Belle~II, respectively.

\clearpage

\section*{Acknowledgments}

This work, based on data collected using the Belle II detector, which was built and commissioned prior to March 2019, was supported by
Science Committee of the Republic of Armenia Grant No.~20TTCG-1C010;
Australian Research Council and Research Grants
No.~DP200101792, 
No.~DP210101900, 
No.~DP210102831, 
No.~DE220100462, 
No.~LE210100098, 
and
No.~LE230100085; 
Austrian Federal Ministry of Education, Science and Research,
Austrian Science Fund
No.~P~31361-N36
and
No.~J4625-N,
and
Horizon 2020 ERC Starting Grant No.~947006 ``InterLeptons'';
Natural Sciences and Engineering Research Council of Canada, Compute Canada and CANARIE;
National Key R\&D Program of China under Contract No.~2022YFA1601903,
National Natural Science Foundation of China and Research Grants
No.~11575017,
No.~11761141009,
No.~11705209,
No.~11975076,
No.~12135005,
No.~12150004,
No.~12161141008,
and
No.~12175041,
and Shandong Provincial Natural Science Foundation Project~ZR2022JQ02;
the Czech Science Foundation Grant No.~22-18469S;
European Research Council, Seventh Framework PIEF-GA-2013-622527,
Horizon 2020 ERC-Advanced Grants No.~267104 and No.~884719,
Horizon 2020 ERC-Consolidator Grant No.~819127,
Horizon 2020 Marie Sklodowska-Curie Grant Agreement No.~700525 ``NIOBE''
and
No.~101026516,
and
Horizon 2020 Marie Sklodowska-Curie RISE project JENNIFER2 Grant Agreement No.~822070 (European grants);
L'Institut National de Physique Nucl\'{e}aire et de Physique des Particules (IN2P3) du CNRS
and
L'Agence Nationale de la Recherche (ANR) under grant ANR-21-CE31-0009 (France);
BMBF, DFG, HGF, MPG, and AvH Foundation (Germany);
Department of Atomic Energy under Project Identification No.~RTI 4002 and Department of Science and Technology (India);
Israel Science Foundation Grant No.~2476/17,
U.S.-Israel Binational Science Foundation Grant No.~2016113, and
Israel Ministry of Science Grant No.~3-16543;
Istituto Nazionale di Fisica Nucleare and the Research Grants BELLE2;
Japan Society for the Promotion of Science, Grant-in-Aid for Scientific Research Grants
No.~16H03968,
No.~16H03993,
No.~16H06492,
No.~16K05323,
No.~17H01133,
No.~17H05405,
No.~18K03621,
No.~18H03710,
No.~18H05226,
No.~19H00682, 
No.~22H00144,
No.~22K14056,
No.~23H05433,
No.~26220706,
and
No.~26400255,
the National Institute of Informatics, and Science Information NETwork 5 (SINET5), 
and
the Ministry of Education, Culture, Sports, Science, and Technology (MEXT) of Japan;  
National Research Foundation (NRF) of Korea Grants
No.~2016R1\-D1A1B\-02012900,
No.~2018R1\-A2B\-3003643,
No.~2018R1\-A6A1A\-06024970,
No.~2019R1\-I1A3A\-01058933,
No.~2021R1\-A6A1A\-03043957,
No.~2021R1\-F1A\-1060423,
No.~2021R1\-F1A\-1064008,
No.~2022R1\-A2C\-1003993,
and
No.~RS-2022-00197659,
Radiation Science Research Institute,
Foreign Large-size Research Facility Application Supporting project,
the Global Science Experimental Data Hub Center of the Korea Institute of Science and Technology Information
and
KREONET/GLORIAD;
Universiti Malaya RU grant, Akademi Sains Malaysia, and Ministry of Education Malaysia;
Frontiers of Science Program Contracts
No.~FOINS-296,
No.~CB-221329,
No.~CB-236394,
No.~CB-254409,
and
No.~CB-180023, and SEP-CINVESTAV Research Grant No.~237 (Mexico);
the Polish Ministry of Science and Higher Education and the National Science Center;
the Ministry of Science and Higher Education of the Russian Federation,
Agreement No.~14.W03.31.0026, and
the HSE University Basic Research Program, Moscow;
University of Tabuk Research Grants
No.~S-0256-1438 and No.~S-0280-1439 (Saudi Arabia);
Slovenian Research Agency and Research Grants
No.~J1-9124
and
No.~P1-0135;
Agencia Estatal de Investigacion, Spain
Grant No.~RYC2020-029875-I
and
Generalitat Valenciana, Spain
Grant No.~CIDEGENT/2018/020
Ministry of Science and Technology and Research Grants
No.~MOST106-2112-M-002-005-MY3
and
No.~MOST107-2119-M-002-035-MY3,
and the Ministry of Education (Taiwan);
Thailand Center of Excellence in Physics;
TUBITAK ULAKBIM (Turkey);
National Research Foundation of Ukraine, Project No.~2020.02/0257,
and
Ministry of Education and Science of Ukraine;
the U.S. National Science Foundation and Research Grants
No.~PHY-1913789 
and
No.~PHY-2111604, 
and the U.S. Department of Energy and Research Awards
No.~DE-AC06-76RLO1830, 
No.~DE-SC0007983, 
No.~DE-SC0009824, 
No.~DE-SC0009973, 
No.~DE-SC0010007, 
No.~DE-SC0010073, 
No.~DE-SC0010118, 
No.~DE-SC0010504, 
No.~DE-SC0011784, 
No.~DE-SC0012704, 
No.~DE-SC0019230, 
No.~DE-SC0021274, 
No.~DE-SC0022350, 
No.~DE-SC0023470; 
and
the Vietnam Academy of Science and Technology (VAST) under Grant No.~DL0000.05/21-23.

These acknowledgements are not to be interpreted as an endorsement of any statement made
by any of our institutes, funding agencies, governments, or their representatives.

We thank the SuperKEKB team for delivering high-luminosity collisions;
the KEK cryogenics group for the efficient operation of the detector solenoid magnet;
the KEK computer group and the NII for on-site computing support and SINET6 network support;
and the raw-data centers at BNL, DESY, GridKa, IN2P3, INFN, and the University of Victoria for offsite computing support.

\end{document}